\documentclass[preprint,3p,times,12pt]{elsarticle}

\usepackage{amssymb}
\usepackage{amsmath}
\usepackage{rotating}
\usepackage{siunitx}
\usepackage{dirtytalk}
\usepackage{graphicx}
\usepackage{dcolumn}
\usepackage{bm}
\usepackage{hyperref}
\usepackage{natbib}
\usepackage{stackengine}
\usepackage{pgfplots}
\usepackage{adjustbox}
\usepackage{makecell, multirow, tabularx}

\usepackage{float}
\usepackage{caption}
\usepackage{subcaption}
\usepackage[normalem]{ulem}
\usepackage{pdfpages}

\pgfplotsset{compat=1.18}

\usepackage{tikz}
\usetikzlibrary{shapes.geometric, shapes, arrows.meta, positioning, fit, shadows, calc, backgrounds}

\usepackage{enumitem} 

\pgfplotsset{compat=1.18}

\journal{}

\begin{document}

\begin{frontmatter}

\title{Intraday Limit Order Price Change Transition Dynamics Across Market Capitalizations Through Markov Analysis}

\author[label1]{Salam Rabindrajit Luwang}
\ead{salamrabindrajit@gmail.com}

\author[label1]{Kundan Mukhia}
\ead{kundanmukhia07@gmail.com}

\author[label1]{Buddha Nath Sharma}
\ead{bnsharma09@yahoo.com}

\author[label1]{Md. Nurujjaman}
\ead{md.nurujjaman@nitsikkim.ac.in}

\author[label2]{Anish Rai}
\ead{anishrai412@gmail.com}

\author[label3]{Filippo Petroni}
\ead{fpetroni@luiss.it}


\affiliation[label1]{organization={Department of Physics, National Institute of Technology Sikkim},
            city={Ravangla},
            postcode={737139}, 
            state={Sikkim},
            country={India}}

\affiliation[label2]{organization={Chennai Mathematical Institute},
            city={Kelambakkam},
            postcode={603103}, 
            state={Tamil Nadu},
            country={India}}

\affiliation[label3]{organization={Department of Economics, University G. d’Annunzio of Chieti-Pescara},
            city={Pescara},
            postcode={65127}, 
            state={Abruzzo},
            country={Italy}}

\begin{abstract}
A quantitative understanding of the stochastic dynamics in limit order price changes is essential for meaningful advances in market microstructure research and effective execution strategy design. This paper presents the first comprehensive empirical analysis of intraday limit order price change transition dynamics, treating ask and bid orders separately across different market capitalization tiers. Using high-frequency tick data from NASDAQ100 stocks, we employ a discrete-time Markov chain framework to analyze the evolution of price adjustments throughout the trading day. We categorize consecutive price changes into nine distinct states and estimate transition probability matrices (TPMs) for six intraday intervals across High ($\mathtt{HMC}$), Medium ($\mathtt{MMC}$), and Low ($\mathtt{LMC}$) market capitalization stocks. Elememt-wise comparison of TPMs reveals systematic intraday patterns:  price inertia i.e. self-transition probability, peaks during opening and closing hours, stabilizing at lower levels during midday. A pronounced capitalization gradient is also observed: $\mathtt{HMC}$ stocks exhibit the strongest price inertia, while $\mathtt{LMC}$ stocks demonstrate significantly lower stability and pronounced bid-ask spread. Markov chain metrics, including spectral gap, entropy rate, and mean recurrence times quantify these dynamics. Clustering analysis identifies three distinct temporal phases on the bid side -- Opening, Midday, and Closing and four phases on the ask side -- Opening, Midday, Pre-Close, and Close, indicating that sellers initiate end-of-day positioning strategies earlier than buyers. Stationary distributions reveal that limit order dynamics are predominantly characterized by neutral and mild price changes. Furthermore, Jensen-Shannon divergence computed between stationary distributions across time-intervals confirms the closing hour as the most distinct phase, with capitalization modulating the intensity of temporal contrasts and the degree of bid-ask asymmetry. These findings advance the understanding of evolving intraday limit order pricing behavior, offering direct applications for capitalization-aware and time-adaptive execution algorithms and risk management frameworks.
\end{abstract}



\end{frontmatter}

\newpage 

\section{Introduction}
\label{sec:Intro}
In equity markets, limit orders represent a critical component of the trading ecosystem, allowing traders to specify precise execution prices while contributing to market liquidity~\cite{bouchaud2002statistical, cont2010stochastic, hautsch2012market, gould2013limit}. Unlike market orders that execute immediately at prevailing prices, limit orders remain active in the order book until matched or canceled. This creates a dynamic environment in which the intraday evolution of limit order prices is driven by traders' continuous strategic adjustments to liquidity conditions, information arrival, and prevailing market sentiment~\cite{madhavan2000market, biais2005market, rocsu2009dynamic, cont2011statistical}. These adjustments manifest as sequences of discrete price revisions, generating complex stochastic dynamics with systematic intraday regularities~\cite{chordia2005evidence, goyenko2009liquidity}. 

The nature of these adjustments is fundamentally asymmetric: buy limit orders compete by improving prices toward the ask, while sell limit orders compete by lowering prices toward the bid, reflecting their opposing economic objectives and inventory considerations~\cite{hasbrouck2007empirical, damico2013semimc}. Understanding how these distinct buy- and sell-side price revision mechanisms evolve over the trading day therefore represents a central challenge in market microstructure research, with direct implications for algorithmic trading design, liquidity provision, and market efficiency.

Beyond intraday temporal effects, market capitalization plays a crucial role in shaping limit order pricing behavior and the associated price change dynamics. High-capitalization stocks typically exhibit frequent but relatively small price adjustments, reflecting dense order books, intense competition among liquidity providers, and narrow bid--ask spreads~\cite{biais2005market, bouchaud2002statistical, cont2011statistical}. In such environments, traders engage in fine price shading to maintain queue priority while minimizing execution costs~\cite{rocsu2009dynamic, gould2013limit}. Medium-capitalization stocks display intermediate liquidity conditions, with more pronounced variability in price changes due to less predictable order flow and reduced market depth~\cite{cont2010stochastic, hautsch2012market}. In contrast, low-capitalization stocks are characterized by sparse order books, wider spreads, and heightened information asymmetry, leading to less frequent but larger price revisions, often triggered by liquidity shocks or information events~\cite{goyenko2009liquidity, chordia2005evidence, madhavan2000market}. These capitalization-dependent features suggest that limit order price changes are governed by distinct stochastic mechanisms across market segments, thereby motivating a stratified analytical approach.

A substantial body of literature has documented intraday regularities in financial markets across multiple dimensions, including trading volume~\cite{kluger2011intraday}, transaction patterns~\cite{ho1991behaviour}, liquidity provision~\cite{brockman1999analysis}, and bid--ask spread dynamics~\cite{lee1993spreads}. Market microstructure studies have further emphasized the role of order types in shaping these patterns~\cite{biais2005market, gould2013limit, cont2010stochastic, hautsch2012market}, stimulating extensive research on order book resilience, transaction costs, and strategic liquidity provision~\cite{bouchaud2002statistical, rocsu2009dynamic}. Within this framework, empirical evidence shows that limit order submission strategies vary systematically over the trading day~\cite{ellul2007order, biais1995empirical}, with distinct regimes at the market open, midday, and close~\cite{admati1988theory, garvey2007intraday}. The rise of high-frequency and algorithmic trading has further amplified these temporal patterns, leading to clustering in price revisions and sequential price improvement behavior~\cite{hautsch2012market, cont2014price, cartea2015algorithmic}. Despite this extensive literature, a critical gap remains. To the best of our knowledge, no existing study has systematically examined the intraday evolution of limit order price change transitions, treating buy and sell orders separately and jointly accounting for both intraday timing and market capitalization effects.

Addressing this gap requires a modeling framework capable of capturing the discrete, state-dependent nature of sequential price adjustments. Stochastic approaches based on Markov chains are particularly well suited for this purpose, having proven effective in the analysis of discrete financial processes~\cite{dewachter2001can, rydberg2003dynamics} and widely applied in market microstructure settings~\cite{cont2012order,d2011semi,damico2013semimc,fitzpatrick2020thermodynamics}. While our prior works~\cite{rabindrajit2024high, luwang2025intraday} employed discrete-time Markov chains to study intraday order transitions, it treated order types in aggregate and did not explicitly focus on the dynamics of price revisions within limit orders. The present study extends this line of research by shifting the analytical focus to the intraday transition dynamics of limit order price changes themselves. Specifically, we investigate how these transitions evolve across trading hours and market capitalization tiers, and how they differ between the bid and ask sides of the order book.

Building on our established methodology~\cite{rabindrajit2024high, luwang2025intraday}, we develop a discrete-time Markov chain (DTMC) framework to analyze limit order price change transitions using high-frequency tick-by-tick data for NASDAQ100 stocks. Consecutive limit order price changes are classified into nine discrete states based on their magnitude and direction, yielding categorical time series that capture the sequential nature of pricing decisions. For each intraday interval $\tau \in \{\mathtt{T_1, T_2, \ldots, T_6}\}$ and market capitalization tier $c \in \{\mathtt{HMC, MMC, LMC}\}$, we estimate transition probability matrices $P_{ij}^{(\tau,c)}$ describing the likelihood of transitioning from state $i$ to state $j$. Treating bid and ask limit orders separately allows us to explicitly account for directional asymmetries in price revision behavior.

Employing this DTMC framework, we conduct a comprehensive empirical analysis of transition probabilities, stationary distributions, and key Markov chain metrics, including spectral gap, entropy rate, and mean recurrence time. We further investigate cross-interval similarities using clustering techniques and Jensen--Shannon divergence. Our results reveal pronounced intraday regularities, with price inertia peaking at the market open and close, and a clear capitalization gradient whereby high-capitalization stocks exhibit the greatest stability and low-capitalization stocks the highest degree of dynamism. Clustering analysis uncovers distinct Opening, Midday, and Closing regimes on both sides of the book, with an additional pre-close phase emerging on the ask side. Stationary distributions indicate that limit order dynamics are dominated by neutral and mild price changes, while the closing hour stands out as the most distinct temporal regime, with capitalization modulating the intensity of intraday contrasts and bid--ask asymmetry.

The primary contribution of this paper is to provide the first data-driven Markov chain analysis of intraday limit order price change transition dynamics across market capitalization tiers. By extending existing intraday order transition frameworks to the granular pricing behavior within limit orders, our study offers new empirical insights into how time of day, market size, and order direction jointly shape limit order price dynamics. These findings have direct implications for time-adaptive and capitalization-aware execution strategies. The remainder of the paper is organized as follows. Section~\ref{sec:Data} describes the data and intraday segmentation. Section~\ref{sec:Method} presents the methodological framework. Section~\ref{sec:Results} reports the empirical findings, while Section~\ref{disscusion} discusses their robustness and implications. Section~\ref{sec:Conc} concludes and outlines directions for future research.

\section{Data}
\label{sec:Data}

\subsection{Data Description}
\label{sec:DD}
The availability of high-frequency, micro-level stock market data has unlocked unprecedented capabilities for granular empirical research in financial markets. We use tick-by-tick order submission data obtained from \href{https://www.algoseek.com}{Algoseek} in this study. The data cover all order types placed from 04{:}00{:}00 to 20{:}00{:}00 Eastern Standard Time (EST) for stocks listed in the NASDAQ100 index. Each trading day typically contains hundreds of millions of records, with raw CSV files of roughly 20--40 GB. Table~\ref{tab:DataSample} presents the dataset structure, with eight columns: Date, Timestamp, Order ID, Event Type, Ticker Symbol, Price, Quantity, and Exchange. 

\begin{table}[H]
\centering
\caption{Sample dataset illustrating high-frequency tick-by-tick order data for stocks listed in the NASDAQ100.}
\label{tab:DataSample}
\begin{adjustbox}{width=\textwidth}
\begin{tabular}{|c|c|c|c|c|c|c|l|} 
\hline 
\textbf{Date} & \textbf{Timestamp} & \textbf{Order Id.} & \textbf{Event Type} & \textbf{Ticker} & \textbf{Price} & \textbf{Quantity} & \textbf{Exchange} \\
\hline 
2018-11-07 & 4:00:00.122 & 11872 & ADD-ASK & AAPL & 173.00 & 500 & NASDAQ \\
\hline 
2018-11-07 & 4:00:00.255 & 12654 & ADD-BID & AAPL & 186.99 & 100 & NASDAQ \\
\hline
2018-11-07 & 4:00:00.123 & 12865 & FILL-BID & XLF  & 0 & 200 & NASDAQ \\
\hline
... & ... & ... & ... & ... & ... & ... & ... \\ 
\hline
2018-11-07 & 9:30:00.145 & 76543 & DELETE-BID & GOOGL & 0 & 400 & NASDAQ \\ 
\hline
2018-11-07 & 9:30:01.678 & 81624 & CANCEL-BID & INTC & 0 & 500 & NASDAQ \\
\hline
... & ... & ... & ... & ... & ... & ... & ... \\ 
\hline
2018-11-07 & 16:00:00.000 & 116752 & EXECUTE-BID & AMD & 0 & 50 & NASDAQ \\ 
\hline
... & ... & ... & ... & ... & ... & ... & ... \\ 
\hline
2018-11-06 & 20:00:00.000 & 547324 & DELETE-ASK & NVDA & 0 & 40 & NASDAQ \\
\hline
\end{tabular}
\end{adjustbox}
\end{table}

Our empirical analysis focuses on the two event types -- ADD-ASK and ADD-BID order types, which correspond to the submission of new ask (selling) and bid (buying) limit orders, respectively. For each order type, we compute price differences between consecutive orders, then categorize these differences into discrete states, as presented in Table~\ref{tab:price_change}, that define our Markov chain framework. The primary objective is to estimate the probabilities of transition between successive limit-order price change states, with a particular focus on their intraday dynamics and variation across different market capitalization tiers. The following subsection explains the segmentation of a trading day for the intraday analysis and selected the stocks for each market capitalization tier.

\subsection{Stock Selection and Time-Interval Division}
\label{sec:SS_TI}

For each market capitalization tier -- High Market Capitalization ($\mathtt{HMC}$), Medium Market Capitalization ($\mathtt{MMC}$), and Low Market Capitalization ($\mathtt{LMC}$), we select five stocks; extract tick data for these five stocks separately from the full raw data shown in Table~\ref{tab:DataSample}. The stock selection follows a stratified approach across the capitalization tiers: for $\mathtt{HMC}$, we consider ranks $1^{st}$–$20^{th}$ and choose five stocks from distinct sectors to limit sector bias; similarly, we select five stocks from ranks $41^{st}$–$60^{th}$ for $\mathtt{MMC}$, and $81^{st}$–$100^{th}$ $\mathtt{LMC}$, as shown in Table~\ref{tab:SelectedStocks}. This analysis spans 12 trading days, balanced between six days where the NASDAQ100 index closed higher than its opening price ($07$-$11$-$2018$, $15$-$11$-$2018$, $28$-$11$-$2018$, $06$-$12$-$2018$, $10$-$12$-$2018$, $26$-$12$-$2018$) and six days where it closed lower ($09$-$11$-$2018$, $12$-$11$-$2018$, $14$-$11$-$2018$, $04$-$12$-$2018$, $07$-$12$-$2018$, $21$-$12$-$2018$), a design that mitigates the risk of our results being driven by a single market trend. The entire process of data extraction and preprocessing was performed efficiently using EmEditor, a tool capable of handling the multi-gigabyte files involved. 

\begin{table}[H]
\centering
\caption{Selected stocks across market capitalization tiers.}
\label{tab:SelectedStocks}
\begin{tabular}{|p{5.0cm}|p{5.0cm}|p{5.9cm}|} 
\hline 
\textbf{$\mathtt{HMC}$} (Ranks $1^{st}-20^{th}$) & \textbf{$\mathtt{MMC}$} (Ranks $41^{st}-60^{th}$) & \textbf{$\mathtt{LMC}$} (Ranks $81^{st}-100^{th}$) \\ 
\hline 
Amazon.com Inc [AMZN]\newline (Consumer Services) & AbbVie Inc [ABBV]\newline (Healthcare) & Broadcom Inc [AVGO]\newline (Information Technology) \\ 
\hline 
Johnson \& Johnson [JNJ]\newline (Healthcare) & HSBC Holdings plc [HSBC]\newline (Finance) & Booking Holdings Inc [BKNG]\newline (Consumer Services) \\ 
\hline
JPMorgan Chase \& Co [JPM]\newline (Finance) & Netflix Inc [NFLX]\newline (Consumer Services) & Bristol-Myers Squibb Co [BMY]\newline (Healthcare) \\ 
\hline
Microsoft Corp [MSFT]\newline (Information Technology) & Oracle Corp [ORCL]\newline (Information Technology) & Nike Inc [NKE]\newline (Consumer Goods) \\ 
\hline
Exxon Mobil Corp [XOM]\newline (Oil \& Gas) & PepsiCo Inc [PEP]\newline (Consumer Goods) & Union Pacific Corp [UNP]\newline (Industrials) \\ 
\hline 
\end{tabular}
\end{table}

To analyze the intraday patterns of limit order price change dynamics for these stocks, we segment the trading hours i.e. $09{:}30{:}00.000 - 16{:}00{:}00.000$, into six distinct intervals as shown in Table~\ref{tab:time_zones}. All intervals span one hour except $\mathtt{T_3}$ and $\mathtt{T_4}$ which are 75 minutes to accommodate the distinctive, often less volatile, patterns of the mid-day trading period.

\begin{table}[H]
\centering
\caption{Time-intervals for intraday limit order price change analysis through Markov chain.}
\label{tab:time_zones}
\newcolumntype{C}[1]{>{\centering\arraybackslash}p{#1}}
\begin{tabular}{|C{2.5cm}|C{5.6cm}|C{2.1cm}|}
\hline
\textbf{Time Interval} & \textbf{Timing (HH:MM:SS.000)} & \textbf{Duration (Minutes)} \\
\hline
$\mathtt{T_1}$ & 09:30:00.000 - 10:29:59.999 & 60  \\
\hline
$\mathtt{T_2}$ & 10:30:00.000 - 11:29:59.999 & 60  \\
\hline
$\mathtt{T_3}$ & 11:30:00.000 - 12:44:59.999 & 75  \\
\hline
$\mathtt{T_4}$ & 12:45:00.000 - 13:59:59.999 & 75  \\
\hline
$\mathtt{T_5}$ & 14:00:00.000 - 14:59:59.999 & 60  \\
\hline
$\mathtt{T_6}$ & 15:00:00.000 - 16:00:00.000 & 60  \\
\hline
\end{tabular}
\end{table}

The combination of time-based segmentation and
capitalization-tiered stock selection across diverse sectors and trading days creates a powerful, two-dimensional framework for comparative analysis. 

\section{Methodology}
\label{sec:Method}

Having established the stock selection and time segmentation protocol, we now
present the methodological framework for analyzing intraday limit order price change dynamics using high-frequency tick-by-tick data. 
The framework is designed to: (i) verify whether consecutive price changes exhibit short-range dependence using the G-test of independence, (ii) conditional on detecting dependencies, model the resulting state-to-state revision mechanism through a discrete-time Markov representation, and (iii) summarize and compare intraday dynamics across time intervals and market capitalization tiers, separately for bid and ask submissions.
Finally, we complement transition-based analysis with metrics, similarity-based comparisons of transition matrices, and stationary behavior to provide both local (one-step) and global (long-run) views of limit order price revision dynamics.

\subsection{G-test of Independence}
\label{subsec:G-test}

To validate the presence of memory effects in high-frequency limit order price changes, we employ the G-test of independence \cite{berrett2021usp,ahad2019applicability}. This likelihood ratio test evaluates whether consecutive price changes exhibit statistical dependence, a prerequisite for Markov chain modeling.
Operationally, the test is applied to the contingency table of consecutive state transitions constructed within each intraday time interval and capitalization tier, and performed separately for the bid and ask sides.

The test examines the null hypothesis of independence against the alternative of dependence:
\begin{align}
H_0&: \text{Consecutive price changes are independent} \\
H_1&: \text{Consecutive price changes exhibit dependence}
\end{align}

The G-statistic quantifies the likelihood ratio between observed and expected frequencies under independence:
\begin{equation}
G = 2 \sum_{i,j} O_{ij} \ln\left(\frac{O_{ij}}{E_{ij}}\right),
\label{eqn:G}
\end{equation}
where $O_{ij}$ represents observed transition frequencies from state $i$ to state $j$, and $E_{ij}$ denotes expected frequencies under independence:
\begin{equation}
E_{ij} = \frac{(\sum_{k} O_{ik})(\sum_{k} O_{kj})}{\sum_{k,l} O_{kl}}.
\end{equation}

Under $H_0$, the G-statistic follows a $\chi^2$ distribution with $(r-1)(c-1)$ degrees of freedom, where $r$ and $c$ are the dimensions of the transition matrix. We reject $H_0$ at the 5\% significance level if the resulting p-value falls below 0.05. 
Rejection of the independence hypothesis indicates that limit order price changes exhibit statistically significant short-range dependence. Consistent with this evidence, we model the sequences using first-order Markov dynamics, where the next state depends only on the current state. This step provides a formal statistical basis for the Markov chain analysis and ensures that the subsequent estimation of transition matrices captures genuine temporal structure rather than sampling noise.

\subsection{Discrete-Time Markov Chain for Limit Order Price Changes}

Markov chain belongs to a category of stochastic processes that are highly effective in describing sequences of categorical events~\cite{rabindrajit2024high,de2019markov}. We employ a discrete-time Markov Chain (DTMC) to model the dynamics of limit order price changes. A DTMC is a stochastic process comprising a series of random variables $X_1, X_2, \ldots, X_n$ that obeys the Markov property~\cite{spedicato2016markovchain}, meaning the probability of transitioning to any future state $X_{n+1}$ depends solely on the current state $X_n$ and is independent of all previous states~\cite{shamshad2005first}.
In our setting, $X_n$ represents the categorical state of the $n$-th consecutive limit order price change for a fixed quote side, intraday interval, and capitalization tier.

To apply this framework, we first categorize limit order price changes into nine distinct states based on their percentage deviation from the previous limit order price, as defined in Table~\ref{tab:price_change}. The set of possible states is therefore $S = {S_1, S_2, \ldots, S_9}$.
This discretization preserves both direction and magnitude of revisions while enabling stable estimation of transition probabilities in a high-frequency setting.

\begin{table}[H]
\centering
\caption{Categorization of limit order price change for Markov chain states.}
\label{tab:price_change}
\newcolumntype{C}[1]{>{\centering\arraybackslash}p{#1}}

\begin{tabular}{|C{1.8cm}|C{3.2cm}|p{4.5cm}|} 
\hline
\textbf{Markov State} & \textbf{Price Change} & \textbf{Price Change Category} \\ \hline
$S_1$ & $>$-5.0\% & A: Very Aggressive Sell \\ \hline
$S_2$ & -5.0\% to -2.0\% & B: Aggressive Sell \\ \hline
$S_3$ & -2.0\% to -1.0\% & C: Moderate Sell \\ \hline
$S_4$ & -1.0\% to -0.01\% & D: Mild Sell \\ \hline
$S_5$ & 0.0\% & E: Neutral \\ \hline
$S_6$ & +0.01\% to +1.0\% & F: Mild Buy \\ \hline
$S_7$ & +1.0\% to +2.0\% & G: Moderate Buy \\ \hline
$S_8$ & +2.0\% to +5.0\% & H: Aggressive Buy \\ \hline
$S_9$ & $>$+5.0\% & I: Very Aggressive Buy \\ \hline
\end{tabular}
\end{table}

The core component of the first-order DTMC is the transition probability $p_{ij}$, which represents the probability of the price change transitioning from the current state $S_i$ to state $S_j$ in the next time step~\cite{rabindrajit2024high}:
\begin{equation}
p_{ij} = P(X_{n+1} = S_j \mid X_n = S_i)
\end{equation}

The complete probability distribution of transitions between all states is concisely represented by a transition probability matrix, $\mathbf{P}$:

\begin{equation}
\label{eqn:transition_matrix}
\mathbf{P} = 
\begin{pmatrix}
p_{11} & p_{12} & \cdots & p_{19} \\
p_{21} & p_{22} & \cdots & p_{29} \\
\vdots  & \vdots  & \ddots & \vdots  \\
p_{91} & p_{92} & \cdots & p_{99} 
\end{pmatrix}
\end{equation}

This matrix is subject to the constraints:
\begin{align}
& 0 \leq p_{ij} \leq 1, \quad \forall i,j \in S, \\
& \sum_{j=1}^{9} p_{ij} = 1, \quad \forall i \in S.
\end{align}

To estimate the elements $p_{ij}$ of the matrix $\mathbf{P}$, we use the Maximum Likelihood Estimation (MLE) method~\cite{shamshad2005first,masseran2015markov}. 
We estimate transition matrices separately for each intraday interval and market capitalization tier, and for each quote side. A modified MATLAB code is used for this estimation, with the original version available in Ref.~\cite{JesseMATLABcode}. These matrices capture the step-by-step dynamics of limit order price adjustments and enable systematic comparisons across time intervals. To translate transition patterns into interpretable market features, we next compute metrics that summarize persistence and randomness.
This stratified estimation design isolates (i) within-day temporal effects, (ii) capitalization effects, and (iii) bid--ask asymmetries in a unified probabilistic framework.

\subsection{Markov Chain Dynamics Metrics}
\label{subsec:metrics}

Beyond transition probabilities, several metrics derived from the transition matrix $\mathbf{P}$ characterize the dynamics of limit order price change processes, as listed below. 
We use these summaries to compare how quickly price revision behavior stabilizes, how predictable it is given the current state, and how frequently different degrees of aggressiveness occur.

\subsubsection{Spectral Gap and Relaxation Time}
The spectral gap $\gamma$ measures the convergence rate to the stationary distribution $\pi$ given later in Eq.~\ref{eq:balance}, defined as the difference between the largest and second-largest eigenvalues of $\mathbf{P}$~\cite{diaconis1991geometric, levin2017markov}:
\begin{equation}
\gamma = 1 - |\lambda_2|,
\end{equation}
where $\lambda_1 = 1$ and $\lambda_2$ is the second-largest eigenvalue in magnitude, with relaxation time $\tau$ defined as:
\begin{equation}
\tau_{\text{rel}} = \frac{1}{\gamma}.
\label{eq:relaxation_time}
\end{equation}
A large spectral gap indicates rapid convergence to equilibrium with short-lived memory effects. Small gaps suggest persistent patterns and slower convergence to steady-state behavior.

\subsubsection{Entropy Rate}
The entropy rate quantifies the average information content per transition for a stationary Markov chain, defined as~\cite{cover1999elements}:
\begin{equation}
H(\mathcal{X}) = -\sum_{i=1}^{9} \pi_i \sum_{j=1}^{9} p_{ij} \log p_{ij}.
\end{equation}
High entropy rates indicate unpredictable price change sequences, while low rates suggest structured, predictable patterns given the current state.

\subsubsection{Mixing Rate}
The mixing rate describes convergence speed to the stationary distribution $\pi$ in Eq.~\ref{eq:balance} from arbitrary initial conditions, bounded by~\cite{levin2017markov,sinclair1992improved}:
\begin{equation}
\| \mathbf{p}^{(n)} - \pi \|_{\text{TV}} \leq C e^{-n / \tau_{\text{rel}}},
\end{equation}
where $\mathbf{p}^{(n)}$ is the state distribution after $n$ steps. Fast mixing implies rapid dissipation of initial shock effects, while slow mixing indicates persistent path-dependency in price formation.

\subsubsection{Mean Recurrence Time}
For each state $S_i$, the mean recurrence time represents the expected return interval~\cite{norris1998markov}:
\begin{equation}
\mu_i = \frac{1}{\pi_i}, \quad i = 1, 2, \ldots, 9.
\end{equation}
$\mu_i$ values reveal typical cycles of order price aggressiveness, with small $\mu_i$ for neutral states indicating frequent orders at current market prices and large values for extreme states reflect rare aggressive adjustments constrained by market impact.

While these metrics effectively quantify the overall dynamic properties, they do not capture the structural similarity between entire transition matrices. To systematically group distinct trading behaviors based on their full probabilistic structure, we reduce the transition data into a lower-dimensional representation suitable for clustering, as detailed in the next subsection. Accordingly, we treat the embedding and clustering steps as complementary tools for comparing entire matrices across time intervals and quote sides, rather than as core modeling contributions.

\subsection{Dimensionality Reduction of the TPMs}
\label{subsec:dim_reduction}

To derive low-dimensional embeddings of the $9 \times 9$ transition probability matrices (TPMs), each TPM is vectorized into $\mathbf{x} \in \mathbb{R}^{81}$ and reduced in two stages. First, we apply Principal Component Analysis (PCA)~\cite{jolliffe2016principal} and retain the top $k=8$ components (capturing over 95\% of the variance), which provides a compact representation and stabilizes the subsequent embedding. Second, the PCA-reduced data are mapped into two dimensions using t-Distributed Stochastic Neighbor Embedding (t-SNE)~\cite{van2008visualizing,luwang2025intraday} to obtain a visualizable representation of similarities across TPMs.


\subsection{Clustering Techniques for Dimension-Reduced TPMs}
\label{subsec:Clustering_techniques}

To summarize latent structures in the dimension-reduced TPM embeddings, we apply two complementary clustering algorithms using the \texttt{scikit-learn} library~\cite{pedregosa2011scikit}. First, we use Agglomerative Hierarchical Clustering with Ward’s minimum-variance linkage~\cite{jaroonchokanan2022dynamics} to obtain a dendrogram that highlights nested similarity patterns between time intervals. Second, we use DBSCAN~\cite{kazemi2017spatio,luwang2025intraday} to identify dense groups and potential outliers in the embedding space. These algorithms allow us to group trading time intervals into naturally occurring behavioral phases.

\subsection{Stationary Distribution Analysis}

In an ergodic Markov chain, the long-run behavior is characterized by a stationary distribution, which specifies the steady-state probability associated with each state. Denoting the long-term probability of being in state $j$ by $\pi_j$, this stationary vector is uniquely determined and must satisfy~\cite{holmes2021discrete,rabindrajit2024high}:

\begin{align}
    \pi_j &= \sum_{i=1}^{9} \pi_i p_{ij} \quad \text{(balance equation)}, \label{eq:balance}\\
    \sum_{j=1}^{9} \pi_j &= 1 \quad \text{(Normalization condition)}.
\end{align}

The stationary probabilities $\pi_j$ are computed using the \texttt{PyDTMC} package~\cite{belluzzo2024pydtmc}. To quantify differences between stationary distributions across time intervals, we employ the Jensen--Shannon Divergence (JSD). For two probability distributions $p$ and $q$ (each non-negative and summing to one), the JSD is defined as~\cite{nielsen2019jensen}:

\begin{equation}
    \text{JSD}(p, q) = 
    \frac{1}{2}
    \left[
        \text{KLD}\left(p \,\middle\|\, \frac{p+q}{2}\right)
        +
        \text{KLD}\left(q \,\middle\|\, \frac{p+q}{2}\right)
    \right],
\end{equation}

where the Kullback--Leibler divergence (KLD) between two discrete distributions $u$ and $v$ is

\begin{equation}
    \text{KLD}(u \| v) = \sum_{i} u_i \log_2 \frac{u_i}{v_i}.
\end{equation}

In this formulation, $u_i$ and $v_i$ correspond to the probabilities assigned to state $i$ by the distributions $u$ and $v$, respectively. The Jensen-Shannon Divergence is symmetric by definition, takes only non-negative values, and becomes zero precisely when the two distributions coincide~\cite{mateos2017detecting}. 
This distributional analysis enables us to characterize the stable, long-run profile of limit order price revisions and quantify the magnitude of behavioral shifts between different trading intervals. With this comprehensive methodological framework established in Section~\ref{sec:Method}, we now proceed to Section~\ref{sec:Results} to detail the empirical findings derived from the NASDAQ100 dataset.

\section{Results}
\label{sec:Results}

How do price adjustments for ask and bid limit orders evolve intraday, and how are these dynamics shaped by market capitalization? To answer this central question, this section presents the empirical results from our discrete-time Markov chain framework. The findings are structured to provide a comparative analysis across the six distinct trading intervals (${\mathtt{T_1}}$--${\mathtt{T_6}}$) and the three market capitalization tiers: High ($\mathtt{HMC}$), Medium ($\mathtt{MMC}$), 
and Low ($\mathtt{LMC}$).

We begin in Subsection~\ref{result:G-test} by applying the G-test of independence to validate the presence of short-term memory in price change sequences, establishing that current price adjustments depend statistically on preceding events, thereby justifying the Markov chain modeling approach. Subsection~\ref{result:TPM_Order} examines the structure and dominant elements of the estimated transition probability matrices (TPMs), conducting systematic comparisons both temporally and cross-sectionally across the capitalization tiers. The analysis in Subsection~\ref{result:DTMC_metrics} quantifies the dynamic properties of these TPMs through key Markov chain metrics including entropy rate, spectral gap, relaxation time, and mean recurrence times, revealing how convergence speed and predictability patterns vary across temporal and capitalization-size dimensions. Subsection~\ref{result:clustering_limitorder} employs dimensionality reduction techniques i.e. Principal Component Analysis and t-distributed Stochastic Neighbor Embedding, followed by clustering analysis with Hierarchical and DBSCAN algorithms to identify latent structural patterns and classifications within the high-dimensional TPMs. Finally, Subsection~\ref{result:SD_LimitOrder} analyzes the stationary distributions of limit order price change states across time-intervals and capitalization tiers, utilizing Jensen-Shannon divergence to quantify distributional differences and assess the stability of long-term adjustment patterns.

\subsection{G-test of independence}
\label{result:G-test}
To validate the use of Markov chains, we test for temporal dependence by applying the G-test of independence and autocorrelation analysis separately to the ask and bid limit order price change sequences. Tables S1 and S2 in the Supplementary material present the average G-test statistics for ask and bid sequences respectively, calculated across all trading days for each market capitalization tier and time-interval combination. The results demonstrate consistently high G-statistics with p-values well below the 0.05 significance threshold ($p \ll 0.05$) across all capitalization tiers and time-intervals. These findings provide strong statistical evidence to reject the null hypothesis of independence, confirming that price changes depend significantly on preceding adjustments. This is corroborated by the autocorrelation analysis, which reveals statistically significant correlations at lags 1 and 2, with magnitudes exceeding the $1/\sqrt{N}$ significance threshold (where $N$ represents the total number of price change events)~\cite{chatfield1978holt}. Although modest, these correlations decay systematically, indicating the presence of short-term memory effects consistent with first-order Markov properties~\cite{beran1992statistical}.

The convergent evidence from both G-test and autocorrelation analyses validates our approach using first-order discrete-time Markov chains. While higher-order dependencies may exist, the predominant lag-1 correlations justify the first-order assumption, which offers computational tractability while capturing the essential temporal dependencies in limit order price adjustment behaviors. Consequently, we proceed with Maximum Likelihood Estimation of first-order transition probability matrices and systematic comparisons of dominant elements temporally and cross-sectionally across the capitalization tiers as detailed in the following subsection. We restrict our analysis to only D, E, and F price changes as the count of price changes for A, B, C, G, H and I are very low, as shown in Fig.S1 in the supplementary material. 


\subsection{Transition Probability Matrix Analysis}
\label{result:TPM_Order}

\begin{figure}[H]
    \centering
    \captionsetup[subfigure]{justification=centering}
    
    \begin{tabular}{ccc}
        \begin{subfigure}{0.315\textwidth}
            \includegraphics[width=\textwidth]{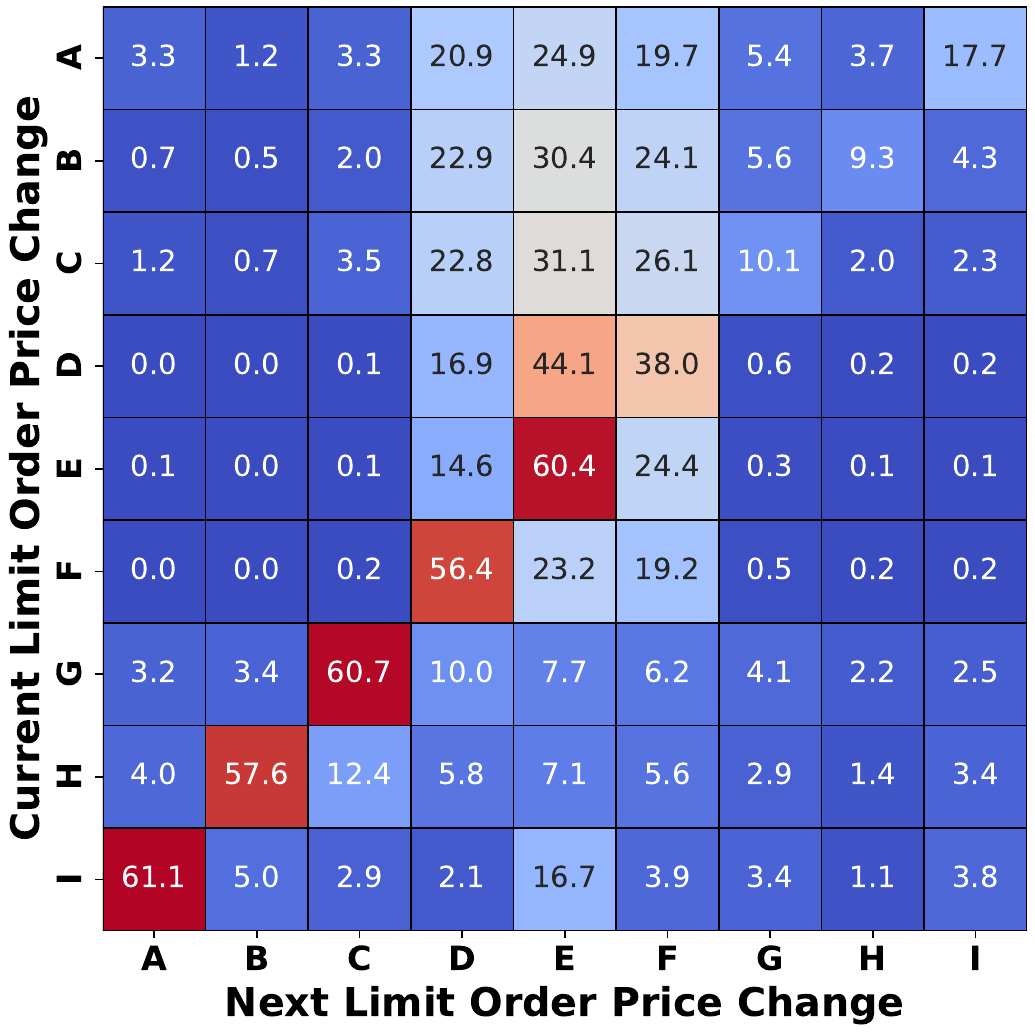}
            \caption{Ask -- $\mathtt{HMC}$}
        \end{subfigure} &
        \begin{subfigure}{0.3\textwidth}
            \includegraphics[width=\textwidth]{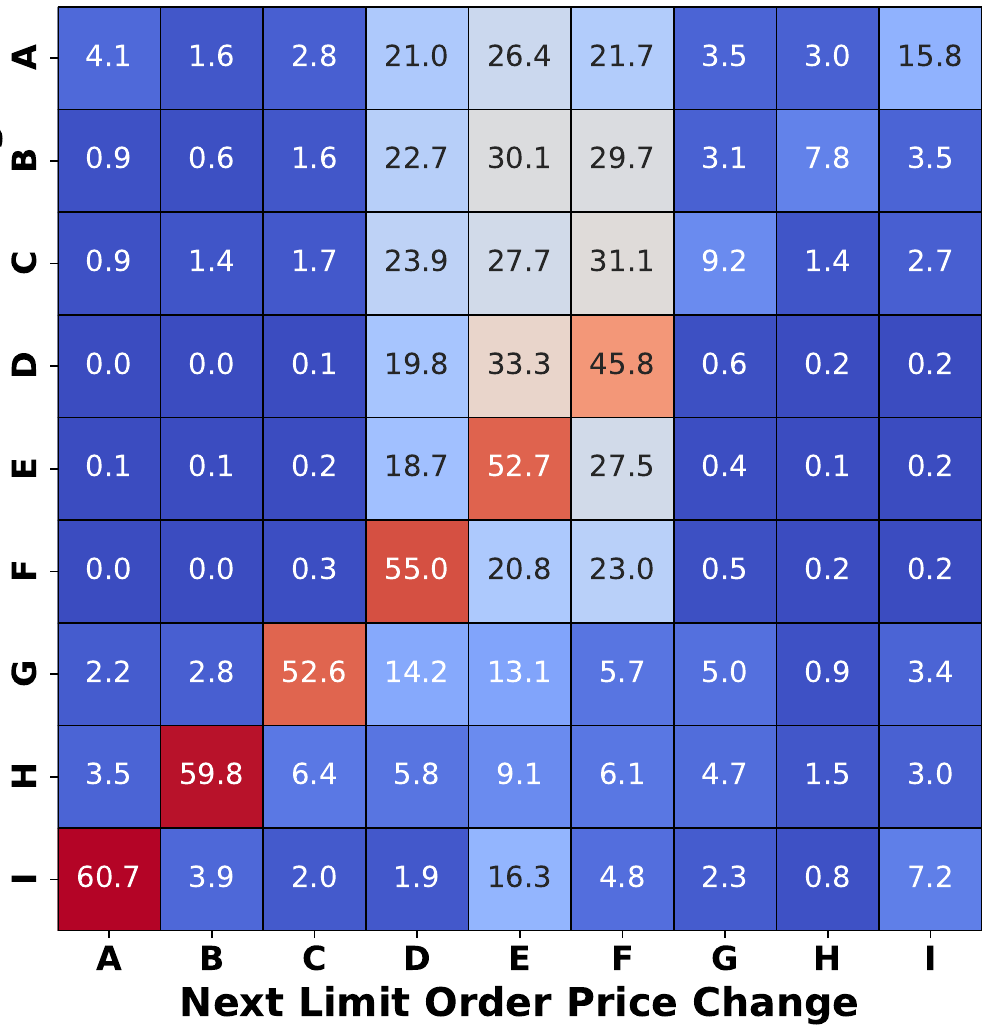}
            \caption{Ask -- $\mathtt{MMC}$}
        \end{subfigure} &
        \begin{subfigure}{0.3\textwidth}
            \includegraphics[width=\textwidth]{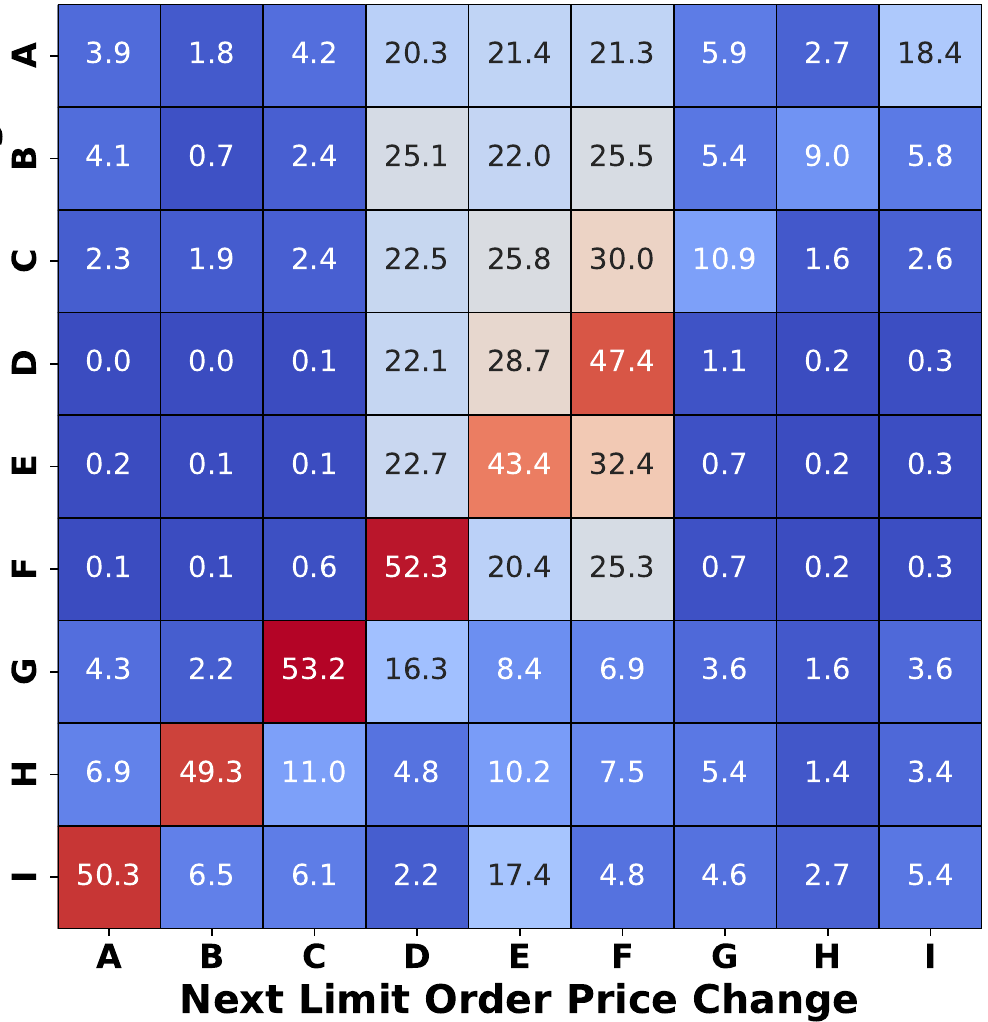}
            \caption{Ask -- $\mathtt{LMC}$}
        \end{subfigure} \\

        \begin{subfigure}{0.315\textwidth}
            \includegraphics[width=\textwidth]{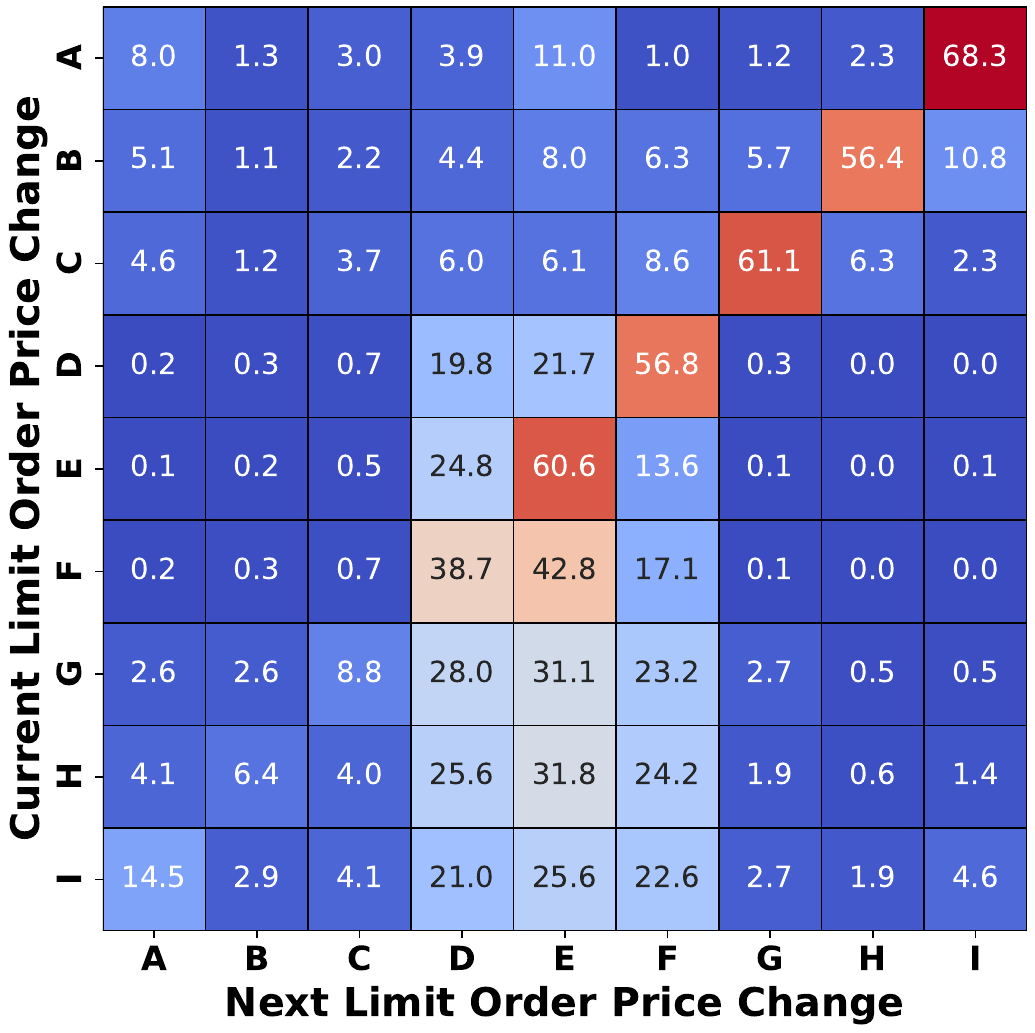}
            \caption{Bid -- $\mathtt{HMC}$}
        \end{subfigure} &
        \begin{subfigure}{0.3\textwidth}
            \includegraphics[width=\textwidth]{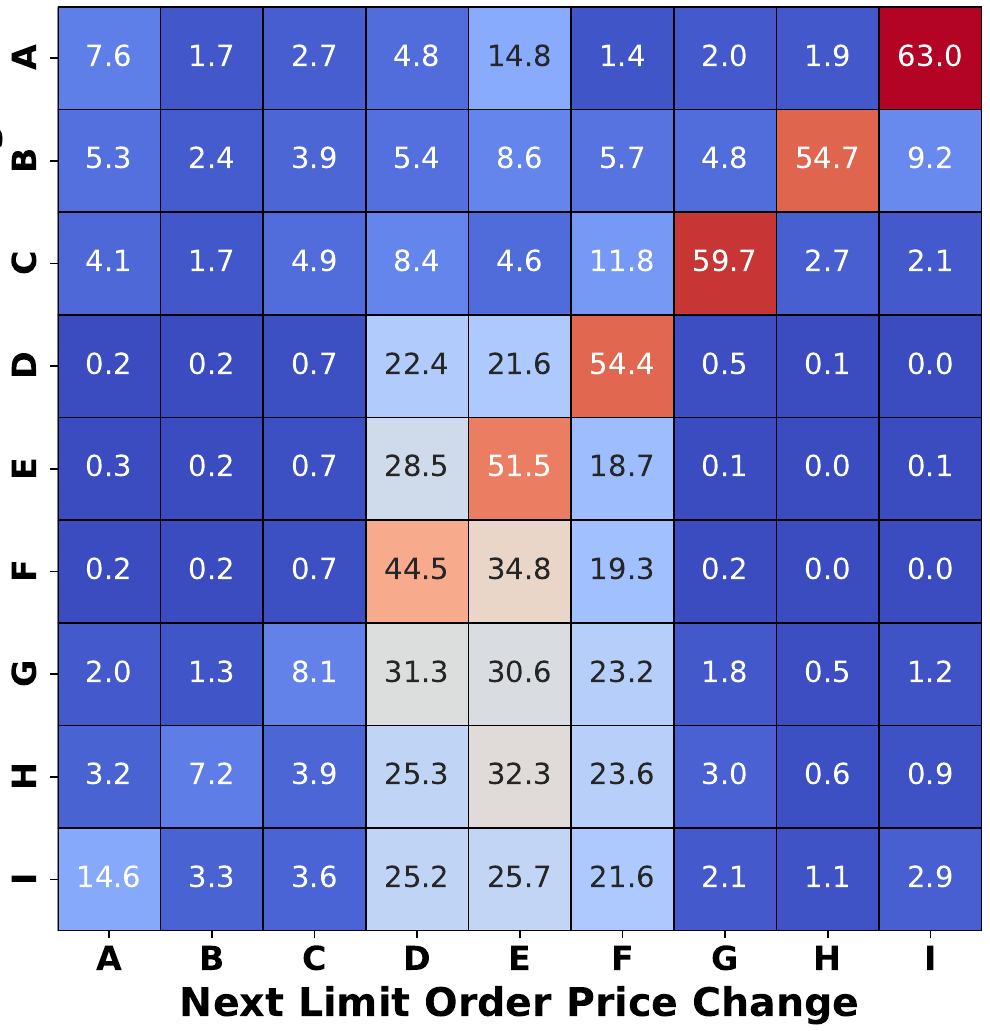}
            \caption{Bid -- $\mathtt{MMC}$}
        \end{subfigure} &
        \begin{subfigure}{0.3\textwidth}
            \includegraphics[width=\textwidth]{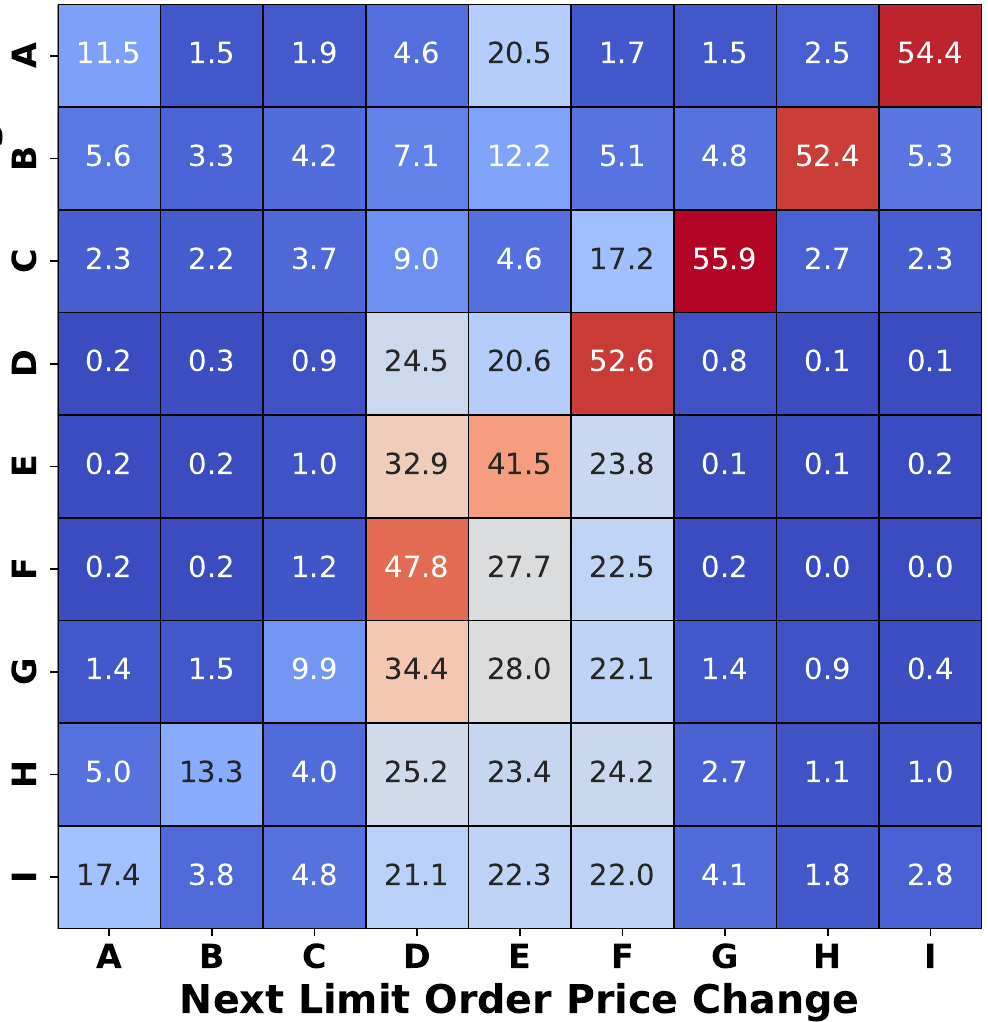}
            \caption{Bid -- $\mathtt{LMC}$}
        \end{subfigure} \\

    \end{tabular}
   \caption{Representative heatmap visualizations of transition probability matrices (TPMs) for the opening interval $\mathtt{T_1}$, displaying Ask (top) and Bid (bottom) limit order price change dynamics across $\mathtt{HMC}$, $\mathtt{MMC}$, and $\mathtt{LMC}$ capitalization tiers.}
    \label{fig:TPM_T1}
\end{figure}

We analyze limit order price changes using a discrete-time Markov chain, with a nine state space defined by the price change magnitudes, as shown in Table~\ref{tab:price_change}. The transition probability matrices (TPMs) for this chain are estimated via Maximum Likelihood Estimation. The analysis encompasses both bid and ask limit orders across six intraday time-intervals for each of the three market capitalization tiers -- $\mathtt{HMC, MMC}$, and $\mathtt{LMC}$ over twelve trading days. This generates 2,160 individual $9 \times 9$ TPMs, which we aggregate to produce 36 representative matrices: 18 for ask-side changes and 18 for bid-side changes, with six matrices per capitalization tier corresponding to the six time-intervals. Fig.~\ref{fig:TPM_T1} presents representative heatmap visualizations of the ask and bid transition probability matrices (TPMs) for the opening interval $\mathtt{T_1}$. In these matrices, rows correspond to the current price change state and columns to the subsequent state, with cell values indicating transition probabilities. For a comprehensive analysis, the complete set of ask-side TPMs across all capitalization tiers ($\mathtt{HMC, MMC, LMC}$) and time intervals ($\mathtt{T_1}$--$\mathtt{T_6}$) is provided in Supplementary Figures S2 and S3. The corresponding complete set of bid-side TPMs is available in Supplementary Figures S4 and S5.

\subsubsection{Inertia of Limit Order Prices}
\label{subsec:inertia}

\begin{figure}[H]
    \centering
    \includegraphics[width=10.5cm]{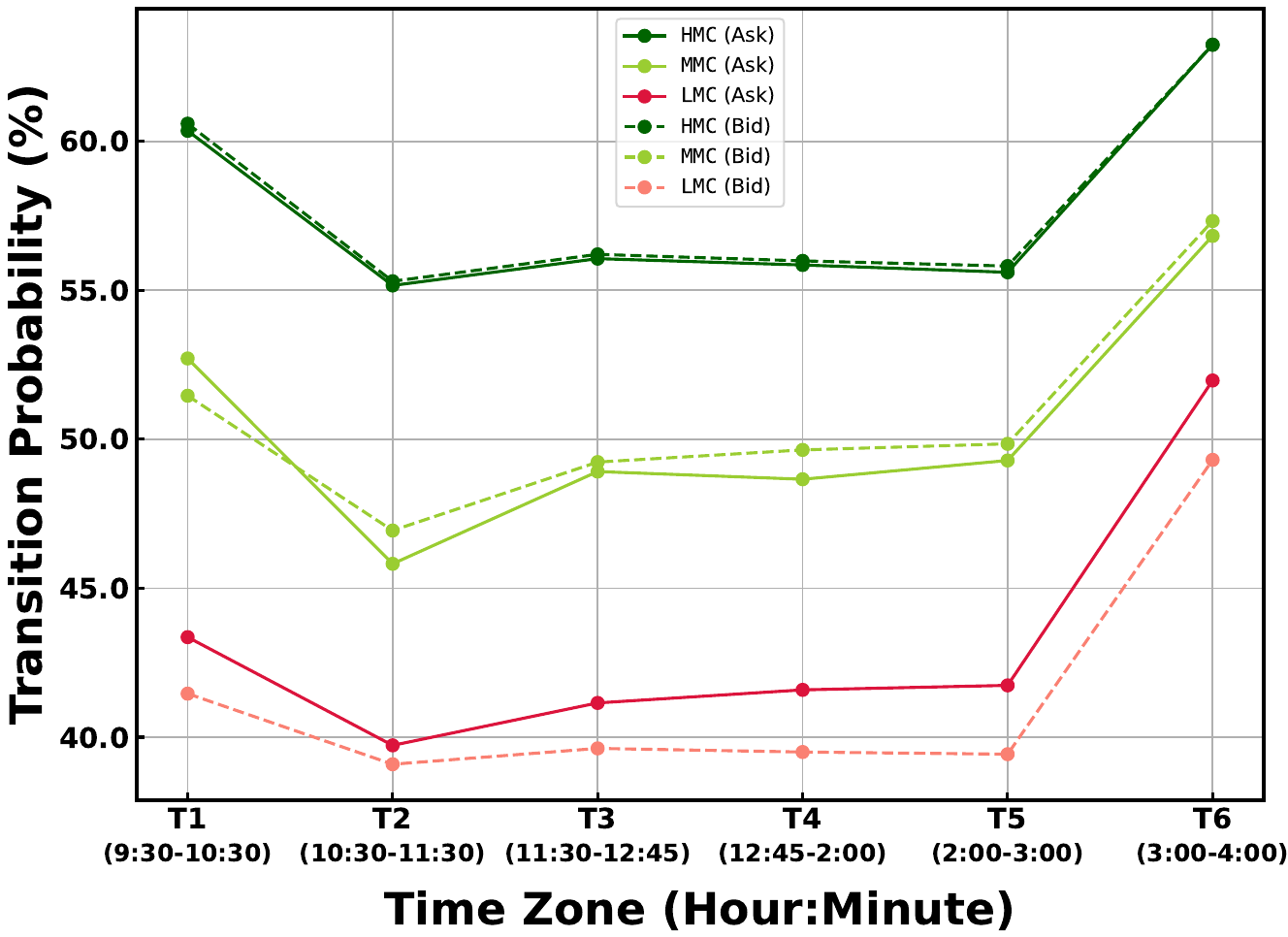}
    \caption{Temporal variation of transition probability for consecutive $0\%$ limit order price changes across market capitalization tiers and time intervals.}
    \label{fig:5_5}
\end{figure}

Figure~\ref{fig:5_5} illustrates the temporal variation in the probability of consecutive $0\%$ limit order price changes i.e. price inertia, on both the bid and ask sides. This metric, which captures the likelihood of limit orders maintaining an identical price across successive submissions, shows systematic variations across different time-intervals and market capitalization tiers. The analysis reveals two key empirical findings: Point~\ref{TPM:55temporal} identifies a shared intraday dynamic common to the $\mathtt{HMC, MMC}$, and $\mathtt{LMC}$, while Point~\ref{TPM:55capitalization} highlights the distinct characteristics for each capitalization tier.

\begin{enumerate}
\item \label{TPM:55temporal} Intraday limit order price change transition dynamics: We observed a consistent temporal pattern in the transition probability for consecutive $0\%$ price changes across all market capitalization tiers for both bid and ask sides, as summarized below:
\begin{enumerate}
    \item Price inertia peaks at market opening ${\mathtt{T_1}}$ indicating a strong tendency for limit orders to be submitted consecutively at the same price. This behavior likely reflects defensive positioning at anticipated support and resistance levels. Heightened opening volatility, driven by the assimilation of overnight news and information asymmetries, creates uncertainty that encourages defensive positioning outside the spread and concentration of orders at specific price points to avoid adverse selection costs~\cite{mcinish1992analysis,glosten1985bid}.

    \item The inertia probability declines sharply post-open and stabilizes at a lower level throughout the midday session i.e. ${\mathtt{T_2}}$ -- ${\mathtt{T_5}}$. The initial decrease aligns with the dynamic price discovery process, where high volatility and fluctuating bid-ask spreads necessitate frequent limit order price adjustments, reducing the repetition of orders at the same price. As the market absorbs new information, volatility subsides and spread narrows~\cite{andersen1997intraday,mcinish1992analysis}. This post-discovery environment fosters more heterogeneous trading conditions where diverse participants employ mixed strategies, leading to dispersed order placement and a stable, lower probability of consecutive submissions at identical prices.
    
    \item Finally, the inertia surges during the closing hour ${\mathtt{T_6}}$, often exceeding the opening peak. This resurgence is driven by intense end-of-day portfolio rebalancing, as the urgency to close positions and mitigate overnight risk prompts market participants to submit orders at key price levels to ensure execution~\cite{pagano2003closing,jegadeesh2022closing}. This strategic shift prioritizes order fulfillment over price optimization, leading to a pronounced reduction in price adjustments and a corresponding peak in the probability of consecutive submissions at the same price.

\end{enumerate}

\item \label{TPM:55capitalization} Differences in price change transition dynamics between $\mathtt{HMC}$, $\mathtt{MMC}$, and $\mathtt{LMC}$ stocks: Cross-capitalization analysis reveals systematic differences in these dynamics that persist across all trading time intervals, reflecting the distinct characteristics of each market tier.

\begin{enumerate}

    \item A strong capitalization gradient is observed; $\mathtt{HMC}$ stocks exhibit the strongest price inertia, followed by $\mathtt{MMC}$ and then $\mathtt{LMC}$ stocks. This hierarchy reflects fundamental differences in their market structure. Greater liquidity and trading volume in $\mathtt{HMC}$ stocks attract sophisticated participants, including market makers and high-frequency traders. These agents provide continuous liquidity by maintaining persistent limit orders at specific prices, resulting in a relatively high self-transition probabilities i.e. higher inertia. Conversely, $\mathtt{LMC}$ stocks exhibit the lowest probability of consecutive unchanged prices. Lower liquidity, relatively higher transaction costs, and greater information asymmetries discourage passive order placement. This necessitates dynamic order management, where participants frequently adjust prices to mitigate risk, resulting in systematically lower inertia.
    
    \item A bid-ask asymmetry gradient is evident, scaling from minimal in $\mathtt{HMC}$ stocks to maximal in $\mathtt{LMC}$ stocks. For $\mathtt{LMC}$ stocks, ask-side transition probabilities consistently exceed bid-side probabilities across all intervals. This pattern is likely driven by three factors specific to illiquid stocks: market makers maintain persistent ask limit orders to manage inventory risk~\cite{ho1981optimal}, greater information asymmetry increases adverse selection costs on the bid side~\cite{easley1996liquidity}, and institutional frictions like short-selling costs disproportionately hinder bid-side liquidity provision~\cite{boehmer2008shorts}. Collectively, these factors heighten the observed price inertia of ask-side limit orders.
    
\end{enumerate}

\end{enumerate}

In summary, these findings demonstrate that limit order price change inertia is not random but follows predictable U-shaped intraday patterns and a distinct capitalization gradient. The results show that high-capitalization stocks exhibit the greatest price stability due to deep, continuous liquidity, whereas low-capitalization stocks are characterized by lower overall stability and significant bid-ask asymmetries. For traders, this implies that execution strategies can be optimized by anticipating higher price persistence during the market opening and closing hours, particularly in $\mathtt{HMC}$ stocks, while accounting for both the lower stability and the dominant ask-side inertia inherent to $\mathtt{LMC}$ stocks.


\subsubsection{Directional Momentum of Limit Order Price Revisions}
\label{subsec:drc_momtm}

\begin{figure}[htbp]
  \centering
  \begin{subfigure}[t]{0.48\textwidth}
    \centering
    \includegraphics[width=\linewidth]{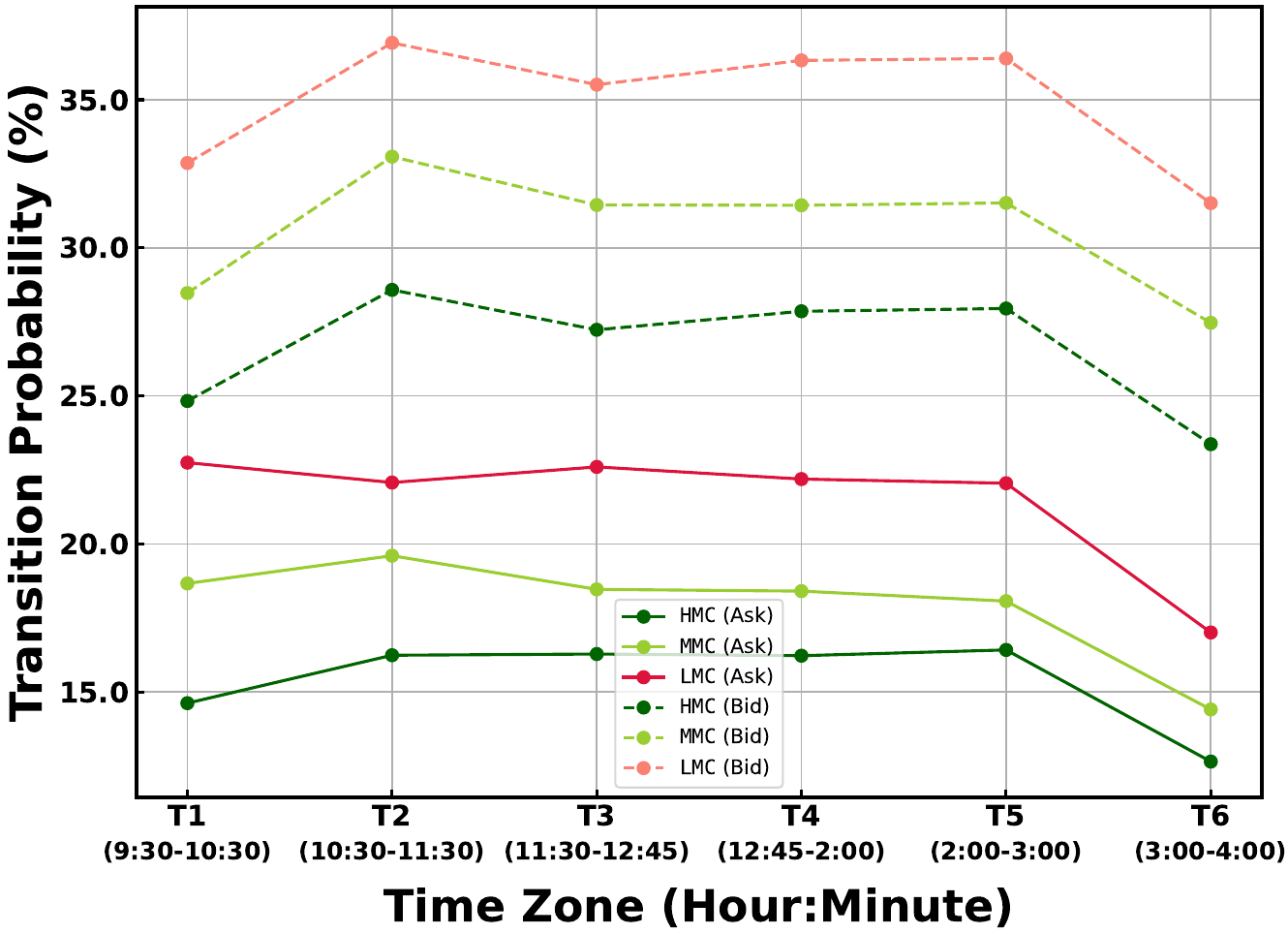}
    \caption{$0\%$ price change to both negative price change.}
    \label{fig:left}
  \end{subfigure}\hfill
  \begin{subfigure}[t]{0.48\textwidth}
    \centering
    \includegraphics[width=\linewidth]{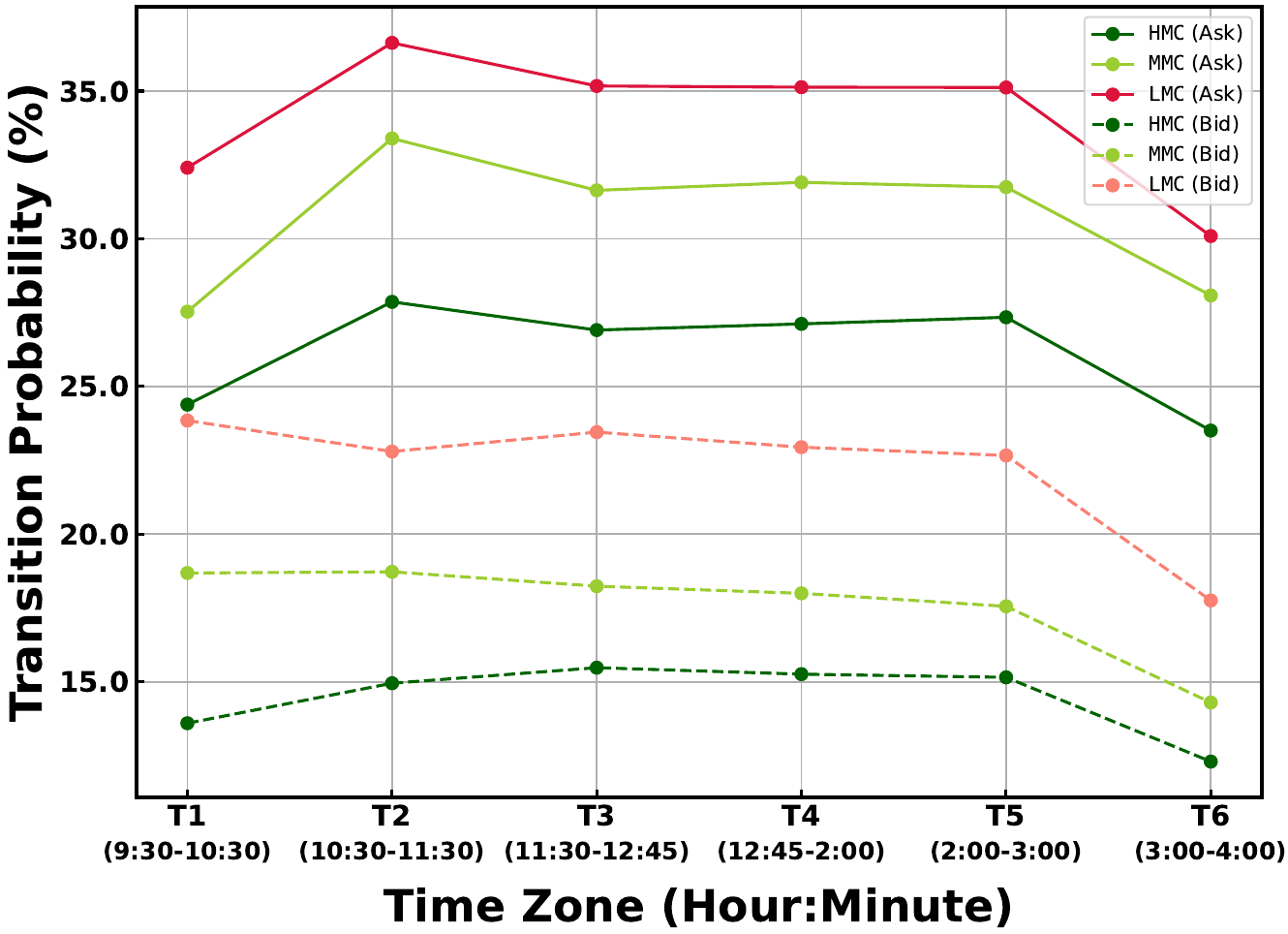}
     \caption{$0\%$ price change to both negative price change.}
    \label{fig:right}
  \end{subfigure}
  \caption{Transition probabilities from neutral i.e. $0\%$ price change, to both negative and positive price changes across market capitalization tiers and intraday intervals.}
  \label{fig:directional_transitions}
\end{figure}

Figure~\ref{fig:directional_transitions} illustrates transition probabilities from neutral i.e. $0\%$ price change, to both negative and positive price changes across market capitalization tiers and intraday intervals. We point out two key empirical findings from this analysis: Point~\ref{TPM:hierarchy} examines the systematic capitalization-based hierarchy in price revision intensity, while Point~\ref{TPM:bidask} analyzes subtle bid-ask asymmetries in directional transition behaviors.

\begin{enumerate}

\item \label{TPM:hierarchy} Hierarchy in Revision Intensity: Cross analysis reveals a systematic ordering in directional transition probabilities that persist across all trading time intervals and capitalization tiers, that reflects the distinct characteristics, as elaborated below.

\begin{enumerate}

    \item Temporally, the intraday evolution patterns reveal distinctive opening hour dynamics where $\mathtt{HMC}$ and $\mathtt{MMC}$ stocks show significant increases in directional transitions: 4.0\%--5.0\% for defensive positioning, and 1.0\%--1.5\% for aggressive positioning, while $\mathtt{LMC}$ stocks exhibit contrarian behavior with slight decreases. This divergence suggests differential responses to opening volatility and information asymmetries, with $\mathtt{LMC}$ traders adopting more conservative adjustment strategies during high-uncertainty hours~\cite{mcinish1992analysis,andersen1997intraday}.

    \item In terms of capitalization asymmetry, we observed that $\mathtt{LMC}$ stocks exhibit the highest directional transition probabilities, demonstrating elevated limit order price revision frequencies. The elevated revision frequency in $\mathtt{LMC}$ stocks stems from their market microstructure: wider bid-ask spreads, higher volatility, limited liquidity, and reduced market maker presence. These conditions create larger gaps between optimal execution prices and current market levels, necessitating more frequent price adjustments to maintain execution viability~\cite{amihud2002illiquidity}. On the other hand, $\mathtt{HMC}$ stocks exhibit the lowest revision frequencies, reflecting their stable market environment characterized by narrower spreads, higher liquidity, and greater institutional presence. The enhanced market maker activity maintains stable pricing conditions, creating extended periods of order competitiveness that reduce revision incentives~\cite{menkveld2013high}. 
\end{enumerate}

\item \label{TPM:bidask} Subtle Bid-Ask Asymmetries in Directional Revisions: We identify two subtle but persistent bid-ask asymmetries, formalized as:

\begin{align*}
    P(\text{Bid}: \text{Neutral} \to \text{Negative}) & \gtrsim P(\text{Ask}: \text{Neutral} \to \text{Negative}) \\
    P(\text{Ask}: \text{Neutral} \to \text{Positive}) & \gtrsim P(\text{Bid}: \text{Neutral} \to \text{Positive})
\end{align*}

where $\gtrsim$ denotes a modest but persistent difference. This pattern reflect nuanced differences in trading behaviors:
\begin{enumerate}
    \item The first inequality reflects a stronger value-seeking motive among buyers, who are more prone to lowering their bids to improve entry prices.
    \item The second inequality reveals a stronger profit-maximizing motive among sellers, who are more prone to raising their asks to capture better execution.
\end{enumerate}

\end{enumerate}

These empirical findings provide guidance for optimizing order management for traders while navigating different market capitalization segments. The dominant capitalization hierarchy implies that static, patient order placement is a viable strategy in $\mathtt{HMC}$ stocks, whereas the high revision frequency in $\mathtt{LMC}$ stocks necessitates dynamic, adaptive strategies to manage execution risk. This contrast is sharpest at the market open, where $\mathtt{LMC}$ traders turn conservative, while $\mathtt{HMC}$ traders actively reposition. The subtle, persistent bid-ask asymmetries can be used to fine-tune price-shading logic, accounting for the slight but predictable value-seeking (bid) and profit-maximizing (ask) tendencies of other participants.


\subsection{Markov Chain Metrics}
\label{result:DTMC_metrics}

We now compute key Markov chain metrics. While the  analysis of individual transition probabilities characterize local, state-to-state transitions, these metrics quantify the global dynamic properties of the price change process. 

\subsubsection{Spectral Gap, Relaxation and Mixing}

The spectral gap $\gamma$ is analyzed to measure the convergence speed to equilibrium, which serves as a proxy for price discovery efficiency. This metric exhibits a clear intraday pattern inverted relative to volatility. As shown in Table~\ref{tab:entropy_metrics}, the gap is smallest at the market open ($\mathtt{T_1}$; $\gamma \approx 0.56$--$0.60$), indicating the slowest convergence during high-information assimilation. It peaks midday ($\mathtt{T_3}$--$\mathtt{T_4}$; $\gamma \approx 0.71$--$0.74$), reflecting the fastest convergence, before declining slightly at the close ($\mathtt{T_6}$; $\gamma \approx 0.65$--$0.69$). In terms of capitalization, $\mathtt{HMC}$ and $\mathtt{MMC}$ stocks exhibit slightly larger spectral gaps than $\mathtt{LMC}$ stocks during active trading ($\mathtt{T_2}$--$\mathtt{T_5}$), consistent with more efficient price discovery in higher-liquidity assets. Further, bid-ask differences in $\gamma$ are modest ($\lesssim 0.05$) and secondary to the primary temporal and capitalization patterns. 

For completeness, we also report complementary measures: Mixing rate, $\lambda_\star = 1-\gamma$ and the relaxation time, $\tau = 1/\gamma$ in Table~\ref{tab:entropy_metrics}. As the relaxation rate is defined as $\kappa \equiv 1/\tau = \gamma$ , $\kappa$ is numerically identical to the spectral gap, while $\lambda_\star$ is a linear reparameterization. Accordingly, these metrics are all complementary: a high mixing rate corresponds to a long relaxation time and a small spectral gap, indicating slow convergence. We include these for readers who prefer those conventions rather than as distinct metrics.

\begin{table}[h]
\centering
\caption{Key Markov chain metrics to quantify the global dynamic properties of the price change process.}
\label{tab:entropy_metrics}
\footnotesize
\begin{tabular}{|c|c|*{8}{S[table-format=1.3]|}}
\hline
\makecell{Market-\\Capitalization\\Tier} & \makecell{Time-\\Interval} & \multicolumn{2}{c|}{Spectral Gap} & \multicolumn{2}{c|}{Relaxation Rate} & \multicolumn{2}{c|}{Entropy Rate} & \multicolumn{2}{c|}{Mixing Rate} \\
\cline{3-10}
& & \multicolumn{1}{c|}{Ask} & \multicolumn{1}{c|}{Bid}
  & \multicolumn{1}{c|}{Ask} & \multicolumn{1}{c|}{Bid}
  & \multicolumn{1}{c|}{Ask} & \multicolumn{1}{c|}{Bid}
  & \multicolumn{1}{c|}{Ask} & \multicolumn{1}{c|}{Bid} \\

\hline
\multirow{6}{*}{\textbf{$\mathtt{HMC}$}} & $\mathtt{T_1}$ & 0.562 & 0.566 & 0.825 & 0.835 & 1.032 & 1.051 & 0.438 & 0.434 \\
\cline{2-10}
                       & $\mathtt{T_2}$ & 0.738 & 0.719 & 1.341 & 1.268 & 1.038 & 1.059 & 0.262 & 0.281 \\
\cline{2-10}
                       & $\mathtt{T_3}$ & 0.740 & 0.695 & 1.345 & 1.188 & 1.035 & 1.054 & 0.260 & 0.305 \\
\cline{2-10}
                       & $\mathtt{T_4}$ & 0.744 & 0.715 & 1.363 & 1.254 & 1.039 & 1.046 & 0.256 & 0.285 \\
\cline{2-10}
                       & $\mathtt{T_5}$ & 0.725 & 0.690 & 1.290 & 1.172 & 1.030 & 1.056 & 0.275 & 0.310 \\
\cline{2-10}
                       & $\mathtt{T_6}$ & 0.692 & 0.651 & 1.177 & 1.053 & 0.984 & 1.021 & 0.308 & 0.349 \\
\hline
\multirow{6}{*}{\textbf{$\mathtt{MMC}$}} & $\mathtt{T_1}$ & 0.590 & 0.595 & 0.891 & 0.904 & 1.089 & 1.112 & 0.410 & 0.405 \\
\cline{2-10}
                       & $\mathtt{T_2}$ & 0.711 & 0.716 & 1.240 & 1.260 & 1.078 & 1.081 & 0.289 & 0.284 \\
\cline{2-10}
                       & $\mathtt{T_3}$ & 0.727 & 0.730 & 1.299 & 1.309 & 1.074 & 1.078 & 0.273 & 0.270 \\
\cline{2-10}
                       & $\mathtt{T_4}$ & 0.708 & 0.709 & 1.229 & 1.235 & 1.084 & 1.077 & 0.292 & 0.291 \\
\cline{2-10}
                       & $\mathtt{T_5}$ & 0.679 & 0.689 & 1.135 & 1.170 & 1.069 & 1.076 & 0.321 & 0.311 \\
\cline{2-10}
                       & $\mathtt{T_6}$ & 0.669 & 0.671 & 1.106 & 1.111 & 1.030 & 1.041 & 0.331 & 0.329 \\
\hline
\multirow{6}{*}{\textbf{$\mathtt{LMC}$}} & $\mathtt{T_1}$ & 0.562 & 0.567 & 0.825 & 0.838 & 1.152 & 1.162 & 0.438 & 0.433 \\
\cline{2-10}
                       & $\mathtt{T_2}$ & 0.693 & 0.705 & 1.181 & 1.222 & 1.124 & 1.108 & 0.307 & 0.295 \\
\cline{2-10}
                       & $\mathtt{T_3}$ & 0.694 & 0.717 & 1.185 & 1.263 & 1.108 & 1.116 & 0.306 & 0.283 \\
\cline{2-10}
                       & $\mathtt{T_4}$ & 0.713 & 0.695 & 1.247 & 1.187 & 1.113 & 1.105 & 0.287 & 0.305 \\
\cline{2-10}
                       & $\mathtt{T_5}$ & 0.698 & 0.680 & 1.196 & 1.140 & 1.105 & 1.115 & 0.302 & 0.320 \\
\cline{2-10}
                       & $\mathtt{T_6}$ & 0.687 & 0.666 & 1.161 & 1.097 & 1.073 & 1.108 & 0.313 & 0.334 \\
\hline
\end{tabular}
\end{table}

\subsubsection{Entropy Rate}

The entropy rate is analyzed to quantify the unpredictability of the price change sequence. As seen from Table~\ref{tab:entropy_metrics}, there is a clear capitalization hierarchy. $\mathtt{LMC}$ stocks consistently exhibit the highest entropy rates, followed by $\mathtt{MMC}$, and $\mathtt{HMC}$. This gradient confirms that price change dynamics are least predictable in $\mathtt{LMC}$ stocks and most structured in $\mathtt{HMC}$ stocks, reflecting underlying differences in market depth and information asymmetry. Temporally, entropy rates are highest at $\mathtt{T_1}$ and generally decline toward $\mathtt{T_6}$, suggesting a shift from information-heavy assimilation to more orderly, execution-driven trading. Further, bid-side entropy rates are marginally higher than ask-side, indicating slightly greater unpredictability in buy-side pricing, though this effect remains secondary.

\begin{sidewaystable}[h]
\centering
\caption{Mean recurrence times for each limit order price change state in the Markov chain.}
\label{tab:mrt_metrics}
\scriptsize
\setlength{\tabcolsep}{2.7pt}
\begin{tabular}{|c|c|*{18}{S[table-format=4.1]|}}
\hline
\textbf{\makecell{Market-\\Capitalization\\Tier}} & \textbf{\makecell{Time-\\Interval}} & \multicolumn{2}{c|}{\textbf{$S_1$}} & \multicolumn{2}{c|}{\textbf{$S_2$}} & \multicolumn{2}{c|}{\textbf{$S_3$}} & \multicolumn{2}{c|}{\textbf{$S_4$}} & \multicolumn{2}{c|}{\textbf{$S_5$}} & \multicolumn{2}{c|}{\textbf{$S_6$}} & \multicolumn{2}{c|}{\textbf{$S_7$}} & \multicolumn{2}{c|}{\textbf{$S_8$}} & \multicolumn{2}{c|}{\textbf{$S_9$}} \\
\cline{3-20}
& & \multicolumn{1}{c|}{Ask} & \multicolumn{1}{c|}{Bid}
  & \multicolumn{1}{c|}{Ask} & \multicolumn{1}{c|}{Bid}
  & \multicolumn{1}{c|}{Ask} & \multicolumn{1}{c|}{Bid}
  & \multicolumn{1}{c|}{Ask} & \multicolumn{1}{c|}{Bid}
  & \multicolumn{1}{c|}{Ask} & \multicolumn{1}{c|}{Bid}
  & \multicolumn{1}{c|}{Ask} & \multicolumn{1}{c|}{Bid}
  & \multicolumn{1}{c|}{Ask} & \multicolumn{1}{c|}{Bid}
  & \multicolumn{1}{c|}{Ask} & \multicolumn{1}{c|}{Bid}
  & \multicolumn{1}{c|}{Ask} & \multicolumn{1}{c|}{Bid} \\

\hline
\multirow{6}{*}{\textbf{$\mathtt{HMC}$}} & $\mathtt{T_1}$ & 487.9 & 297.4 & 598.2 & 306.1 & 215.1 & 139.2 & 3.8 & 3.7 & 2.2 & 2.3 & 3.8 & 3.8 & 190.0 & 159.7 & 549.3 & 352.1 & 482.1 & 294.0 \\
\cline{2-20}
                   & $\mathtt{T_2}$ & 750.1 & 380.4 & 607.8 & 345.3 & 216.9 & 134.3 & 3.6 & 3.6 & 2.3 & 2.4 & 3.6 & 3.6 & 214.1 & 135.1 & 597.0 & 346.6 & 697.2 & 377.8 \\
\cline{2-20}
                   & $\mathtt{T_3}$ & 753.6 & 378.4 & 623.2 & 393.6 & 215.3 & 143.1 & 3.7 & 3.7 & 2.2 & 2.3 & 3.7 & 3.7 & 208.5 & 146.0 & 647.1 & 380.9 & 713.5 & 380.0 \\
\cline{2-20}
                   & $\mathtt{T_4}$ & 641.3 & 463.1 & 495.8 & 570.2 & 224.8 & 158.6 & 3.8 & 3.6 & 2.2 & 2.3 & 3.7 & 3.7 & 216.3 & 161.8 & 540.6 & 526.6 & 604.0 & 463.6 \\
\cline{2-20}
                   & $\mathtt{T_5}$ & 887.4 & 368.4 & 641.6 & 484.4 & 284.4 & 140.9 & 3.7 & 3.6 & 2.3 & 2.3 & 3.7 & 3.7 & 277.2 & 142.4 & 738.0 & 442.4 & 751.0 & 377.9 \\
\cline{2-20}
                   & $\mathtt{T_6}$ & 620.1 & 305.1 & 531.7 & 278.0 & 328.8 & 161.2 & 4.1 & 4.1 & 2.0 & 2.0 & 4.1 & 4.2 & 327.7 & 146.0 & 552.4 & 273.3 & 605.1 & 292.6 \\
\hline
\multirow{6}{*}{\textbf{$\mathtt{MMC}$}} & $\mathtt{T_1}$ & 364.3 & 263.2 & 540.1 & 377.4 & 192.3 & 121.4 & 3.3 & 3.2 & 2.8 & 2.8 & 3.2 & 3.3 & 165.1 & 133.0 & 560.6 & 443.6 & 359.7 & 274.6 \\
\cline{2-20}
                   & $\mathtt{T_2}$ & 696.5 & 386.2 & 446.5 & 483.9 & 163.8 & 158.9 & 3.0 & 3.1 & 3.2 & 3.0 & 3.0 & 3.1 & 167.0 & 156.3 & 478.3 & 474.5 & 685.8 & 392.2 \\
\cline{2-20}
                   & $\mathtt{T_3}$ & 809.2 & 402.6 & 429.9 & 449.2 & 177.4 & 179.4 & 3.2 & 3.2 & 2.9 & 2.9 & 3.1 & 3.2 & 183.3 & 173.7 & 444.5 & 403.6 & 760.6 & 399.6 \\
\cline{2-20}
                   & $\mathtt{T_4}$ & 845.3 & 428.4 & 410.6 & 447.8 & 140.5 & 194.4 & 3.2 & 3.2 & 2.9 & 2.8 & 3.1 & 3.2 & 152.3 & 180.8 & 480.3 & 392.1 & 751.9 & 490.5 \\
\cline{2-20}
                   & $\mathtt{T_5}$ & 860.3 & 456.9 & 531.8 & 510.4 & 178.9 & 165.0 & 3.2 & 3.2 & 2.9 & 2.8 & 3.1 & 3.3 & 190.8 & 151.6 & 615.4 & 490.2 & 760.1 & 513.9 \\
\cline{2-20}
                   & $\mathtt{T_6}$ & 675.6 & 360.0 & 577.3 & 414.6 & 252.8 & 208.1 & 3.5 & 3.5 & 2.5 & 2.4 & 3.4 & 3.5 & 241.9 & 223.5 & 622.9 & 399.0 & 645.7 & 362.1 \\
\hline
\multirow{6}{*}{\textbf{$\mathtt{LMC}$}} & $\mathtt{T_1}$ & 224.4 & 267.2 & 380.5 & 303.3 & 108.0 & 84.1 & 3.1 & 2.9 & 3.3 & 3.5 & 2.9 & 3.0 & 101.2 & 90.2 & 386.6 & 319.1 & 208.1 & 266.9 \\
\cline{2-20}
                   & $\mathtt{T_2}$ & 329.5 & 1051.3 & 376.5 & 498.4 & 110.7 & 119.6 & 2.9 & 2.8 & 3.5 & 3.6 & 2.9 & 2.9 & 115.6 & 116.0 & 419.8 & 438.6 & 295.4 & 1003.7 \\
\cline{2-20}
                   & $\mathtt{T_3}$ & 419.2 & 869.1 & 468.2 & 399.2 & 151.0 & 112.2 & 3.0 & 2.9 & 3.3 & 3.4 & 2.9 & 3.0 & 151.6 & 108.5 & 520.9 & 401.5 & 356.3 & 727.9 \\
\cline{2-20}
                   & $\mathtt{T_4}$ & 456.1 & 723.3 & 443.3 & 394.0 & 125.9 & 144.8 & 3.0 & 2.9 & 3.3 & 3.5 & 2.9 & 2.9 & 129.2 & 132.2 & 517.7 & 352.9 & 375.0 & 666.3 \\
\cline{2-20}
                   & $\mathtt{T_5}$ & 441.8 & 581.1 & 497.6 & 381.7 & 141.3 & 120.8 & 3.0 & 2.9 & 3.3 & 3.5 & 2.9 & 2.9 & 151.0 & 105.9 & 576.0 & 356.3 & 343.4 & 534.4 \\
\cline{2-20}
                   & $\mathtt{T_6}$ & 373.8 & 362.1 & 506.7 & 291.3 & 189.3 & 112.2 & 3.2 & 3.1 & 2.8 & 2.9 & 3.2 & 3.2 & 175.9 & 115.3 & 505.8 & 306.9 & 353.6 & 346.9 \\
\hline
\end{tabular}
\end{sidewaystable}

\subsubsection{Mean Recurrence Time}

Finally, we analyze the mean recurrence times (MRT), which measures the average number of steps required for the process to return to a given state. As shown in Table~\ref{tab:mrt_metrics}, the MRTs clearly distinguish between neutral, mild, and extreme limit-order price change states across intraday intervals, market-capitalization tiers, and order side. The intraday dynamics show opposing trends at the market close: neutral-state $S_5$ MRTs compress, for example $\mathtt{HMC}$ shortens from ~2.3 to 2.0 steps, due to execution urgency, while mild-states $S_4$ and $S_6$ MRTs lengthen, for example $\mathtt{HMC}$ from ~3.7 to 4.1 steps, as price fine-tuning subsides. Extreme states are most frequent, having the shortest MRT, at the open $\mathtt{T_1}$, aligning with high initial volatility, a pattern most pronounced in $\mathtt{LMC}$ stocks, such as the Ask state $S_1$ MRT of 224.4.

Neutral changes, represented by state~$S_5$, recur most frequently with a capitalization hierarchy: $\mathtt{HMC:}$ 2.0–2.4 steps, $\mathtt{MMC:}$ 2.4–3.2, and $\mathtt{LMC:}$ 2.8–3.6. This pattern reflects more stable order maintenance in liquid stocks, whereas thinner $\mathtt{LMC}$ order books require more frequent non-neutral revisions. Mild adjustments, i.e. states~$S_4$ and~$S_6$, show an inverted pattern, occurring most often in $\mathtt{LMC}$ stocks at 2.9–3.2 steps, consistent with continuous fine-tuning under wider spreads. Moderate states i.e. $S_3$ and~$S_7$ lie between these benchmarks. Extreme states $S_1, S_2, S_8$, and $S_9$ are rare, with mean recurrence measured in hundreds of steps, in all the cases. Further, a pronounced bid-ask asymmetry appears, primarily in $\mathtt{LMC}$ extremes. For example, at $\mathtt{T_2}$, an extreme negative revision i.e. state~$S_1$ recurs in 329.5 steps on the ask side versus 1051.3 steps on the bid side. This is consistent with traders posting less aggressive bid-side limit orders to avoid accumulating long inventory in illiquid stocks.

In summary, these Markov chain metrics quantify the global dynamics of limit order price change process, confirming a clear capitalization hierarchy. $\mathtt{HMC}$ stocks are characterized by efficient price discovery with high $\gamma$, relatively higher predictability from low $H$, and stable persistence from low $S_5$ MRT, supporting patient, queue-joining trading strategies. Conversely, $\mathtt{LMC}$ stocks show a turbulent, less efficient regime with lower $\gamma$ and higher $H$, and a rapid recurrence of mild adjustments, evident from low $S_4$ and $S_6$ MRT, which necessitates dynamic, price-adaptive order management. The extreme bid-side scarcity in $\mathtt{LMC}$ further warrants more conservative bidding, while the universal compression of neutral-state recurrence times at market close requires intensified monitoring across all tiers as execution urgency increases.


\subsection{Clustering Analysis of Transition Dynamics}
\label{result:clustering_limitorder}

\begin{figure}[h]
\centering
\begin{subfigure}[t]{0.45\textwidth}
\centering
\includegraphics[width=\textwidth]{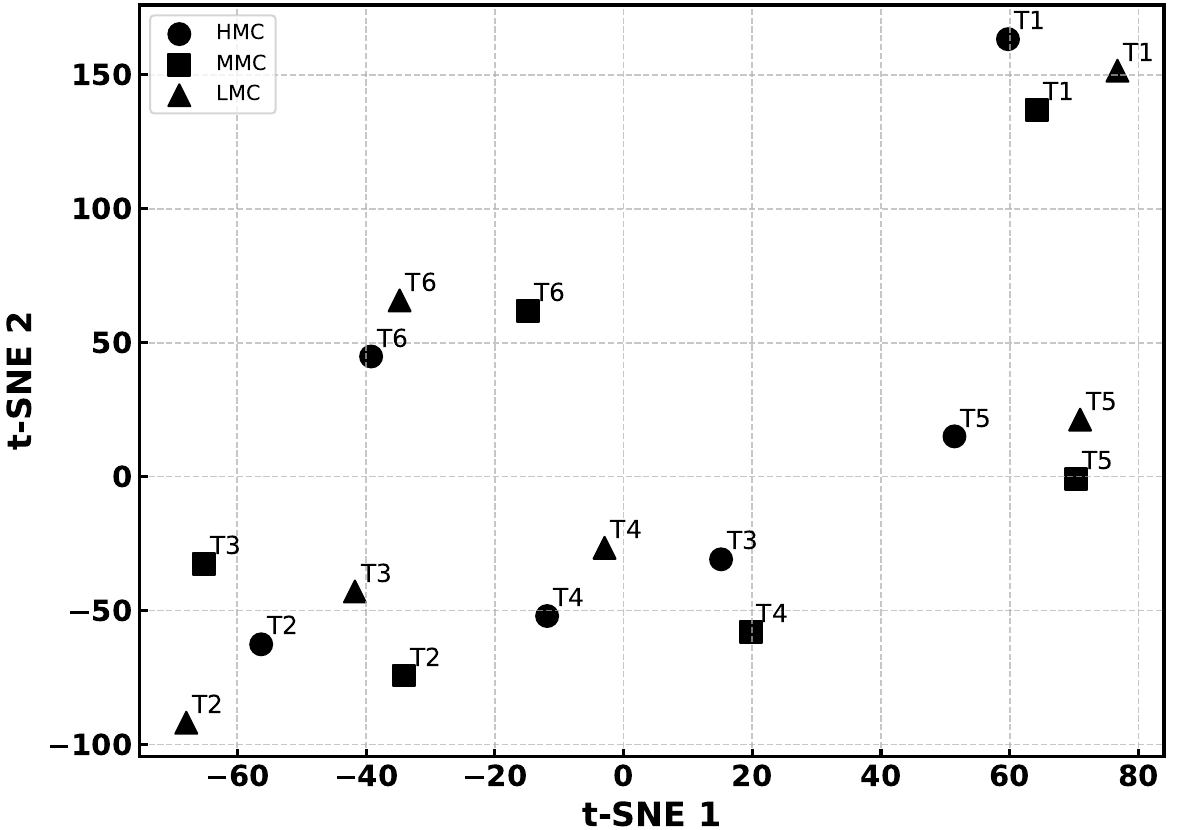}
\caption{Ask-side PCA--t-SNE}
\label{fig:Ask_PCA_tSNE}
\end{subfigure}
\hfill
\begin{subfigure}[t]{0.44\textwidth}
\centering
\includegraphics[width=\textwidth]{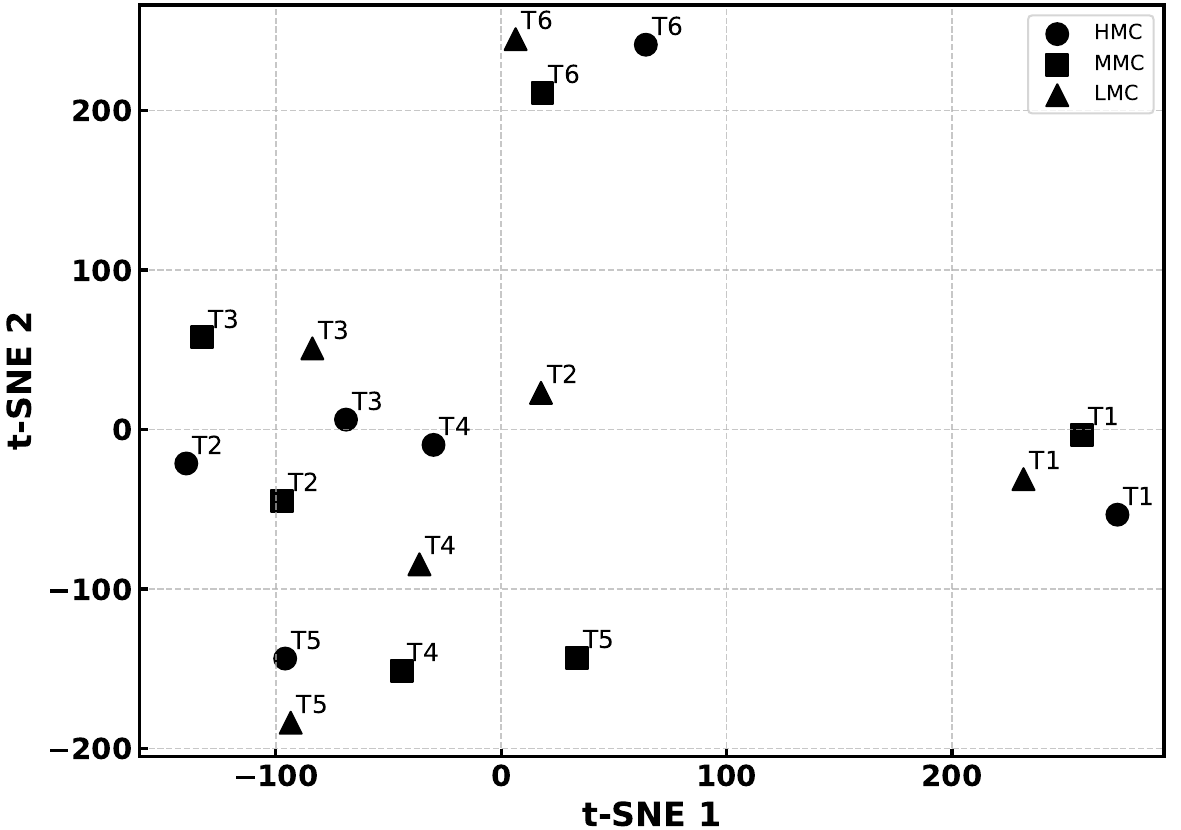}
\caption{Bid-side PCA--t-SNE}
\label{fig:Bid_PCA_tSNE}
\end{subfigure}
\caption{Two-dimensional embeddings of transition probability matrices (TPM) obtained by PCA (9 components) followed by t-SNE, for (a) ask-side and (b) bid-side limit orders. Each point represents a TPM for one of the six intraday intervals, T1, T2, .., T6 and one of the three market capitalization tiers, High, Medium and Low.}
\label{fig:pca_tsne_combined}
\end{figure}

We now examine the collective similarity of transition probability matrices (TPMs) through clustering. This examination identifies natural groupings in limit order price change behaviors across time intervals and market capitalization tiers. Each $9\times 9$ TPM is flattened into an 81-dimensional vector, yielding 18 vectors per side i.e., 6 intervals $\times$ 3 tiers. We then reduce the dimensionality of the TPMs using principal component analysis (PCA) where only 9 components are retain, followed by t-distributed stochastic neighbor embedding (t-SNE) for 2D visualization as shown in Fig.~\ref{fig:Ask_PCA_tSNE} and~\ref{fig:Bid_PCA_tSNE}, for ask and bid sides, respectively. We then apply hierarchical agglomerative clustering and DBSCAN to these embeddings.

\begin{figure}[h]
\centering
\begin{subfigure}[t]{0.45\textwidth}
\centering
\includegraphics[width=\textwidth]{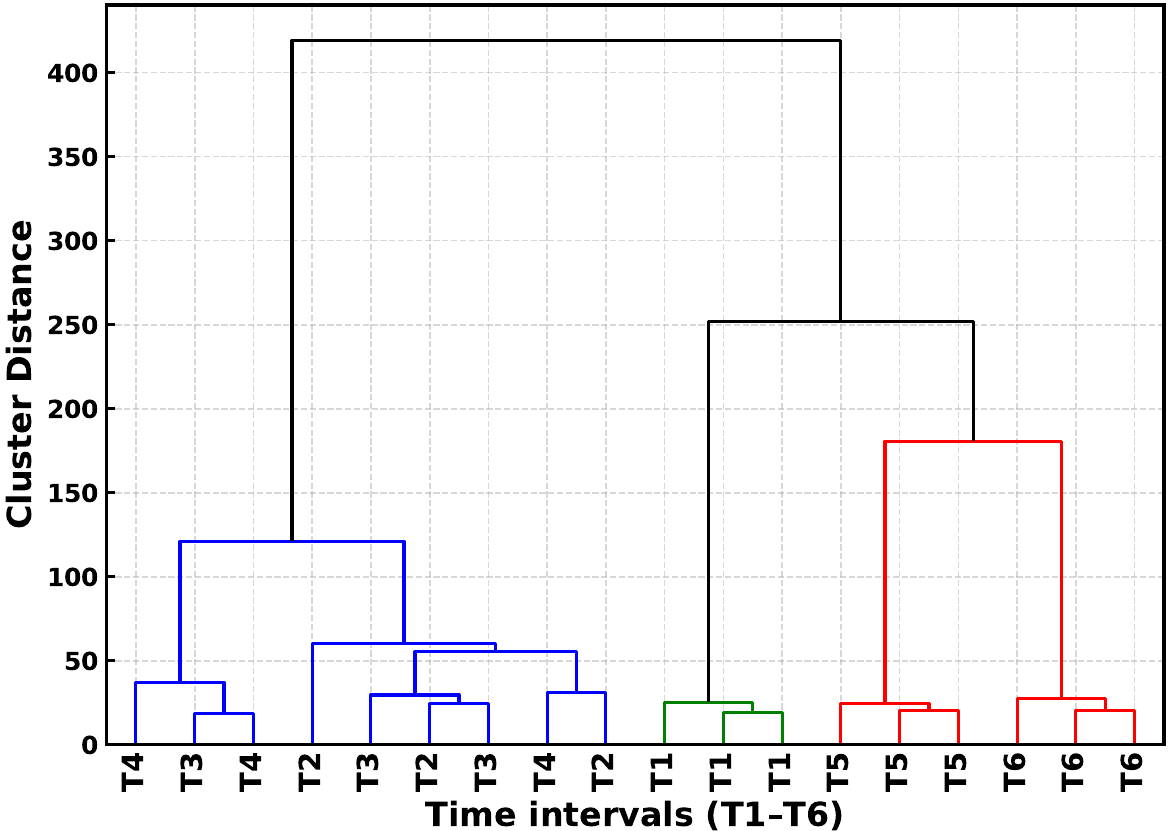}
\caption{Ask-side dendrogram}
\label{fig:Ask_Hierarchical_Dendrogram}
\end{subfigure}
\hfill
\begin{subfigure}[t]{0.45\textwidth}
\centering
\includegraphics[width=\textwidth]{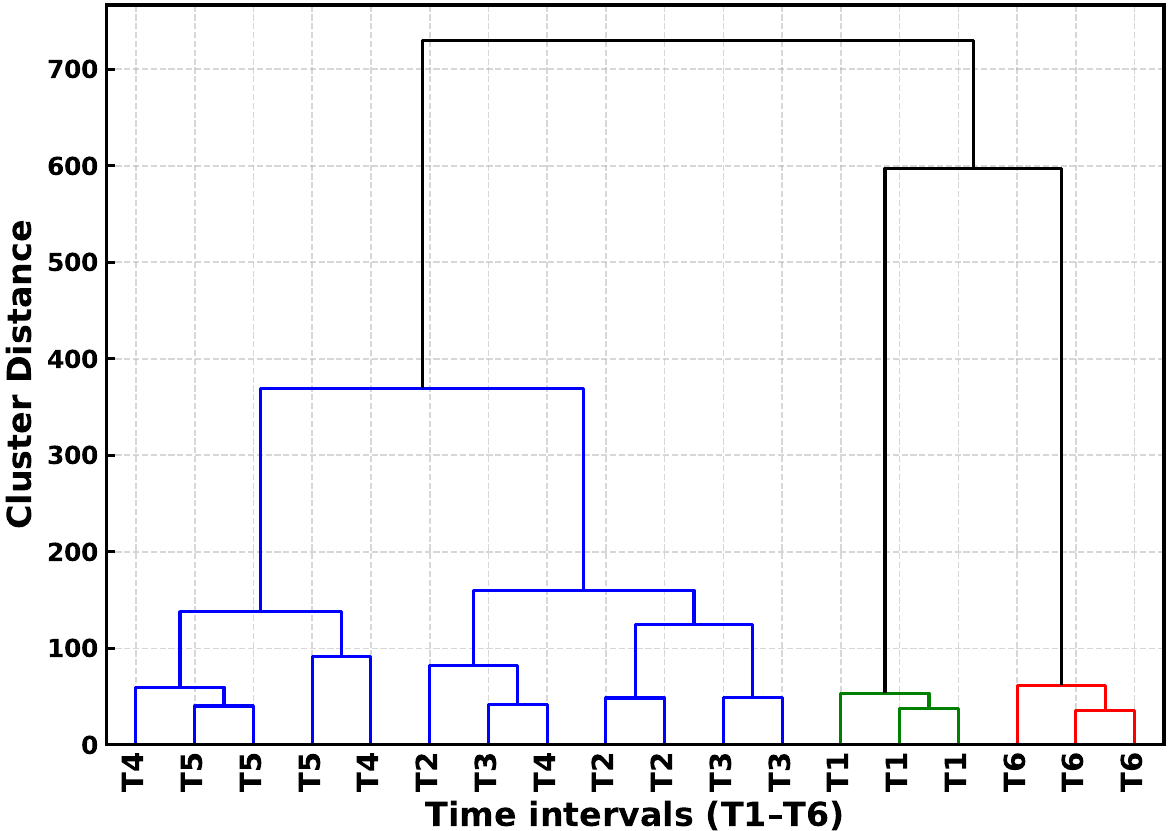}
\caption{Bid-side dendrogram}
\label{fig:Bid_Hierarchical_Dendrogram}
\end{subfigure}
\caption{Hierarchical clustering dendrograms. The Bid tree spans a larger distance range (\(\approx 0\)–\(\approx 700\)) than Ask (\(\approx 0\)–\(\approx 400\)), indicating greater heterogeneity on Bid. In both trees, \(T_1\), \(T_5\), \(T_6\) appear as tight triplets. Ask shows heavier interleaving among \(T_2\), \(T_3\), \(T_4\); Bid forms cleaner adjacent sub-blocks for \(T_2\) and \(T_3\), with \(T_4\) bridging.}
\label{fig:dendro_combined}
\end{figure}

Hierarchical clustering reveals significant ask-bid side differences in temporal structure, as shown in Figs.~\ref{fig:Ask_Hierarchical_Dendrogram} and~\ref{fig:Bid_Hierarchical_Dendrogram}. The bid side exhibits greater heterogeneity, with linkage distances ranging $0-700$ versus $0-400$ on the ask side. Despite this, the bid side shows clear temporal segmentation. Opening $\mathtt{T_1}$ and closing $\mathtt{T_6}$ hours form distinct clusters. Midday trading hours i.e., $\mathtt{T_2}$--$\mathtt{T_5}$ consolidates into a single cluster. In contrast, the ask side displays more complex temporal dynamics. Midday intervals ($\mathtt{T_2}$--$\mathtt{T_4}$) show extensive interleaving. The pre-closing hour i.e. $\mathtt{T_5}$ separates from midday to cluster with the closing hour $\mathtt{T_6}$. This suggests that sellers initiate end-of-day positioning strategies earlier than the formal closing hour.

DBSCAN analysis reinforces this key temporal asymmetry. The bid side produces three well-separated, dense clusters corresponding to the traditional trading sessions: Opening $\mathtt{T_1}$, Midday from $\mathtt{T_2}$ to $\mathtt{T_5}$, and Closing $\mathtt{T_6}$. In contrast, the ask side yields four distinct clusters: Opening $\mathtt{T_1}$, Midday from $\mathtt{T_2}$ to $\mathtt{T_4}$, and two separate end-of-day clusters: Pre-Close $\mathtt{T_5}$ and Close $\mathtt{T_6}$. The clear separation of $\mathtt{T_5}$ on the ask side provides data-driven evidence that sellers begin strategic preparations for the closing open positions significantly earlier than buyers. 

\begin{figure}[H]
\centering
\begin{subfigure}[t]{0.45\textwidth}
\centering
\includegraphics[width=\textwidth]{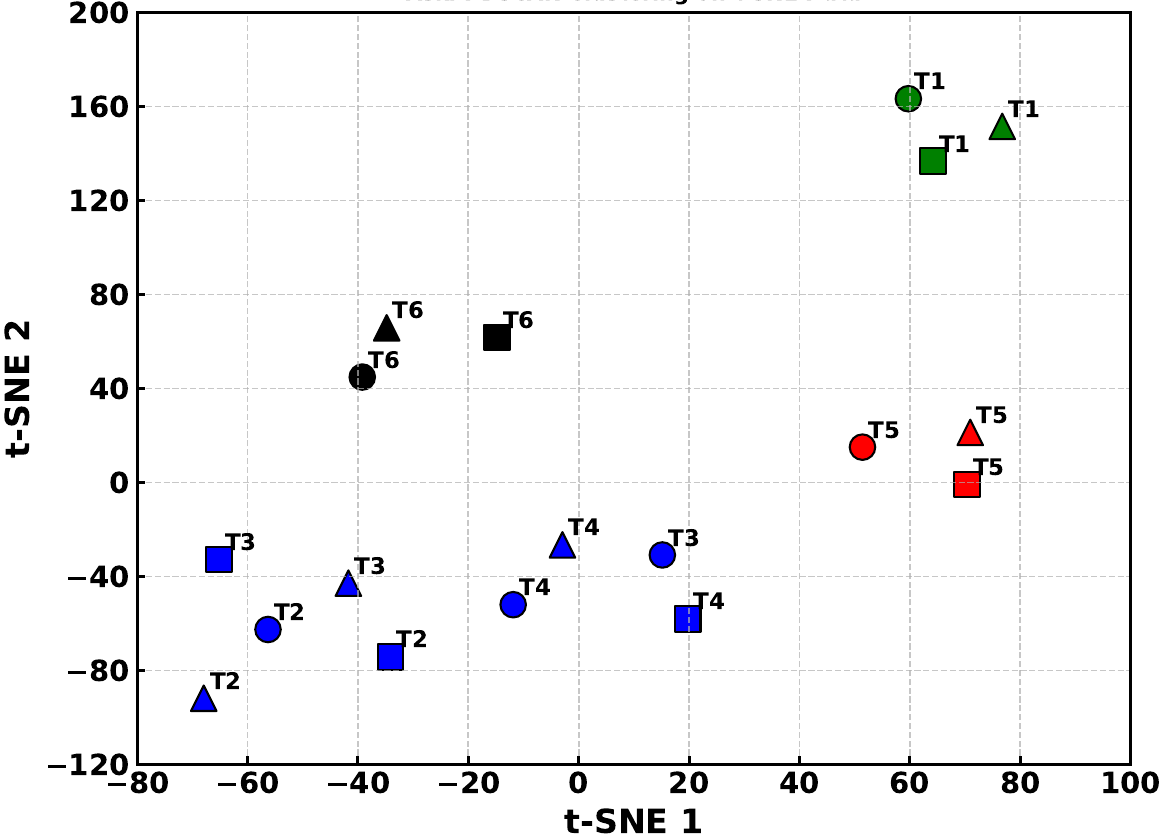}
\caption{Ask-side DBSCAN}
\label{fig:Ask_DBSCAN}
\end{subfigure}
\hfill
\begin{subfigure}[t]{0.45\textwidth}
\centering
\includegraphics[width=\textwidth]{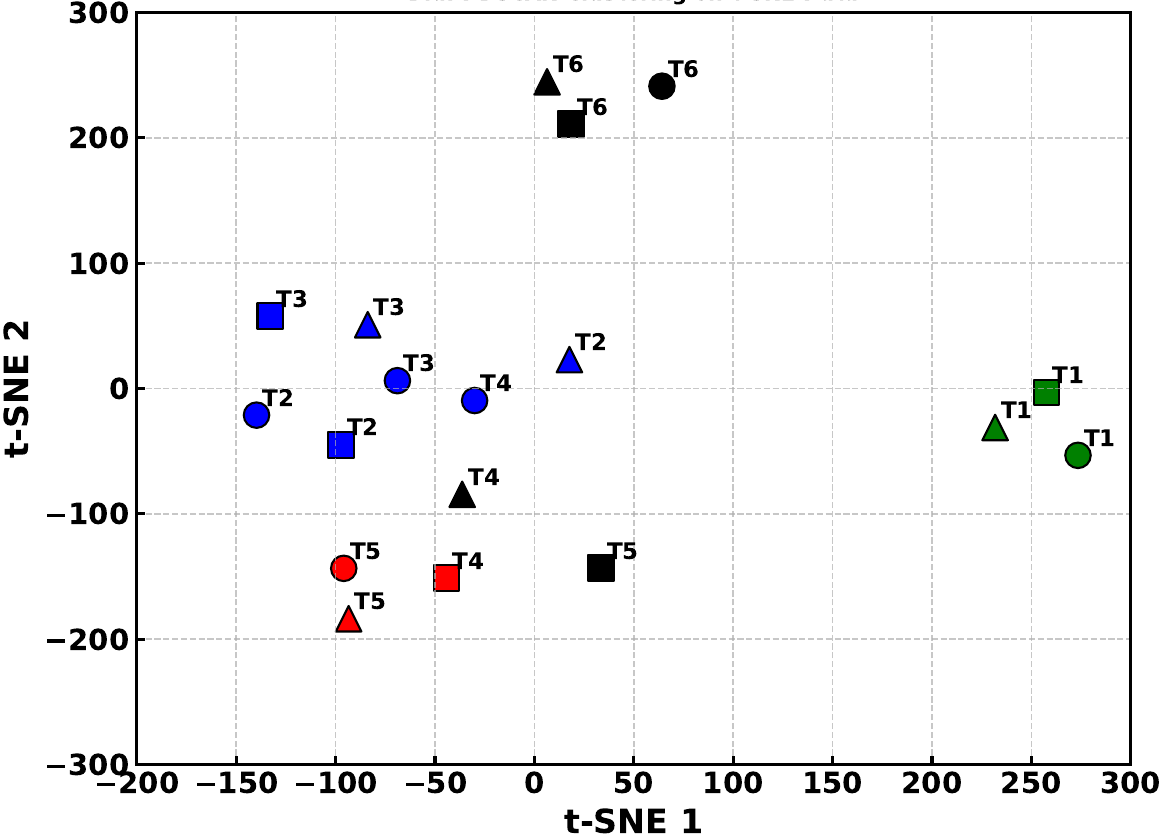}
\caption{Bid-side DBSCAN}
\label{fig:Bid_DBSCAN}
\end{subfigure}
\caption{DBSCAN on the PCA--t-SNE embeddings. With a common \(\varepsilon\)/minPts, Bid’s broader, more separable layout yields clearer Opening/Midday/Closing islands; Ask’s tighter layout produces more mid-interval blending unless \(\varepsilon\) is reduced.}
\label{fig:dbscan_combined}
\end{figure}

In summary, these clustering patterns reveal that market dynamics are primarily driven by time-of-day and order side, with market capitalization playing a secondary role. A key finding is the greater heterogeneity of the bid side, which nonetheless organizes into a clear, three-regime structure: Opening $\mathtt{T_1}$, Midday $\mathtt{T_2}$--$\mathtt{T_5}$, and Closing $\mathtt{T_6}$, that supports structured execution algorithms with predictable transition points. Conversely, the ask side, while more homogeneous, displays a more complex, four-regime structure defined by the early emergence of a $\mathtt{T_5}$ pre-closing dynamic, implying sellers begin position unwinding earlier. For traders, this necessitates asymmetric timing: bid-side logic can follow the standard intraday pattern, while ask-side logic must anticipate this early shift to closing dynamics.


\subsection{Stationary distribution of limit-order price changes}
\label{result:SD_LimitOrder}

Having identified distinct temporal regimes through clustering, we now examine the stationary distribution $\pi$ to characterize the long-term equilibrium behavior of limit order price change states. This distribution, representing the equilibrium state probabilities, reveals the predominant price change tendencies and is reported in Tables S3 and S4 of the Supplementary Material. Across all configurations, the distribution is heavily concentrated with over 97\% probability in three states: mild negative change $\pi_4$, zero change $\pi_5$, and mild positive change $\pi_6$. A consistent capitalization gradient emerges -- moving from $\mathtt{HMC}$ to $\mathtt{LMC}$, $\pi_5$ declines substantially while $\pi_4$ and $\pi_6$ increase. This confirms that lower-capitalization stocks exhibit less price inertia and require more frequent small revisions. Furthermore, the probabilities of mild upward and downward changes remain nearly symmetric, with $\pi_4 \approx \pi_6$ within each tier.

\begin{figure}[H]
\centering
\begin{subfigure}[t]{0.4\textwidth}
\centering
\includegraphics[width=\textwidth]{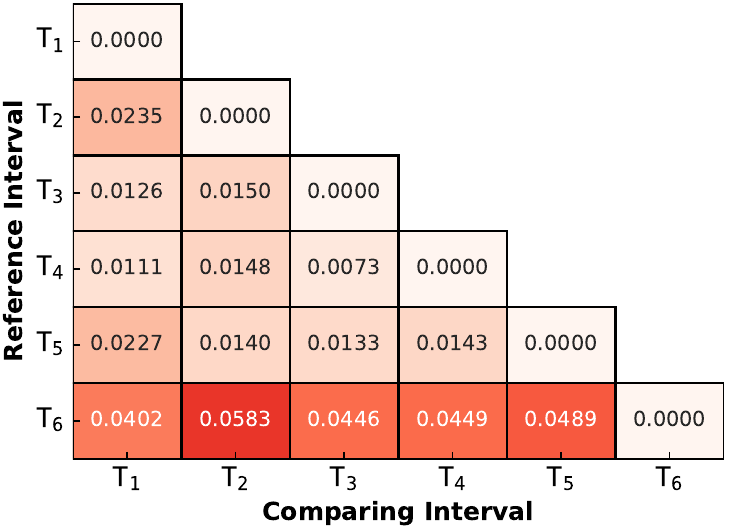}
\end{subfigure}
\hfill
\begin{subfigure}[t]{0.47\textwidth}
\centering
\includegraphics[width=\textwidth]{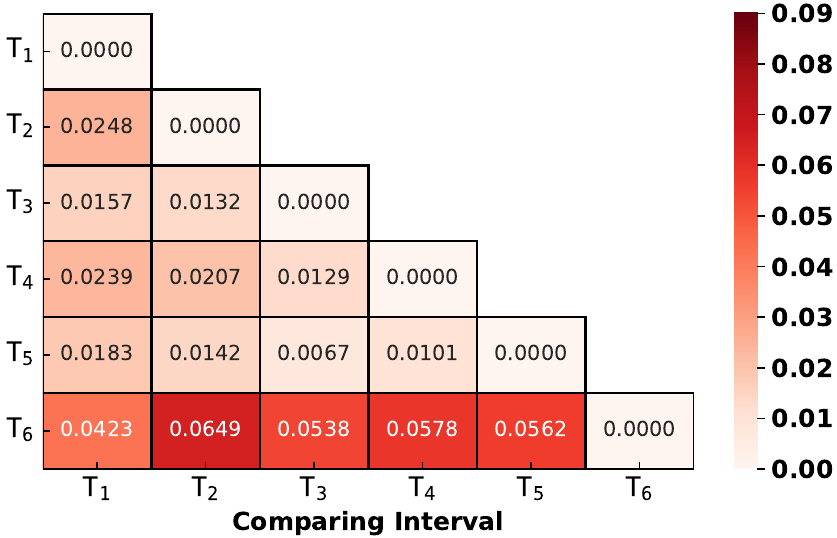}
\end{subfigure}
\caption{Jensen–Shannon Divergence between $\pi_s$ of different time intervals for $\mathtt{HMC}$ limit order price changes: \textbf{Ask} on left and \textbf{Bid} on right.}
\label{tab:JSD_HMC}
\end{figure}

\begin{figure}[H]
\centering
\begin{subfigure}[t]{0.4\textwidth}
\centering
\includegraphics[width=\textwidth]{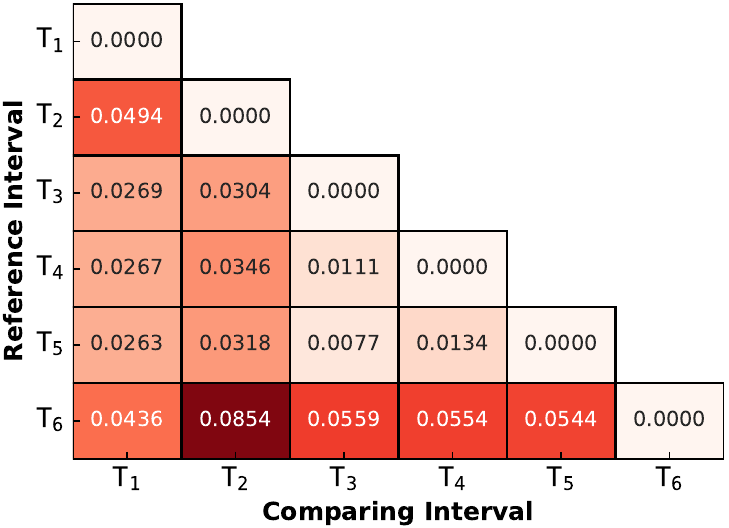}
\end{subfigure}
\hfill
\begin{subfigure}[t]{0.47\textwidth}
\centering
\includegraphics[width=\textwidth]{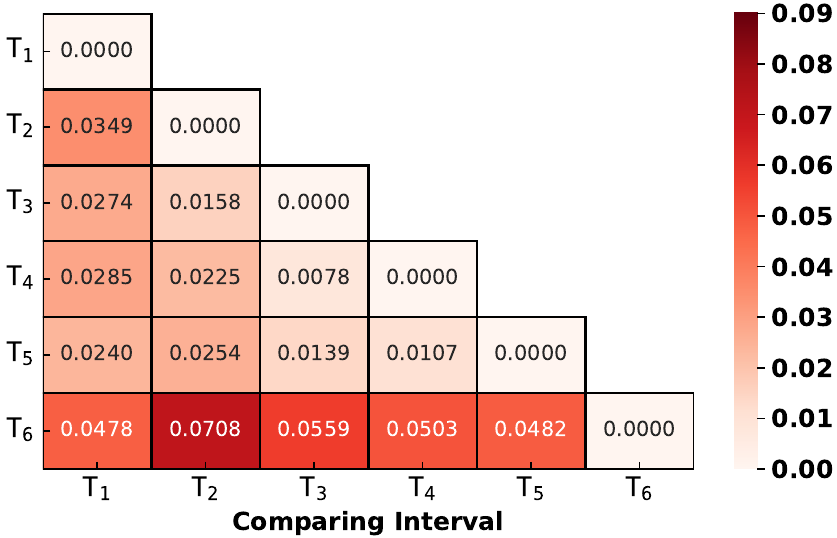}

\end{subfigure}
\caption{Jensen–Shannon Divergence between $\pi_s$ of different time intervals for $\mathtt{MMC}$ limit order price changes: \textbf{Ask} on left and \textbf{Bid} on right.}
\label{tab:JSD_MMC}
\end{figure}

\begin{figure}[h]
\centering
\begin{subfigure}[t]{0.4\textwidth}
\centering
\includegraphics[width=\textwidth]{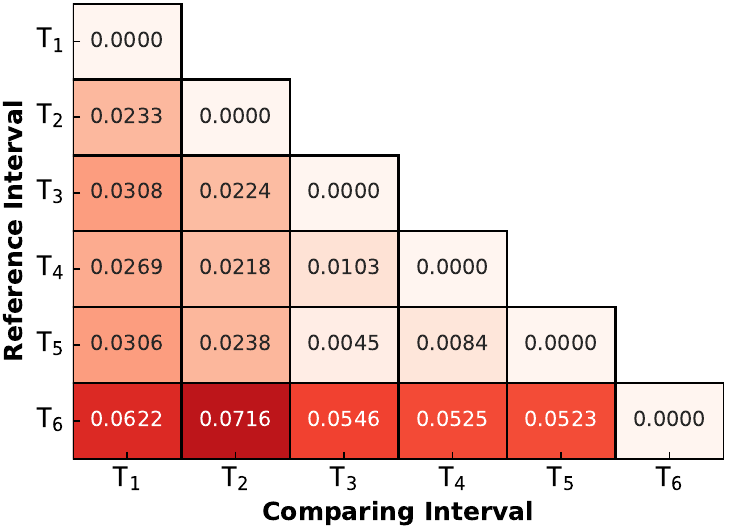}
\end{subfigure}
\hfill
\begin{subfigure}[t]{0.46\textwidth}
\centering
\includegraphics[width=\textwidth]{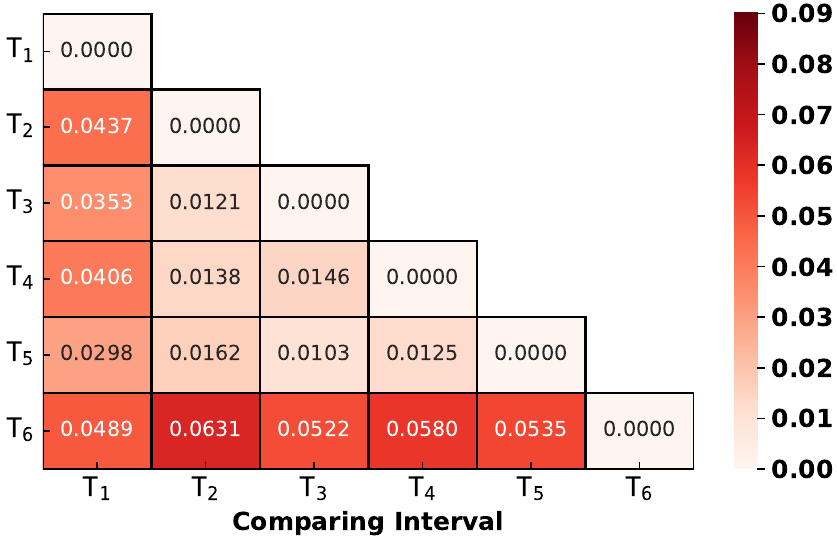}
\end{subfigure}
\caption{Jensen–Shannon Divergence between $\pi_s$ of different time intervals for $\mathtt{LMC}$ limit order price changes: \textbf{Ask} on left and \textbf{Bid} on right.}
\label{tab:JSD_LMC}
\end{figure}

To quantify the dissimilarity between these stationary distributions across different time intervals, we compute the Jensen-Shannon Divergence (JSD), as visualized in the heatmaps in Figs.~\ref{tab:JSD_HMC}--\ref{tab:JSD_LMC}. The JSD analysis reveals a robust temporal structure common to both ask and bid sides. The closing hour $\mathtt{T_6}$ consistently emerges as the most distinct, showing the largest divergences from the midday block $\mathtt{T_2}$--$\mathtt{T_5}$ across all tiers. This pronounced divergence signals a fundamental shift in market participants' objectives: as the trading session concludes, the primary incentive transitions from value-seeking price discovery to deadline-driven inventory management. The urgency to square positions and mitigate overnight risk forces a structural reconfiguration of limit order submission probabilities, creating a unique equilibrium state that differs significantly from the steady liquidity provision characterizing the midday regime. A secondary distinct shift occurs immediately post-open, with $\mathtt{T_2}$ diverging significantly from $\mathtt{T_1}$, likely driven by the resolution of initial information asymmetry. In contrast, the midday period spanning $\mathtt{T_2}$ to $\mathtt{T_5}$ remains relatively stable, characterized by low pairwise JSD values between adjacent midday intervals. Comparing sides, the bid generally exhibits larger JSD values, particularly at the close and post-open, confirming previous findings from clustering and metrics that bid-side behavior varies more and changes more significantly between periods, while the ask side is more consistently stable during midday.

Market capitalization further modulates the temporal divergences. $\mathtt{HMC}$ stocks exhibit the smallest overall JSD differences with $\mathtt{T_6}$ departures in the range of 0.040–0.049 on ask and 0.042–0.056 on bid. $\mathtt{T_2}$–$\mathtt{T_1}$ breaks measure approximately 0.02–0.03, reflecting greater stability across trading phases. Whereas, $\mathtt{MMC}$ stocks demonstrate the most pronounced closing differentiation. This pattern is particularly strong on the ask side where $\mathtt{T_6}$ versus $\mathtt{T_2}$ reaches 0.085. Bid-side $\mathtt{T_6}$ gaps also remain elevated. Finally, $\mathtt{LMC}$ stocks feature sharp post-open adjustments. The bid-side $\mathtt{T_2}$ versus $\mathtt{T_1}$ divergence equals 0.0437. Closing-hour divergences are substantial though remain below $\mathtt{MMC}$ peaks. These systematic variations confirm that intraday phase remains the primary organizing force, while capitalization modulates both the intensity of temporal contrasts and the degree of ask–bid asymmetry. These findings necessitate capitalization-aware parameterization: tighter monitoring thresholds for $\mathtt{LMC}$ stocks during the post-open period i.e. between $\mathtt{T_1}$ \& $\mathtt{T_2}$, and the lead-up to the close i.e. between $\mathtt{T_5}$ \& $\mathtt{T_6}$, and more conservative bid-side risk management during phase shifts across all tiers.

In summary, the stationary distribution analysis confirms that limit order dynamics are predominantly characterized by neutral and mild price changes, with the lower probability of neutral states in $\mathtt{LMC}$ stocks underscoring their need for more frequent revisions. The JSD analysis provides strong quantitative evidence for the three distinct intraday phases—Open, Midday, and Close—previously identified via clustering. Crucially, it demonstrates that market capitalization systematically modulates the intensity of divergences between these phases, with $\mathtt{HMC}$ stocks showing the greatest stability and $\mathtt{LMC}$ exhibiting amplified shifts, especially post-open. These findings collectively reinforce the necessity of phase-aware and capitalization-aware trading strategies. Furthermore, the consistently higher JSD values on the bid side confirm its greater heterogeneity, warranting tighter risk controls for buy-side liquidity provision during these critical phase transitions.

\begin{figure}[htbp]
    \centering
    \resizebox{1.0\textwidth}{!}{%
    \begin{tikzpicture}[
        node distance=0.8cm and 1.2cm,
        base/.style={rectangle, rounded corners, fill=white, thick, font=\footnotesize},
        process_data/.style={base, draw=cyan!60!black, align=left},
        process_model/.style={base, draw=orange!60!black, align=left},
        process_analysis/.style={base, draw=green!60!black, align=left},
        arrow/.style={-{Latex[length=3mm]}, thick, gray!50!black},
        line/.style={thick, gray!50!black},
        phase label/.style={font=\bfseries\footnotesize, text=black!70}
    ]
        \pgfdeclarelayer{background}
        \pgfsetlayers{background,main}

        \node (raw) [process_data, text width=7.5cm, align=center] {\centering \textbf{High-Frequency Data}\\(NASDAQ 100 Tick-by-tick order submissions)};
        
        \node (prep) [process_data, below=0.6cm of raw, text width=7.2cm, inner sep=5pt] {%
            \centering \textbf{Data Pre-processing}
            \begin{itemize}[leftmargin=1.2em, nosep, topsep=3pt]
                \item Extract limit orders: ADD-ASK \& ADD-BID.
                \item Select 5 stocks per capitalization tier $c \in \{\mathtt{HMC, MMC, LMC}\}$.
                \item Segment trading time $\tau$ into $\mathtt{T_1, \dots, T_6}$.
            \end{itemize}
        };

        \node (discretize) [process_data, below=0.6cm of prep, text width=5.0cm, inner sep=5pt] {%
            \centering \textbf{State Discretization}\\[2pt]
            Discretize consecutive limit order price changes -- 9 states.
        };

        \node (gtest) [process_data, below=0.6cm of discretize, text width=5.3cm, inner sep=5pt, align=center] {\centering \textbf{G-Test of Independence}\\Assess memory for Markov validity to limit order price changes.};

        \node (tpm) [process_model, below=1.2cm of gtest, text width=5.8cm, inner sep=5pt] {%
            \centering \textbf{Discrete-Time Markov Chain}
            \begin{itemize}[leftmargin=1.2em, nosep, topsep=3pt]
                \item $1^{\text{st}}$ Order time-homogeneous DTMC.
                \item Estimate TPMs $\mathbf{P}^{(\tau,c)}$ via MLE.
                \item Compare $p_{ij}$ across $\tau$ and $c$.
            \end{itemize}
        };

        \coordinate (fork) at ($(tpm.south) + (0,-0.8cm)$);

        \node (metrics) [process_analysis, below=1.8cm of tpm, xshift=-5.5cm, text width=4.5cm, inner sep=5pt] {%
            \centering \textbf{Markov Chain Metrics}
            \begin{itemize}[leftmargin=1.2em, nosep, topsep=2pt]
                \item Spectral gap.
                \item Entropy rate.
                \item Mean recurrence time.
            \end{itemize}
        };

        \node (dimred) [process_analysis, below=1.8cm of tpm, text width=4.3cm, inner sep=5pt] {%
            \centering \textbf{Dimensionality Reduction}
            \begin{itemize}[leftmargin=0.9em, nosep, topsep=2pt]
                \item Flatten TPMs: ($10 \times 10$) $\to$ ($1 \times 100$).
                \item PCA (Noise reduction) $\to$ t-SNE (2D Embedding).
            \end{itemize}
        };
        
        \node (cluster) [process_analysis, below=0.6cm of dimred, text width=4.5cm, inner sep=5pt, align=center] {%
            \centering \textbf{Clustering Analysis}\\[2pt]
            Hierarchical \& DBSCAN
        };

        \node (stat) [process_analysis, below=1.8cm of tpm, xshift=5.8cm, text width=5.7cm, inner sep=5pt] {%
            \centering \textbf{Equilibrium Analysis}
            \begin{itemize}[leftmargin=0.9em, nosep, topsep=2pt]
                \item Compute stationary distribution $\pi^{(\tau,c)}$.
                \item Compare $\pi_s$ across $\tau$ via JS divergence.
            \end{itemize}
        };

        \draw[arrow] (raw) -- (prep);
        \draw[arrow] (prep) -- (discretize);
        \draw[arrow] (discretize) -- (gtest);
        \draw[arrow] (gtest) -- (tpm);
        
        \draw[line] (tpm.south) -- (fork);
        \draw[arrow] (fork) -| (metrics.north);
        \draw[arrow] (fork) -- (dimred.north);
        \draw[arrow] (dimred) -- (cluster);
        \draw[arrow] (fork) -| (stat.north);

        \begin{pgfonlayer}{background}
            \node[fit=(raw)(prep)(discretize)(gtest), fill=cyan!5, rounded corners, draw=cyan!30, dashed, inner sep=12pt] (p1) {};
            \node[phase label, anchor=north] at (p1.north) {Data Preparation};

            \node[fit=(tpm), fill=orange!5, rounded corners, draw=orange!30, dashed, inner sep=10pt] (p2) {};
            \node[phase label, anchor=north, xshift=-1.4cm, yshift=0.1cm] at (p2.north) {Modeling};

            \node[fit=(metrics)(dimred)(cluster)(stat), fill=green!5, rounded corners, draw=green!30, dashed, inner sep=10pt] (p3) {};
            \node[phase label, anchor=north, xshift=-2.0cm, yshift=0.1cm] at (p3.north) {Analysis};
        \end{pgfonlayer}
    \end{tikzpicture}
    }
    \caption{Integrated framework flowchart illustrating: (1) Data preparation and Markov validation, (2) Transition probability estimation, and (3) Comparative analysis via Markov chain metrics, clustering and equilibrium (stationary) states.}
    \label{fig:method-flowchart}
\end{figure}
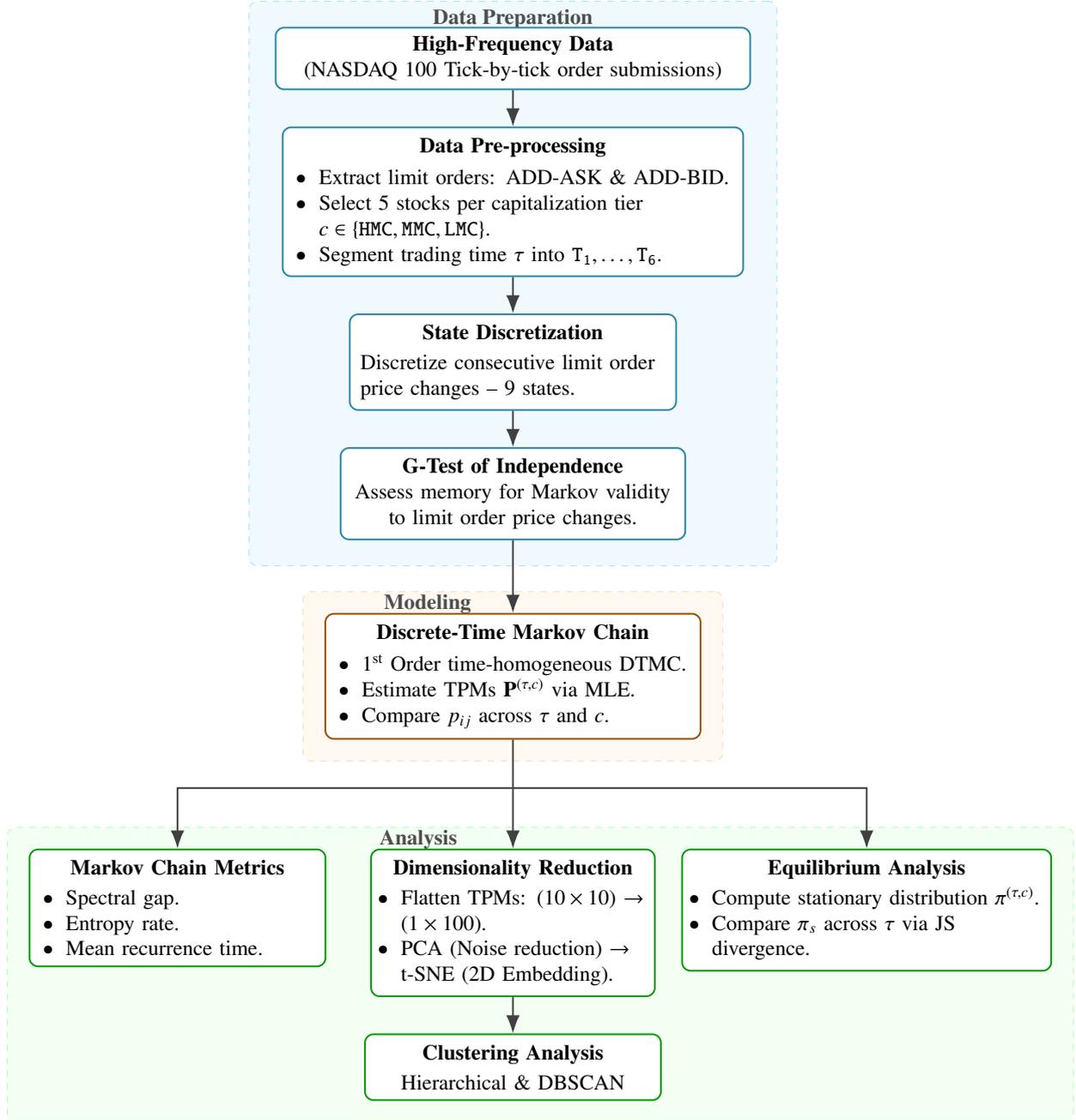

To provide a holistic overview before concluding in Section~\ref{sec:Conc}, we present the overall study workflow in Fig.~\ref{fig:method-flowchart} alongside a structured summary of the key empirical findings from Section~\ref{sec:Results}.

\begin{itemize}
    \item \textbf{~\ref{result:TPM_Order}: Transition Probability Matrix Analysis}
    \begin{itemize}
        \item \textit{\textbf{~\ref{subsec:inertia} Inertia of Limit Order Prices:}} Price inertia follows a U-shaped intraday pattern; Capitalization gradient exists where $\mathtt{HMC}$ stocks exhibit the strongest inertia, indicating higher price stability, while $\mathtt{LMC}$ stocks display lower stability.
        
        \item \textit{\textbf{~\ref{subsec:drc_momtm} Directional Momentum:}} $\mathtt{LMC}$ stocks show the highest price revision frequency. Directional asymmetry is evident at the open, where $\mathtt{HMC}$ stocks increase revision activity for positioning, while $\mathtt{LMC}$ traders adopt more conservative strategies.
        
    \end{itemize}

    \item \textbf{~\ref{result:DTMC_metrics}: Markov Chain Metrics}
    \begin{itemize}
        \item \textit{\textbf{Convergence \& Predictability:}} Spectral gap is lowest at the market open and peaks during midday; $\mathtt{HMC}$ stocks show lower entropy rates, indicating higher predictability and faster convergence to equilibrium compared to the more stochastic $\mathtt{LMC}$ stocks.
        \item \textit{\textbf{Recurrence Times:}} Neutral price changes recur most frequently in $\mathtt{HMC}$ stocks, whereas mild price adjustments recur most frequently in $\mathtt{LMC}$ stocks, indicating continuous fine-tuning.
    \end{itemize}

\newpage

    \item \textbf{~\ref{result:clustering_limitorder}: Clustering Analysis of Transition Dynamics}
    \begin{itemize}
        \item \textit{\textbf{Temporal Regimes:}} Data-driven clustering reveals that the trading time intervals organize into naturally occurring behavioral phases.
        
        \item \textit{\textbf{Bid-Ask Asymmetry:}} Bid side follows a three-regime structure -- Opening, Midday, Closing, while the ask side exhibits four regimes -- Opening, Midday, Pre-Close, Closing, providing evidence that sellers initiate end-of-day positioning strategies earlier than buyers.
    \end{itemize}

    \item \textbf{~\ref{result:SD_LimitOrder}: Stationary Distribution Analysis}
    \begin{itemize}
        \item \textit{\textbf{Long-term Equilibrium:}} Equilibrium probabilities are heavily concentrated in neutral and mild price changes; the probability of neutral price change declines substantially from $\mathtt{HMC}$ to $\mathtt{LMC}$, confirming the higher revision necessity in lower-capitalization tiers.

        \item \textit{\textbf{Closing Dynamics:}} Closing hour $\mathtt{T_6}$ forms the most distinct regime, driven by a structural shift to deadline-driven inventory management. This divergence from the stable midday phase is sharpest in $\mathtt{LMC}$ stocks.
        
    \end{itemize}
\end{itemize}

\section{Discussion}
\label{disscusion}

While this study provides evidence on intraday limit order price change transitions, we recognize that the empirical findings can be sensitive to modeling and sampling choices.
Our analysis focuses on NASDAQ100 stocks during a twelve-day period, employs first-order Markov chains for modeling state transitions, and divides the trading day into six unequal-duration intervals. Each of these choices, while empirically motivated, represents one among several plausible approaches. Accordingly, we explicitly assess whether the main conclusions are stable under alternative specifications, so that the documented patterns are not driven by arbitrary design decisions.

To ensure our findings reflect genuine market patterns rather than methodological artifacts, we conducted four comprehensive robustness analyses that assess the sensitivity of our conclusions. Detailed methodologies and complete results are provided in Supplementary Material Section S3. 
For clarity, we summarize here the objective and outcome of each test, emphasizing the practical (rather than purely statistical) magnitude of deviations. First, we validated the time-homogeneity assumption by subdividing each interval into four sub-periods. 
While formal likelihood ratio tests detected minor fluctuations (as expected under large sample sizes),
mean absolute differences remained well below the $10\%$ statistical threshold, confirming practical stability of transition probabilities within intervals. 
This supports the interpretation of each intraday segment as approximately stationary in transition dynamics. Second, we evaluated our first-order Markov specification against a second-order alternative. Although the latter was statistically favored, conditional mutual information was negligible, out-of-sample predictive gains were minimal, and the second-order specification required nearly four times as many parameters, supporting our parsimonious first-order choice. In other words, higher-order dependence exists statistically but contributes little incremental explanatory or predictive content relative to the complexity it introduces in a regime-comparison setting. Third, we assessed sensitivity to interval duration by comparing pooled one-hour versus two-hour aggregations. 
he resulting transition probability matrices were numerically indistinguishable at the level relevant for the metric, clustering, and divergence analyses. Fourth, we examined our unequal-duration interval specification against uniform 65-minute segmentation. The transition probabilities showed low mean absolute differences, with no statistically significant differences after multiple testing correction. Taken together, these exercises indicate that the main intraday patterns---including the capitalization gradient, the temporal clustering structure, and the bid--ask asymmetries---are not artifacts of interval design. Overall, the robustness checks support the conclusion that the documented transition dynamics reflect stable features of the underlying order submission and revision process within the scope of our dataset.

Beyond statistical validity, the results also carry direct practical implications for market participants. For execution algorithms, the distinct capitalization gradient we document suggests that static, patient execution logic is well suited for $\mathtt{HMC}$ stocks, where price adjustments are infrequent and stable, whereas $\mathtt{LMC}$ stocks require more dynamic, price-adaptive strategies to accommodate higher revision intensities. With respect to intraday regime detection, our clustering results show that market phases are better characterized by behavioral shifts such as the early onset of ask-side closing dynamics, rather than by fixed clock-based partitions, offering a data-driven approach to regime switching. Finally, liquidity provision strategies can be improved by incorporating the documented bid–ask asymmetries -- in particular, market makers may tighten risk controls on the bid side for $\mathtt{LMC}$ stocks, where extreme negative revisions are structurally less frequent, thereby enabling more efficient inventory and spread management. The stability of the transition dynamics across alternative temporal aggregations, interval specifications, and Markov orders suggests that the behaviors documented in this study are robust within modern electronic limit order book trading.

While our methodological choices are validated within the current scope, several extensions remain valuable. Applying the framework to other markets, asset classes, and longer time horizons would enable assessment of generalizability and detection of structural changes. In particular, extending the sample beyond a short window would clarify the extent to which the identified regimes persist across volatility states, macro announcements, and market-wide liquidity conditions.
Exploring non-Markovian specifications may capture additional dependence in sequential price adjustments. For example, semi-Markov or variable-length specifications could directly account for state-dependent sojourn times and heterogeneous memory, potentially refining the characterization of intraday persistence. Finally, although our nine-state discretization captures key dynamics, alternative state definitions or continuous-state approaches may reveal further nuances in limit order price revisions. These may include state definitions anchored to tick-size constraints, queue-position proxies, or spread-conditioned revisions, which could improve interpretability in specific microstructure settings.
These extensions would help generalize and deepen the insights developed in this study.

\section{Conclusion}
\label{sec:Conc}

This study provides the first systematic examination of intraday limit order price change transition dynamics, addressing a critical gap in market microstructure research by analyzing ask and bid orders separately across High ($\mathtt{HMC}$), Medium ($\mathtt{MMC}$), and Low ($\mathtt{LMC}$) market capitalization stocks. By employing a discrete-time Markov chain (DTMC) framework on high-frequency tick-by-tick NASDAQ-100 data, we have uncovered fundamental patterns that govern how traders adjust limit order prices throughout the trading day, revealing complex interactions between temporal dynamics, market capitalization, and order side.

Time-interval wise comparison of transition probability matrices reveals systematic intraday patterns in price change inertia. The probability of consecutive zero price changes exhibits a distinct temporal structure: peaking at market opening due to defensive positioning, declining and stabilizing during midday as price discovery progresses, and surging again at the close—often exceeding the opening peak—driven by execution urgency. A pronounced capitalization gradient emerges in these dynamics. $\mathtt{HMC}$ stocks exhibit the strongest price inertia reflecting deep liquidity, while $\mathtt{LMC}$ stocks demonstrate lower stability and pronounced bid-ask asymmetries. In extreme price change states, ask-side transition probabilities consistently exceed bid-side probabilities in $\mathtt{LMC}$ stocks, with extreme negative bid-side revisions recurring less frequently. Markov chain metrics quantify the global dynamic properties of these processes. The spectral gap, smallest at market open and peaking midday, indicates that $\mathtt{HMC}$ and $\mathtt{MMC}$ stocks converge faster to equilibrium than $\mathtt{LMC}$ stocks. The entropy rate reveals a clear capitalization hierarchy, with $\mathtt{LMC}$ stocks exhibiting the highest unpredictability. Mean recurrence times show that neutral changes recur most frequently in $\mathtt{HMC}$ stocks, while mild adjustments occur most often in $\mathtt{LMC}$ stocks. Extreme states are rare across all configurations, with pronounced bid-ask asymmetry in $\mathtt{LMC}$ stocks. These findings enable execution strategy optimization: patient approaches in high-capitalization stocks during midday versus dynamic strategies in low-capitalization stocks during phase shifts.

Clustering analysis reveals distinct temporal regimes differing fundamentally between order sides. The bid side organizes into three regimes—Opening, Midday, and Closing—while the ask side exhibits four—Opening, Midday, Pre-Close, and Close—providing data-driven evidence that sellers begin strategic positioning earlier than buyers. Stationary distributions show heavy concentration with over 97\% probability in mild negative, zero, and mild positive change states. Zero-change probability declines substantially from $\mathtt{HMC}$ to $\mathtt{LMC}$ stocks, confirming more frequent small revisions in lower-capitalization stocks. Jensen-Shannon divergence computed between stationary distributions across time intervals reveals robust temporal structure. The closing hour emerges as the most distinct phase with largest divergences from midday, while a secondary shift occurs post-open. The bid side exhibits larger divergence values, confirming greater variability. Market capitalization modulates these contrasts: $\mathtt{HMC}$ stocks show smallest differences reflecting stability, MMC stocks demonstrate pronounced closing differentiation, and $\mathtt{LMC}$ stocks feature sharp post-open adjustments. These variations confirm intraday phase as the primary organizing force, with capitalization modulating the intensity of temporal contrasts and the degree of bid-ask asymmetry. The early emergence of ask-side closing dynamics necessitates asymmetric timing in algorithmic strategies.

This paper demonstrates that limit order price change dynamics are neither random nor homogeneous but follow systematic patterns shaped by the interplay of intraday timing, market capitalization, and order side. The DTMC framework proves effective in capturing both short-term sequential dependencies through transition probabilities and long-term equilibrium behaviors through stationary distributions. By revealing the fundamental asymmetries between ask and bid orders and quantifying how these dynamics vary across liquidity regimes, this research advances understanding of modern equity markets and provides a foundation for developing context-aware trading strategies. Future research could extend this framework by incorporating order book depth and liquidity measures to analyze how price changes interact with available liquidity at different price levels, and by examining the impact of specific news events or macroeconomic releases on transition dynamics to illuminate how information shocks propagate through limit order pricing behavior. As electronic trading continues to dominate financial markets, the insights from this study become increasingly relevant for all market participants seeking to navigate the complex, high-frequency environment of contemporary limit order books.


\section*{Acknowledgements}

We extend our gratitude to Chris Bartlett, Jodhie Cabarles, and the technical support staff of \href{https://www.algoseek.com}{Algoseek} for generously providing the data and offering assistance with data preprocessing necessary for our analysis. The authors, SR Luwang, K Mukhia, and BN Sharma, would also like to thank the Director of our institute for allocating doctoral research fellowship.

\section*{Declarations}

\medskip
\noindent\textbf{Competing interests} \\
The authors declare no competing interests.

\medskip

\noindent\textbf{Data Availability Statement} \\
The full dataset that support the findings of this study are available from \href{https://www.algoseek.com}{Algoseek}. Due to licensing restrictions, the underlying tick-level data of these TPMs cannot be redistributed by the authors and requests for access should be directed to \href{https://www.algoseek.com}{Algoseek}.

\medskip
\noindent\textbf{Author contribution} \\
SR Luwang: Conceptualization, Methodology, Data curation, Formal analysis, Visualization, Writing - original draft. K Mukhia: Data curation, Writing - review \& editing. BN Sharma: Formal analysis, Writing - review \& editing. Md. Nurujjaman: Supervision, Writing - review \& editing. Anish Rai: Software, Writing - review \& editing. Filippo Petroni: Writing - review \& editing, Validation. All the authors discussed the results and approved the final manuscript.

\bibliographystyle{elsarticle-num} 
\bibliography{References} 

\clearpage 


\section*{Supplementary material}

# Supplementary Material: Intraday Limit Order Price Change Transition Dynamics Across Market Capitalizations Through Markov Analysis

This document provides the supplementary material associated with the main article *Intraday Limit Order Price Change Transition Dynamics Across Market Capitalizations Through Markov Analysis*. The content is organized as follows, presented in the order referenced in the main manuscript.

**Section S1. G-Test**
S1 presents supplementary Tables S1 and S2. These tables report the G-statistics and *p*-values that assess the Markov property by testing and rejecting the null hypothesis of independence between price change states. The average count of limit order price change categories are also presented in Fig. S1.

**Section S2. Transition Probability Matrices**
S2 contains supplementary Figures S2–S5. These figures display the complete set of transition probability matrices for both the ask and bid sides, across all market capitalization tiers and intraday time intervals.

**Section S3. Stationary Distribution of Price Change States**
S3 provides Tables S3 and S4. These tables present the estimated stationary distributions for all market capitalization tiers, for both ask and bid sides, and for each intraday time interval.

**Section S4. Robustness Analyses**
S4 details the robustness analyses conducted to validate the study's methodology and findings. This section is further divided into two subsections:

- **S4(a) Methodologies.** This subsection provides the formal mathematical descriptions for all robustness and sensitivity checks, including tests for time-homogeneity, comparisons of first- and second-order Markov models, and alternative interval segmentation schemes.

- **S4(b) Results.** This subsection reports the findings of the robustness and sensitivity analyses, including interpretation of how these results support the main empirical findings.

## S1. G-TEST

Supplementary Tables S1 and S2 report the average G-statistics and corresponding p-values for each stock in the HMC, MMC and LMC tiers across all intraday time-intervals, where the averages are computed over 12 trading days. In both tables, G-statistics remain high and all p-values satisfy $p \ll 0.05$, confirming that the null hypothesis of independence is decisively rejected for every scenario. These results imply that each limit order price change depends significantly on the preceding change, indicating clear temporal dependence in the sequences. Consequently, the application of the Markov property to analyze the limit order price change dynamics is strongly validated.

**Table S1.** Average G-statistic and p-value for ask-side limit order price change sequences of HMC, MMC and LMC stocks across intraday time-intervals.

| Market Capitalization | Stocks | Average G-Statistic ($\times 10^3$) | | | | | | P-Value |
|---|---|---|---|---|---|---|---|---|
| | | $T_1$ | $T_2$ | $T_3$ | $T_4$ | $T_5$ | $T_6$ | |
| HMC | AMZN | 40.044 | 19.625 | 22.893 | 22.326 | 17.078 | 24.656 | $\ll 0.05$ |
| | JNJ | 2.349 | 2.151 | 2.598 | 2.804 | 2.355 | 4.677 | $\ll 0.05$ |
| | JPM | 10.973 | 9.076 | 8.852 | 8.829 | 8.079 | 11.048 | $\ll 0.05$ |
| | MSFT | 46.267 | 43.583 | 47.107 | 39.035 | 30.544 | 42.886 | $\ll 0.05$ |
| | XOM | 10.340 | 10.215 | 11.045 | 9.256 | 8.068 | 12.979 | $\ll 0.05$ |
| MMC | ABBV | 2.645 | 2.413 | 2.519 | 2.236 | 2.153 | 3.853 | $\ll 0.05$ |
| | HSBC | 4.731 | 5.437 | 1.970 | 1.541 | 1.359 | 2.286 | $\ll 0.05$ |
| | NFLX | 24.316 | 12.574 | 14.149 | 11.283 | 8.884 | 14.789 | $\ll 0.05$ |
| | ORCL | 18.503 | 20.603 | 22.207 | 20.736 | 17.139 | 24.866 | $\ll 0.05$ |
| | PEP | 2.974 | 3.471 | 4.046 | 4.106 | 3.339 | 6.704 | $\ll 0.05$ |
| LMC | AVGO | 9.674 | 6.371 | 7.426 | 7.534 | 5.745 | 10.178 | $\ll 0.05$ |
| | BKNG | 1.282 | 1.063 | 1.292 | 1.058 | 0.838 | 1.548 | $\ll 0.05$ |
| | BMY | 2.557 | 2.809 | 3.143 | 3.040 | 2.991 | 6.197 | $\ll 0.05$ |
| | NKE | 2.476 | 2.468 | 3.109 | 3.137 | 2.465 | 3.943 | $\ll 0.05$ |
| | UNP | 2.259 | 2.511 | 2.649 | 2.871 | 2.765 | 4.349 | $\ll 0.05$ |

**Table S2.** Average G-statistic and p-value for bid-side limit order price change sequences of HMC, MMC and LMC stocks across intraday time-intervals.

| Market Capitalization | Stocks | Average G-Statistic ($\times 10^3$) | | | | | | P-Value |
|---|---|---|---|---|---|---|---|---|
| | | $T_1$ | $T_2$ | $T_3$ | $T_4$ | $T_5$ | $T_6$ | |
| HMC | AMZN | 41.863 | 21.829 | 23.483 | 21.734 | 17.677 | 27.929 | $\ll 0.05$ |
| | JNJ | 3.578 | 3.874 | 4.039 | 3.719 | 3.440 | 5.499 | $\ll 0.05$ |
| | JPM | 11.439 | 10.738 | 10.935 | 9.668 | 8.358 | 12.564 | $\ll 0.05$ |
| | MSFT | 59.585 | 52.452 | 62.233 | 50.127 | 40.676 | 55.766 | $\ll 0.05$ |
| | XOM | 10.538 | 10.931 | 12.455 | 9.973 | 9.355 | 14.144 | $\ll 0.05$ |
| MMC | ABBV | 2.839 | 2.525 | 3.209 | 2.717 | 2.273 | 4.842 | $\ll 0.05$ |
| | HSBC | 5.190 | 5.129 | 2.084 | 1.888 | 1.476 | 2.403 | $\ll 0.05$ |
| | NFLX | 20.176 | 13.424 | 13.727 | 11.200 | 9.725 | 14.805 | $\ll 0.05$ |
| | ORCL | 18.664 | 18.739 | 22.395 | 19.823 | 17.201 | 25.429 | $\ll 0.05$ |
| | PEP | 3.806 | 4.331 | 5.349 | 5.285 | 4.429 | 8.365 | $\ll 0.05$ |
| LMC | AVGO | 7.570 | 4.713 | 6.631 | 5.463 | 4.133 | 9.337 | $\ll 0.05$ |
| | BKNG | 1.273 | 0.846 | 1.059 | 0.943 | 0.867 | 1.481 | $\ll 0.05$ |
| | BMY | 2.872 | 3.830 | 4.711 | 4.131 | 3.674 | 7.407 | $\ll 0.05$ |
| | NKE | 2.419 | 2.562 | 3.622 | 2.930 | 2.341 | 5.010 | $\ll 0.05$ |
| | UNP | 2.375 | 2.794 | 3.428 | 2.851 | 2.484 | 5.156 | $\ll 0.05$ |

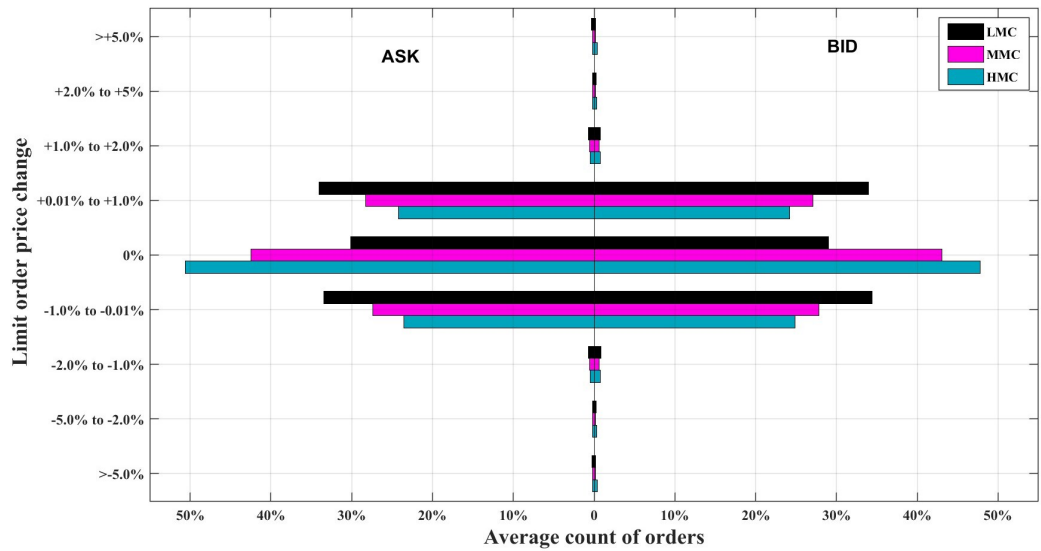

**Fig. S1.** Average count of ask and bid limit order price change states for HMC, MMC, and LMC stocks.

## S2. TRANSITION PROBABILITY MATRICES

Figures S2 & S3 and S4 & S5 present the transition probability matrices of limit order price change states for the ask and bid sides, evaluated at different intraday time-intervals across the three market capitalization tiers.

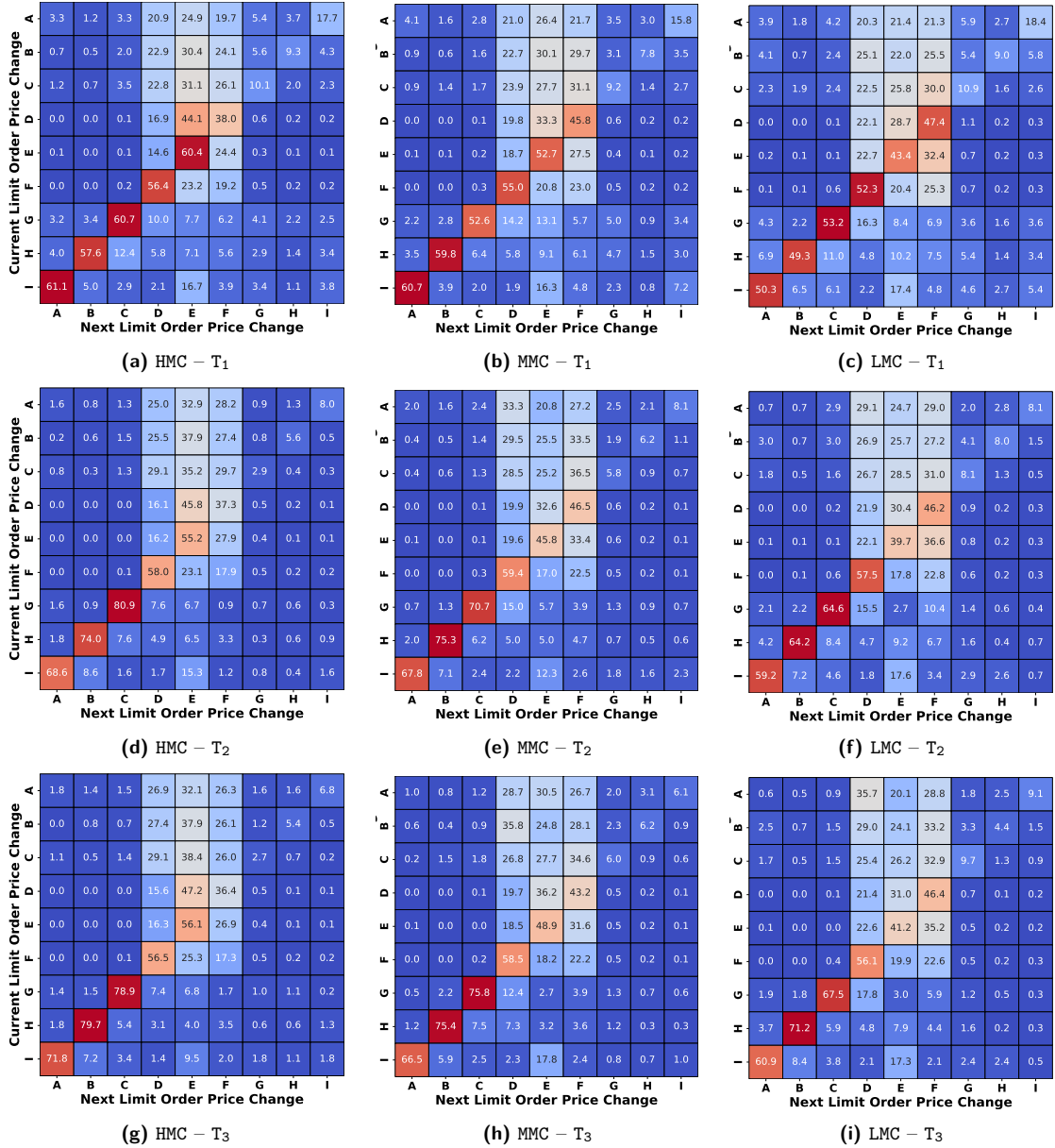

**Fig. S2.** Transition probability matrices for ask-side limit order price change states across intraday time-intervals for HMC, MMC and LMC capitalization stocks.

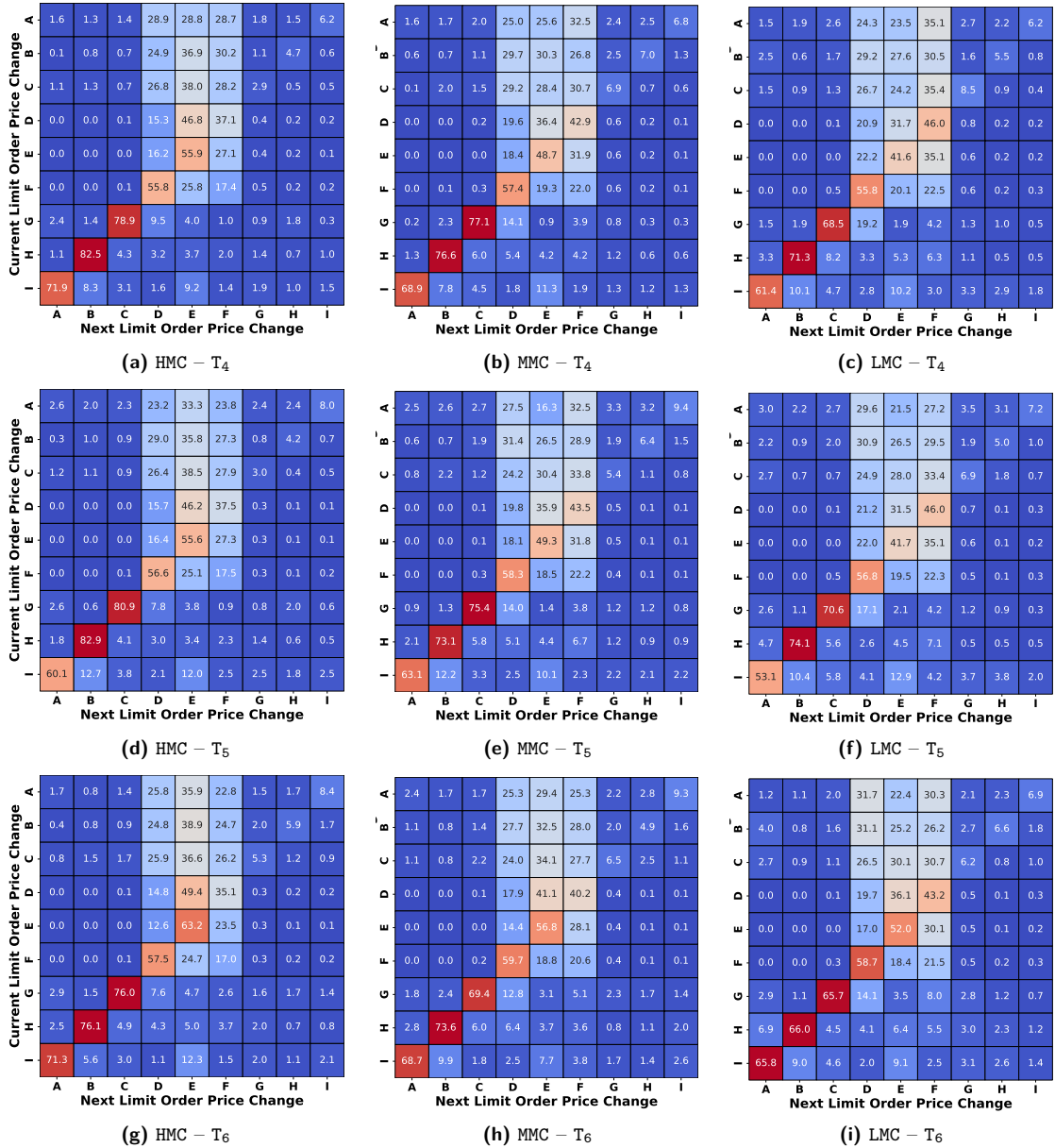

**Fig. S3.** Transition probability matrices for ask-side limit order price change states across intraday time-intervals for HMC, MMC and LMC capitalization stocks.

5

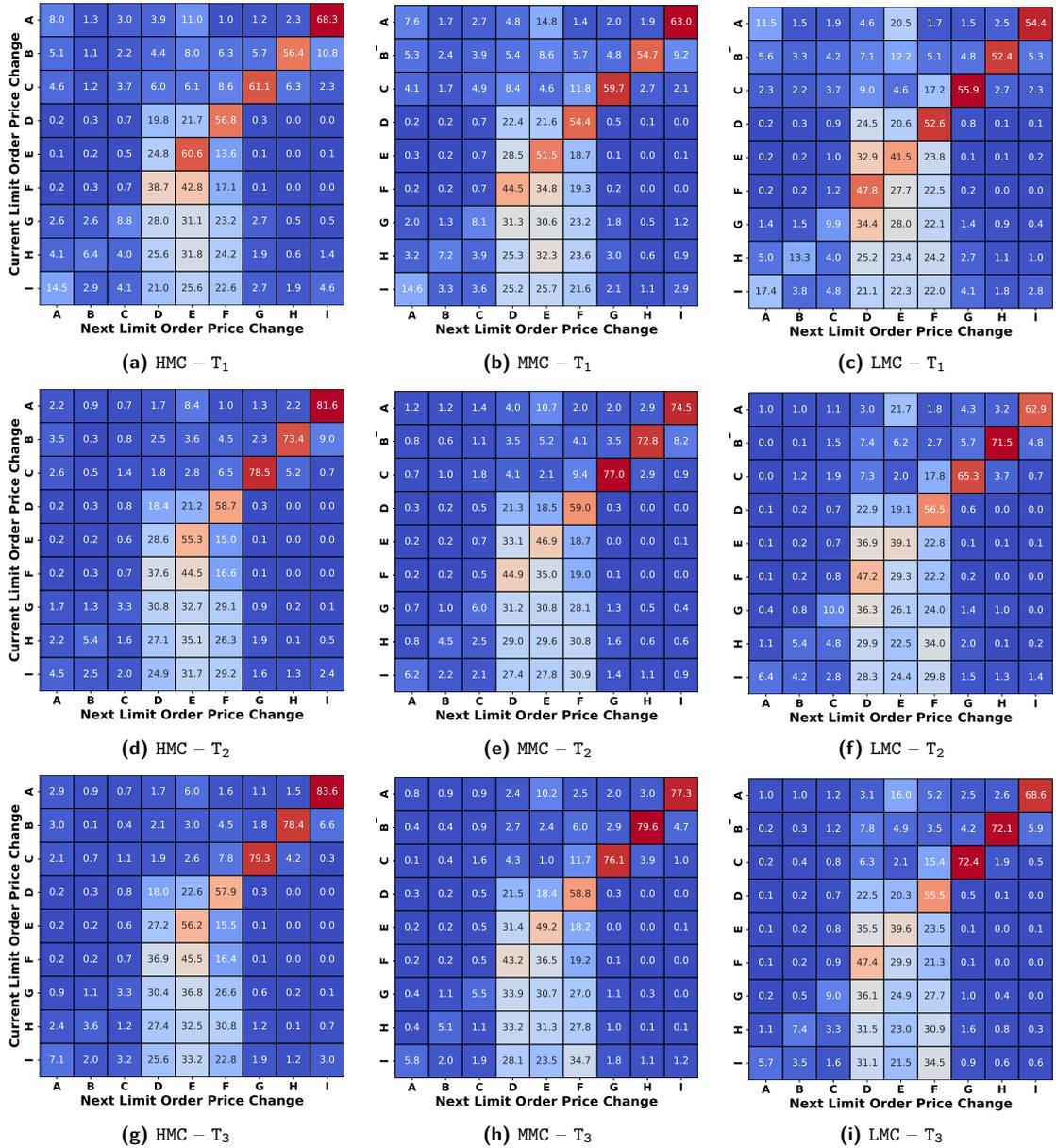

**Fig. S4.** Transition probability matrices for bid-side limit order price change states across intraday time-intervals for HMC, MMC and LMC capitalization stocks.

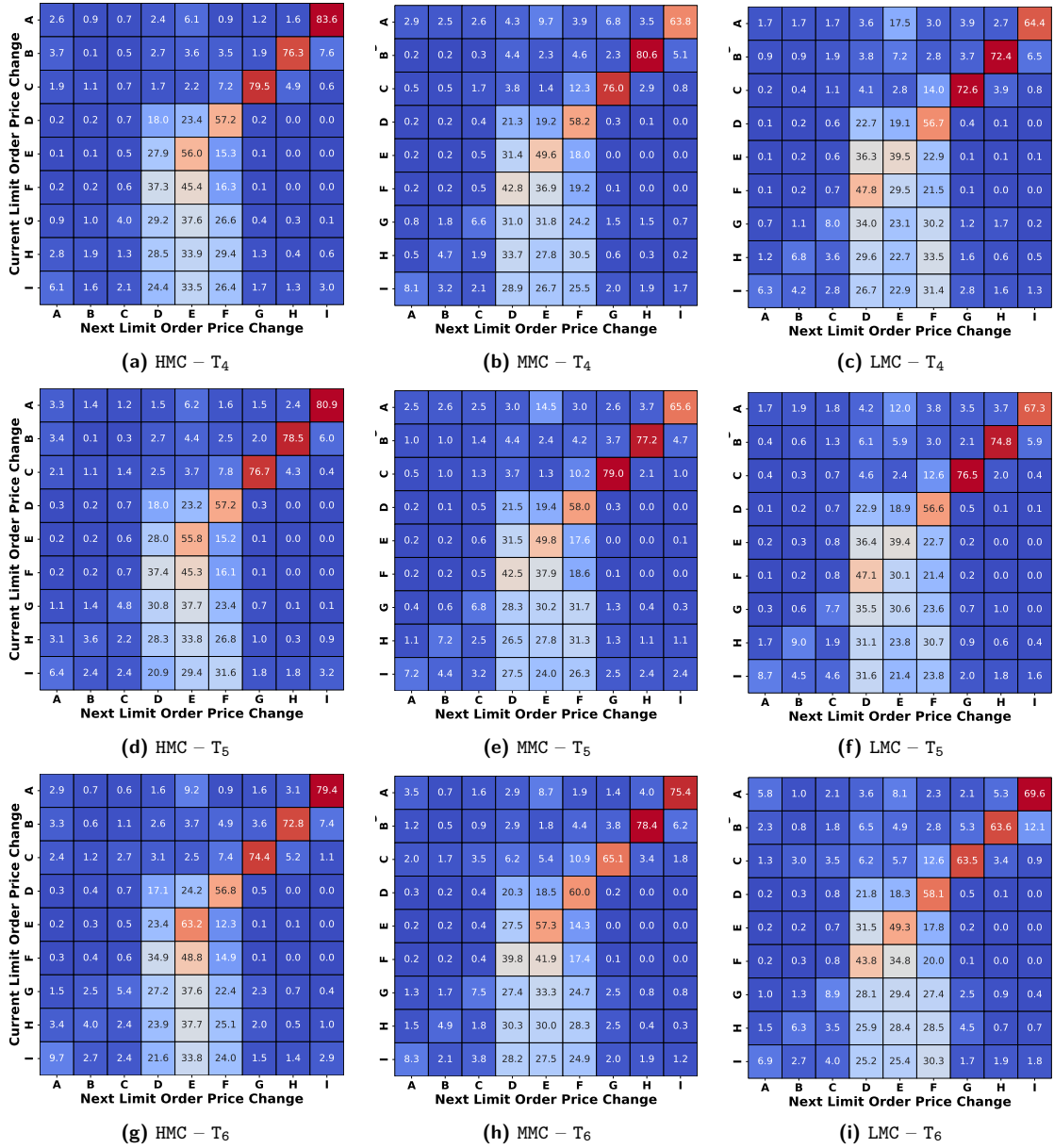

**Fig. S5.** Transition probability matrices for bid-side limit order price change states across intraday time-intervals for HMC, MMC and LMC capitalization stocks.

## S4. STATIONARY DISTRIBUTION OF PRICE CHANGE STATES

Tables S3 and S4 present the stationary distributions of limit order price change states at different intraday time-intervals for the three market capitalization tiers, separately for the ask and bid quotes. The long-run probability distributions are compared across time-intervals and capitalization tiers for both quotes using the Jensen–Shannon divergence metric, as described in the main manuscript.

**Table S3.** Stationary distributions of ask-side limit order price change states across intraday time-intervals for HMC, MMC and LMC capitalization stocks.

| Time Interval | Market Tier | Stationary Distribution | | | | | | | | |
|---|---|---|---|---|---|---|---|---|---|---|
| | | $\pi_A$ | $\pi_B$ | $\pi_C$ | $\pi_D$ | $\pi_E$ | $\pi_F$ | $\pi_G$ | $\pi_H$ | $\pi_I$ |
| $T_1$ | HMC | 0.0016 | 0.0018 | 0.0041 | 0.2653 | 0.4492 | 0.2661 | 0.0047 | 0.0017 | 0.0018 |
| | MMC | 0.0020 | 0.0020 | 0.0062 | 0.3109 | 0.3497 | 0.3179 | 0.0063 | 0.0020 | 0.0022 |
| | LMC | 0.0027 | 0.0025 | 0.0077 | 0.3321 | 0.3010 | 0.3389 | 0.0082 | 0.0023 | 0.0030 |
| $T_2$ | HMC | 0.0015 | 0.0018 | 0.0041 | 0.2684 | 0.4463 | 0.2664 | 0.0046 | 0.0016 | 0.0016 |
| | MMC | 0.0014 | 0.0020 | 0.0060 | 0.3187 | 0.3423 | 0.3201 | 0.0060 | 0.0020 | 0.0019 |
| | LMC | 0.0022 | 0.0023 | 0.0078 | 0.3396 | 0.2971 | 0.3416 | 0.0078 | 0.0022 | 0.0027 |
| $T_3$ | HMC | 0.0015 | 0.0017 | 0.0042 | 0.2676 | 0.4475 | 0.2683 | 0.0046 | 0.0017 | 0.0017 |
| | MMC | 0.0014 | 0.0019 | 0.0058 | 0.3169 | 0.3460 | 0.3197 | 0.0057 | 0.0020 | 0.0018 |
| | LMC | 0.0022 | 0.0022 | 0.0075 | 0.3385 | 0.2982 | 0.3415 | 0.0073 | 0.0022 | 0.0027 |
| $T_4$ | HMC | 0.0016 | 0.0019 | 0.0042 | 0.2655 | 0.4488 | 0.2685 | 0.0046 | 0.0017 | 0.0017 |
| | MMC | 0.0015 | 0.0020 | 0.0057 | 0.3155 | 0.3492 | 0.3186 | 0.0058 | 0.0021 | 0.0018 |
| | LMC | 0.0022 | 0.0023 | 0.0075 | 0.3379 | 0.2996 | 0.3420 | 0.0076 | 0.0022 | 0.0027 |
| $T_5$ | HMC | 0.0016 | 0.0018 | 0.0042 | 0.2672 | 0.4477 | 0.2674 | 0.0046 | 0.0017 | 0.0017 |
| | MMC | 0.0015 | 0.0019 | 0.0057 | 0.3165 | 0.3478 | 0.3192 | 0.0058 | 0.0020 | 0.0018 |
| | LMC | 0.0022 | 0.0023 | 0.0074 | 0.3377 | 0.2998 | 0.3421 | 0.0076 | 0.0022 | 0.0027 |
| $T_6$ | HMC | 0.0016 | 0.0019 | 0.0042 | 0.2639 | 0.4509 | 0.2672 | 0.0047 | 0.0018 | 0.0018 |
| | MMC | 0.0016 | 0.0020 | 0.0056 | 0.3148 | 0.3498 | 0.3179 | 0.0058 | 0.0020 | 0.0019 |
| | LMC | 0.0023 | 0.0024 | 0.0075 | 0.3380 | 0.2995 | 0.3422 | 0.0076 | 0.0023 | 0.0028 |

**Table S4.** Stationary distributions of bid-side limit order price change states across intraday time-intervals for HMC, MMC and LMC capitalization stocks.

| Time Interval | Market Tier | Stationary Distribution | | | | | | | | |
|---|---|---|---|---|---|---|---|---|---|---|
| | | $\pi_A$ | $\pi_B$ | $\pi_C$ | $\pi_D$ | $\pi_E$ | $\pi_F$ | $\pi_G$ | $\pi_H$ | $\pi_I$ |
| $T_1$ | HMC | 0.0034 | 0.0033 | 0.0072 | 0.2685 | 0.4439 | 0.2613 | 0.0063 | 0.0028 | 0.0034 |
| | MMC | 0.0038 | 0.0027 | 0.0082 | 0.3105 | 0.3628 | 0.2986 | 0.0075 | 0.0023 | 0.0036 |
| | LMC | 0.0037 | 0.0033 | 0.0119 | 0.3442 | 0.2885 | 0.3305 | 0.0111 | 0.0031 | 0.0037 |
| $T_2$ | HMC | 0.0026 | 0.0029 | 0.0074 | 0.2789 | 0.4185 | 0.2767 | 0.0074 | 0.0029 | 0.0026 |
| | MMC | 0.0026 | 0.0021 | 0.0063 | 0.3268 | 0.3316 | 0.3196 | 0.0064 | 0.0021 | 0.0025 |
| | LMC | 0.0010 | 0.0020 | 0.0084 | 0.3517 | 0.2812 | 0.3438 | 0.0086 | 0.0023 | 0.0010 |
| $T_3$ | HMC | 0.0026 | 0.0025 | 0.0070 | 0.2706 | 0.4330 | 0.2721 | 0.0068 | 0.0026 | 0.0026 |
| | MMC | 0.0025 | 0.0022 | 0.0056 | 0.3170 | 0.3478 | 0.3142 | 0.0058 | 0.0025 | 0.0025 |
| | LMC | 0.0012 | 0.0025 | 0.0089 | 0.3466 | 0.2899 | 0.3378 | 0.0092 | 0.0025 | 0.0014 |
| $T_4$ | HMC | 0.0022 | 0.0018 | 0.0063 | 0.2744 | 0.4344 | 0.2707 | 0.0062 | 0.0019 | 0.0022 |
| | MMC | 0.0023 | 0.0022 | 0.0051 | 0.3151 | 0.3546 | 0.3105 | 0.0055 | 0.0026 | 0.0020 |
| | LMC | 0.0014 | 0.0025 | 0.0069 | 0.3509 | 0.2838 | 0.3426 | 0.0076 | 0.0028 | 0.0015 |
| $T_5$ | HMC | 0.0027 | 0.0021 | 0.0071 | 0.2748 | 0.4317 | 0.2697 | 0.0070 | 0.0023 | 0.0026 |
| | MMC | 0.0022 | 0.0020 | 0.0061 | 0.3139 | 0.3593 | 0.3060 | 0.0066 | 0.0020 | 0.0019 |
| | LMC | 0.0017 | 0.0026 | 0.0083 | 0.3489 | 0.2848 | 0.3395 | 0.0094 | 0.0028 | 0.0019 |
| $T_6$ | HMC | 0.0033 | 0.0036 | 0.0062 | 0.2433 | 0.4916 | 0.2381 | 0.0069 | 0.0037 | 0.0034 |
| | MMC | 0.0028 | 0.0024 | 0.0048 | 0.2866 | 0.4107 | 0.2830 | 0.0045 | 0.0025 | 0.0028 |
| | LMC | 0.0028 | 0.0034 | 0.0089 | 0.3179 | 0.3393 | 0.3129 | 0.0087 | 0.0033 | 0.0029 |

## S4. ROBUSTNESS ANALYSES

### S4 (a). Robustness Analyses Methodologies

This subsection presents detailed methodologies for robustness analyses to validate the main findings of our study on intraday limit order price change transitions. We perform four key robustness checks: (a1) comparison with second-order Markov chains, (a2) time-homogeneity validation within intervals, (a3) sensitivity to interval duration, and (a4) sensitivity to equal versus unequal interval segmentation, as given below.

### S4 (a1). Comparison with Second-Order Markov Chains

While we employ first-order Markov chains based on statistical validation via G-test and autocorrelation analysis, it is important to assess whether second-order dependencies provide substantial improvements in model fit and predictive performance. This analysis justifies the parsimony of the first-order assumption.

With state space $\{1, \ldots, 9\}$ and sequence $(X_t)_{t=1}^T$, the two Markov orders are defined as:

$$\text{First-order:} \quad P(X_t = j \mid X_{t-1} = i) = p_{ij}^{(1)}, \tag{S1}$$

$$\text{Second-order:} \quad P(X_t = k \mid X_{t-1} = j, X_{t-2} = i) = p_{ijk}^{(2)}. \tag{S2}$$

For maximum-likelihood estimation (MLE), we define the following transition and history counts:

$$c_{ij} = \sum_{t=2}^{T} \mathbb{I}\{X_{t-1} = i, X_t = j\}, \tag{S3}$$

$$c_{ijk} = \sum_{t=3}^{T} \mathbb{I}\{X_{t-2} = i, X_{t-1} = j, X_t = k\}, \tag{S4}$$

and the row totals are defined by $c_{i\cdot} = \sum_j c_{ij}$ and $c_{ij\cdot} = \sum_k c_{ijk}$. The MLEs are

$$\hat{p}_{ij}^{(1)} = \frac{c_{ij}}{c_{i\cdot}} \quad \text{for } c_{i\cdot} > 0, \tag{S5}$$

$$\hat{p}_{ijk}^{(2)} = \frac{c_{ijk}}{c_{ij\cdot}} \quad \text{for } c_{ij\cdot} > 0. \tag{S6}$$

**In-sample performance:** The log-likelihood for the first-order model is:

$$\ell_1 = \sum_{i=1}^{9}\sum_{j=1}^{9} c_{ij}\log\bigl(\hat{p}_{ij}^{(1)}\bigr) = \sum_{i=1}^{9}\sum_{j=1}^{9}\left[c_{ij}\log(c_{ij}) - c_{ij}\log\left(\sum_{j'} c_{ij'}\right)\right], \tag{S7}$$

and the log-likelihood for the second-order model is:

$$\ell_2 = \sum_{i=1}^{9}\sum_{j=1}^{9}\sum_{k=1}^{9} c_{ijk}\log\bigl(\hat{p}_{ijk}^{(2)}\bigr) = \sum_{i=1}^{9}\sum_{j=1}^{9}\sum_{k=1}^{9}\left[c_{ijk}\log(c_{ijk}) - c_{ijk}\log\left(\sum_{k'} c_{ijk'}\right)\right], \tag{S8}$$

with the usual convention that terms with $c_{ij} = 0$ or $c_{ijk} = 0$ contribute zero.

We compute Akaike Information Criterion (AIC) and Bayesian Information Criterion (BIC) for model comparison [1], [2], [3]:

$$\text{AIC}_1 = 2k_1 - 2\ell_1, \quad \text{AIC}_2 = 2k_2 - 2\ell_2, \tag{S9}$$

$$\text{BIC}_1 = k_1\log(N) - 2\ell_1, \quad \text{BIC}_2 = k_2\log(N) - 2\ell_2, \tag{S10}$$

where $N = \sum_{i=1}^{9}\sum_{j=1}^{9}\sum_{k=1}^{9} c_{ijk}$ is the total number of triples, and $k_1$, $k_2$ are the effective numbers of parameters:

$$k_1 = \sum_{i \in \mathcal{I}_1} \max\bigl(m_i^{(1)} - 1,\, 0\bigr), \tag{S11}$$

$$k_2 = \sum_{(i,j) \in \mathcal{I}_2} \max\bigl(m_{ij}^{(2)} - 1,\, 0\bigr), \tag{S12}$$

where $\mathcal{I}_1$ and $\mathcal{I}_2$ are the sets of origin states (or state pairs) with non-zero transitions, and $m_i^{(1)}$, $m_{ij}^{(2)}$ are the numbers of destination states with positive counts.

The likelihood ratio test statistic for comparing nested models is [4], [5]:

$$G^2 = 2(\ell_2 - \ell_1) \sim \chi^2_{\text{df}}, \quad \text{df} = k_2 - k_1. \tag{S13}$$

**Information-theoretic measures:** The conditional mutual information quantifies the additional information provided by $X_{t-2}$ about $X_t$ given $X_{t-1}$ [6]:

$$I(X_{t-2}; X_t \mid X_{t-1}) = \frac{\ell_2 - \ell_1}{N}, \tag{S14}$$

expressed in nats per event. To convert to bits per event:

$$I_{\text{bits}} = \frac{I}{\log 2}. \tag{S15}$$

**Mathematical equivalence tests:** A key property is that the first-order transition probabilities should equal the marginal of the second-order probabilities [7], [8]. We verify:

$$\hat{p}_{jk}^{(1)} \stackrel{?}{=} \sum_{i=1}^{9} w_{i|j}\,\hat{p}_{ijk}^{(2)}, \tag{S16}$$

where the weights are:

$$w_{i|j} = \frac{c_{ij\cdot}}{\sum_{i'} c_{i'j\cdot}}, \quad c_{ij\cdot} = \sum_{k=1}^{9} c_{ijk}. \tag{S17}$$

10

The maximum absolute difference is:

$$\Delta_{\max} = \max_{j,k} \left| \hat{p}_{jk}^{(1)} - \sum_{i=1}^{9} w_{i|j}\, \hat{p}_{ijk}^{(2)} \right|. \quad (S18)$$

**Out-of-sample validation:** To assess predictive performance, we split each sequence into training (first 80%) and testing (last 20%) subsets. We employ Dirichlet smoothing with hyperparameter $\alpha$ to handle zero counts [9]:

$$\tilde{p}_{ij}^{(1)} = \frac{c_{ij} + \alpha}{\sum_{j'}(c_{ij'} + \alpha)}, \quad (S19)$$

$$\tilde{p}_{ijk}^{(2)} = \frac{c_{ijk} + \alpha}{\sum_{k'}(c_{ijk'} + \alpha)}. \quad (S20)$$

The out-of-sample log-likelihood per event is:

$$\ell_1^{\text{test}} = \frac{1}{N_{\text{test}}} \sum_{t \in \text{test}} \log\left(\tilde{p}_{X_{t-1},X_t}^{(1)}\right), \quad (S21)$$

$$\ell_2^{\text{test}} = \frac{1}{N_{\text{test}}} \sum_{t \in \text{test}} \log\left(\tilde{p}_{X_{t-2},X_{t-1},X_t}^{(2)}\right), \quad (S22)$$

where $N_{\text{test}}$ is the number of test-set transitions used for each model. The difference $\Delta\ell^{\text{test}} = \ell_2^{\text{test}} - \ell_1^{\text{test}}$ quantifies the predictive improvement of the second-order model. We report the mean, minimum, and maximum across multiple smoothing parameters $\alpha \in \{0.1, 0.5, 1.0\}$.

The first-order assumption ($M_1$) is justified by a convergence of evidence, prioritizing parsimony and practical performance. $M_1$ is validated if: (1) information criteria, particularly the more stringent $\text{BIC}_1$, favor it ($\text{BIC}_1 < \text{BIC}_2$); (2) the Conditional Mutual Information (**CMI**) is practically zero (e.g., $< 0.01$ bits/event), confirming $X_{t-2}$ adds negligible information; (3) the marginal maximum absolute difference check ($\Delta_{\max}$) is near machine-zero, confirming both models capture identical one-step dynamics; and (4) the out-of-sample (OOS) predictive gain (mean$_\alpha$ OOS $\Delta\ell^{\text{test}}$) is negligible or negative. A statistically significant $G^2$ ($p < 0.05$) is expected with large datasets [5] and does not invalidate $M_1$ if these practical, complexity-penalized, and OOS metrics support it [3], [10].

### S4 (a2). *Time-Homogeneity*

The discrete-time Markov chain framework assumes time-homogeneity within each intraday time interval, meaning that transition probabilities remain constant throughout the interval. This assumption is critical for valid inference. We test this assumption by dividing each interval into sub-intervals and comparing the resulting transition probability matrices.

For each trading interval $\tau \in \{\texttt{T}_1, \texttt{T}_2, \ldots, \texttt{T}_6\}$ and market capitalization tier $c \in \{\texttt{HMC}, \texttt{MMC}, \texttt{LMC}\}$, we load the sequence of price change states $\{X_t\}_{t=1}^T$ where $X_t \in \{1, 2, \ldots, 9\}$ represents the discretized price change category at time $t$. Each sequence is partitioned into $K$ sub-intervals of approximately equal length using the following procedure:

$$\text{Sub-interval } k: \quad \{X_t\}_{t=\ell_k}^{u_k}, \quad k = 1, 2, \ldots, K, \quad (S23)$$

where $\ell_k$ and $u_k$ denote the lower and upper indices of sub-interval $k$, determined by:

$$\ell_k = \left\lfloor \frac{(k-1)T}{K} \right\rfloor + 1, \quad u_k = \left\lfloor \frac{kT}{K} \right\rfloor. \quad (S24)$$

We use $K = 4$ sub-intervals as the default specification, ensuring each sub-interval contains sufficient transitions for stable estimation while providing adequate temporal resolution. For each sub-interval $k$, we construct the transition count matrix $\mathbf{C}^{(k)} = [c_{ij}^{(k)}]$ where

$$c_{ij}^{(k)} = \sum_{t=\ell_k}^{u_k - 1} \mathbb{I}\{X_t = i, X_{t+1} = j\}. \quad (S25)$$

The corresponding transition probability matrix is obtained via row normalization:

$$p_{ij}^{(k)} = \frac{c_{ij}^{(k)}}{\sum_{j'=1}^{9} c_{ij'}^{(k)}}, \quad \text{if } \sum_{j'=1}^{9} c_{ij'}^{(k)} > 0, \quad (S26)$$

and the pooled transition probability matrix across all sub-intervals is computed as:

$$\mathbf{C}^{\text{pool}} = \sum_{k=1}^{K} \mathbf{C}^{(k)}, \quad p_{ij}^{\text{pool}} = \frac{c_{ij}^{\text{pool}}}{\sum_{j'=1}^{9} c_{ij'}^{\text{pool}}}. \tag{S27}$$

**Mean Absolute Difference Metric:** We quantify practical homogeneity using the mean absolute difference (MAD) of transition probabilities across all pairs of sub-intervals:

$$\text{MAD} = \frac{2}{K(K-1)} \sum_{k=1}^{K-1} \sum_{k'=k+1}^{K} \frac{1}{81} \sum_{i=1}^{9} \sum_{j=1}^{9} \left| p_{ij}^{(k)} - p_{ij}^{(k')} \right|. \tag{S28}$$

A threshold of MAD $< 0.10$ is used to support practical time-homogeneity, following established guidelines for transition probability stability [11].

**Likelihood Ratio Test:** We perform a formal likelihood ratio test for homogeneity of transition matrices across sub-intervals. The test statistic is [12], [13]:

$$G = 2 \sum_{i=1}^{9} \sum_{j=1}^{9} \sum_{k=1}^{K} c_{ij}^{(k)} \log \left( \frac{c_{ij}^{(k)}}{E_{ij}^{(k)}} \right), \tag{S29}$$

where the expected count under the homogeneity hypothesis is:

$$E_{ij}^{(k)} = n_i^{(k)} p_{ij}^{\text{pool}}, \quad n_i^{(k)} = \sum_{j=1}^{9} c_{ij}^{(k)}. \tag{S30}$$

The degrees of freedom are computed as:

$$\text{df} = \sum_{i \in \mathcal{I}} (K-1)(m_i - 1), \tag{S31}$$

where $\mathcal{I} = \{i : \sum_{j=1}^{9} c_{ij}^{\text{pool}} > 0\}$ is the set of origin states with non-zero transitions in the pooled data, and $m_i = |\{j : c_{ij}^{\text{pool}} > 0\}|$ is the number of destination states with positive pooled counts for origin state $i$.

Under the null hypothesis of time-homogeneity, $G \sim \chi_{\text{df}}^2$. The *p*-value is computed as:

$$p = P(\chi_{\text{df}}^2 \geq G) = 1 - F_{\chi_{\text{df}}^2}(G). \tag{S32}$$

For each $(\tau, c)$ we apply the following decision rules: (i) if MAD $< 0.10$, the interval is considered practically time-homogeneous; (ii) the LRT *p*-value is reported for completeness but, given large-sample sensitivity [14], [15], it does not overturn the practical-homogeneity conclusion in (i) unless accompanied by MAD $\geq 0.10$; (iii) rows with zero pooled transitions are excluded from $G$ and df, and expected counts use pooled row-wise MLEs.

### S4 (a3). Sensitivity to Interval Duration

Our primary analysis uses six unequal-duration intervals based on established market microstructure patterns. To assess robustness, we compare transition probability matrices estimated from pooled one-hour intervals versus pooled two-hour intervals formed by concatenating consecutive one-hour sequences.

To compare the different aggregation durations, we define two distinct pooling methods:

- **One-Hour Pooling:** Estimate transition counts separately for each of the six one-hour files, then sum across all files:

$$\mathbf{C}^{(1h)} = \sum_{h=1}^{6} \mathbf{C}^{(h)}, \quad \mathbf{P}^{(1h)} = \text{RowNorm}(\mathbf{C}^{(1h)}). \tag{S33}$$

- **Two-Hour Pooling:** Concatenate consecutive hour pairs $\{(1,2), (3,4), (5,6)\}$ to form three two-hour sequences, estimate counts for each, then sum:

$$\mathbf{C}^{(2h)} = \sum_{p=1}^{3} \mathbf{C}^{(p)}, \quad \mathbf{P}^{(2h)} = \text{RowNorm}(\mathbf{C}^{(2h)}). \tag{S34}$$

We quantify the similarity between $\mathbf{P}^{(1h)}$ and $\mathbf{P}^{(2h)}$ using:

**Frobenius Norm:**

$$d_F = \left\|\mathbf{P}^{(1h)} - \mathbf{P}^{(2h)}\right\|_F = \sqrt{\sum_{i=1}^{9}\sum_{j=1}^{9}\left(p_{ij}^{(1h)} - p_{ij}^{(2h)}\right)^2}. \tag{S35}$$

**Maximum Absolute Difference:**

$$d_{\max} = \max_{i,j}\left|p_{ij}^{(1h)} - p_{ij}^{(2h)}\right|. \tag{S36}$$

When concatenating hour pairs, the total number of transitions increases by the number of cross-boundary transitions. For a pair $(h, h+1)$ with sequences $\{X_t^{(h)}\}$ and $\{X_t^{(h+1)}\}$, concatenation adds exactly one transition: $(X_{T_h}^{(h)}, X_1^{(h+1)})$. The expected increase in total transitions is:

$$\Delta N = \sum_{p=1}^{3}\mathbb{I}\bigl(\text{both hours in pair } p \text{ are non-empty}\bigr). \tag{S37}$$

Our interpretation of the distance metrics is as follows [16], [17]: if $d_F < 10^{-4}$ and $d_{\max} < 10^{-4}$, the TPMs are considered numerically indistinguishable. If $d_F < 10^{-3}$ and $d_{\max} < 10^{-3}$, they are considered extremely similar and the aggregation choice is immaterial. Otherwise, noticeable differences exist, and the sensitivity should be reported.

### S4 (a4). Sensitivity to Equal versus Unequal Interval Segmentation

Our baseline analysis utilizes six unequal-duration intervals specifically designed to capture distinct market microstructure phases. To ensure our findings are not artifacts of this specific segmentation, we assess their sensitivity by comparing them against a uniform grid of equal-duration intervals spanning the same trading day.

To compare the different segmentation schemes, we define the following interval specifications:

- **Unequal Intervals (Baseline):** Intervals are 60 minutes long, except for the midday periods $\mathtt{T_3}$ (11:30–12:45) and $\mathtt{T_4}$ (12:45–14:00), which are extended to 75 minutes.

- **Equal Intervals:** The trading day is strictly partitioned into six equal 65-minute intervals.

We focus our sensitivity analysis on three specific transitions from the neutral state (state 5): $p_{5,5}$ (price inertia), $p_{5,4}$ (defensive positioning), and $p_{5,6}$ (aggressive positioning). For each interval $i$ and focal transition $(s, s')$, we use the following statistical measures to compare $\hat{p}_{ss'}^{\text{unequal},i}$ versus $\hat{p}_{ss'}^{\text{equal},i}$:

**Standard Errors (Binomial Approximation):**

$$\text{SE}(\hat{p}_{ss'}) = \sqrt{\frac{\hat{p}_{ss'}(1-\hat{p}_{ss'})}{n_s}}, \tag{S38}$$

where $n_s$ is the number of observed transitions out of state $s$ in the corresponding setting.

**Test Statistic ($z$) and p-value:**

$$z_{ss'}^i = \frac{\hat{p}_{ss'}^{\text{unequal},i} - \hat{p}_{ss'}^{\text{equal},i}}{\sqrt{\left[\text{SE}\bigl(\hat{p}_{ss'}^{\text{unequal},i}\bigr)\right]^2 + \left[\text{SE}\bigl(\hat{p}_{ss'}^{\text{equal},i}\bigr)\right]^2}}, \quad p_{ss'}^i = 2\Phi\bigl(-|z_{ss'}^i|\bigr), \tag{S39}$$

where $\Phi(\cdot)$ denotes the standard normal CDF.

**Effect Size (Cohen's $h$):**

$$h_{ss'}^i = 2\left[\arcsin\left(\sqrt{\hat{p}_{ss'}^{\text{unequal},i}}\right) - \arcsin\left(\sqrt{\hat{p}_{ss'}^{\text{equal},i}}\right)\right]. \tag{S40}$$

Since this analysis involves 18 simultaneous comparisons (3 transitions × 6 intervals), we apply the Benjamini-Hochberg procedure to control the False Discovery Rate (FDR) at $\alpha = 0.05$ [18].

Our interpretation of these sensitivity metrics is as follows [16], [17]: we deem the baseline results robust if the mean absolute differences are small ($< 0.01$), few significant differences remain after FDR correction, and effect sizes are negligible (mean $|h| < 0.2$). Substantial differences would indicate sensitivity to interval definition, necessitating cautious interpretation of temporal patterns.

### S4 (b). Robustness Analyses Results

The subsection presents the results of the robustness analyses that validates the findings of our study on limit order price change transition dynamics.

*S4 (b1). Comparison with Second-Order Markov Chains*

We assessed whether incorporating second-order dependencies provides a substantial improvement over the parsimonious first-order model. Tables S5 and S6 present representative results for Amazon (AMZN) on November 7, 2018.

Consistent with the high-frequency nature of the data with large $N_{\text{triples}}$, formal in-sample statistical tests heavily favor the more complex second-order model. The likelihood ratio test statistics ($G^2$) are all highly significant ($p < 0.001$), and information criteria (AIC and BIC) generally confirm this statistical preference due to the sheer volume of observations. However, multiple alternative metrics demonstrate that these statistically significant gains are negligible in practice, justifying the sufficiency of the first-order assumption:

1. **Marginal Identity:** The maximum absolute difference ($\Delta_{\max}$) between the first-order MLE and the marginalized second-order mixture is extremely small, typically $< 10^{-3}$, and often $< 10^{-4}$. This confirms that the one-step transition dynamics are effectively identical under both models.

2. **Information Gain:** The Conditional Mutual Information (CMI) is very low, averaging approximately 0.057 bits per event for Ask and 0.056 bits per event for Bid sequences. This indicates that $X_{t-2}$ provides minimal additional predictive information once $X_{t-1}$ is known.

3. **Predictive Performance:** The out-of-sample (OOS) log-likelihood gains ($\text{mean}_\alpha \Delta \ell^{\text{test}}$) are minor, averaging only $\approx 0.03$ to $0.04$ nats per event.

4. **Complexity Cost:** The second-order model requires estimating approximately 4 times as many parameters ($k_2/k_1 \approx 3.8$ to $4.0$ on average), significantly increasing model complexity for marginal predictive return.

This convergence of evidence—where formal statistical significance is driven by sample size rather than effect size—was observed consistently across all 15 stocks and 12 trading days in our dataset. Therefore, we retain the first-order Markov chain as the preferred model due to its parsimony and demonstrated practical sufficiency.

**Table S5.** Second-Order vs. First-Order Comparison (AMZN, Ask Side, 2018-11-07)

| Interval | $N_{\text{triples}}$ | $G^2$ | CMI (bits) | $\Delta_{\max}$ | OOS $\Delta \ell$ | $k_2/k_1$ |
|---|---|---|---|---|---|---|
| $T_1$ | 74,085 | 7479.9 | 0.073 | 0.00111 | 0.021 | 5.6 |
| $T_2$ | 47,216 | 3489.0 | 0.053 | 0.00005 | 0.025 | 3.4 |
| $T_3$ | 48,445 | 3069.8 | 0.046 | 0.00003 | 0.028 | 2.7 |
| $T_4$ | 38,768 | 2503.0 | 0.047 | 0.00006 | 0.025 | 2.8 |
| $T_5$ | 32,408 | 2161.3 | 0.048 | 0.00006 | 0.017 | 3.0 |
| $T_6$ | 47,741 | 4858.9 | 0.073 | 0.00006 | 0.060 | 5.6 |

All $G^2$ values are significant at $p < 0.001$. OOS $\Delta \ell$ is the mean across $\alpha$.

**Table S6.** Second-Order vs. First-Order Comparison (AMZN, Bid Side, 2018-11-07)

| Interval | $N_{\text{triples}}$ | $G^2$ | CMI (bits) | $\Delta_{\max}$ | OOS $\Delta\ell$ | $k_2/k_1$ |
|:---:|:---:|:---:|:---:|:---:|:---:|:---:|
| $\mathtt{T_1}$ | 83,710 | 9435.0 | 0.081 | 0.00123 | 0.024 | 6.2 |
| $\mathtt{T_2}$ | 45,069 | 3061.3 | 0.049 | 0.00006 | 0.029 | 3.3 |
| $\mathtt{T_3}$ | 50,159 | 3457.4 | 0.050 | 0.00005 | 0.026 | 3.2 |
| $\mathtt{T_4}$ | 41,307 | 2292.8 | 0.040 | 0.00003 | 0.021 | 2.9 |
| $\mathtt{T_5}$ | 32,448 | 1920.0 | 0.043 | 0.00003 | 0.021 | 2.7 |
| $\mathtt{T_6}$ | 37,536 | 3874.4 | 0.074 | 0.00003 | 0.129 | 6.2 |

All $G^2$ values are significant at $p < 0.001$. OOS $\Delta\ell$ is the mean across $\alpha$.

### S4 (b2). Time-Homogeneity

We validated the assumption of time-homogeneity within each of the six intraday intervals $\mathtt{T_1}$–$\mathtt{T_6}$ by subdividing them into four smaller periods and comparing their transition structures. As a representative example, Tables S7 and S8 present the results for Amazon (AMZN) on a single trading day (November 7, 2018). This specific day and stock were selected from our wider dataset of 15 stocks across 12 trading days.

For this representative sample, every interval exhibited statistically significant deviations from perfect homogeneity according to the Likelihood Ratio Test ($G$). As shown in the tables, $p$-values are consistently less than 0.01. This statistical rejection is expected given the high frequency of the data, where intervals often contain within the range of 30,000 – 80,000 transitions ($N$), rendering the chi-squared test highly sensitive to minor fluctuations.

Crucially, however, these deviations are negligible in practice. The Mean Absolute Difference (MAD) between sub-interval transition matrices remained well below our conservative threshold of 0.10 for all intervals. For the Ask sequence in Table S7, the average MAD was 0.035. For the Bid sequence as shown in Table S8, it was 0.030.

This pattern—statistical significance due to large sample sizes but practical homogeneity indicated by low MAD scores—was consistent across the other remaining days and stocks in our dataset. Consequently, we conclude that the transition probabilities are sufficiently stable within each 60–75 minute trading interval to justify the use of time-homogeneous Markov chains for our primary analysis.

**Table S7.** Time-Homogeneity Test Results (AMZN, Ask Side, 2018-11-07)

| Interval | $N$ (Transitions) | $G$ Statistic | $p$-value | MAD | Result |
|:---:|---:|---:|:---:|:---:|:---:|
| $\mathtt{T_1}$ (09:30–10:30) | 74,083 | 8883.5 | $< 0.001^*$ | 0.051 | Valid |
| $\mathtt{T_2}$ (10:30–11:30) | 47,214 | 533.7 | $< 0.001^*$ | 0.023 | Valid |
| $\mathtt{T_3}$ (11:30–12:45) | 48,443 | 301.8 | $< 0.001^*$ | 0.022 | Valid |
| $\mathtt{T_4}$ (12:45–14:00) | 38,766 | 589.9 | $< 0.001^*$ | 0.026 | Valid |
| $\mathtt{T_5}$ (14:00–15:00) | 32,406 | 847.2 | $< 0.001^*$ | 0.039 | Valid |
| $\mathtt{T_6}$ (15:00–16:00) | 47,739 | 2761.8 | $< 0.001^*$ | 0.048 | Valid |

* Indicates statistical significance ($p < 0.05$). Valid if MAD $< 0.10$.

**Table S8.** Time-Homogeneity Test Results (AMZN, Bid Side, 2018-11-07)

| Interval | $N$ (Transitions) | $G$ Statistic | $p$-value | MAD | Result |
|---|---:|---:|---:|---:|---|
| $T_1$ (09:30–10:30) | 83,708 | 10673.6 | $< 0.001^*$ | 0.045 | Valid |
| $T_2$ (10:30–11:30) | 45,067 | 472.8 | $< 0.001^*$ | 0.020 | Valid |
| $T_3$ (11:30–12:45) | 50,157 | 504.0 | $< 0.001^*$ | 0.020 | Valid |
| $T_4$ (12:45–14:00) | 41,305 | 332.0 | $< 0.001^*$ | 0.024 | Valid |
| $T_5$ (14:00–15:00) | 32,446 | 185.5 | $0.007^*$ | 0.022 | Valid |
| $T_6$ (15:00–16:00) | 37,534 | 2027.0 | $< 0.001^*$ | 0.050 | Valid |

\* Indicates statistical significance ($p < 0.05$). Valid if MAD $< 0.10$.

### S4 (b3). Sensitivity to Interval Duration

We assessed the robustness of our baseline intraday segmentation comprising 60- and 75-minute intervals by comparing the resulting pooled Transition Probability Matrices (TPMs) against those derived from concatenating consecutive interval pairs.

Using AMZN on November 7, 2018, as a representative example, the differences between the TPMs estimated from the baseline single intervals ($\mathbf{P}^{(1h)}$) versus paired intervals ($\mathbf{P}^{(2h)}$) pooled data were found to be negligible in practice. For the Ask sequence, the Frobenius norm of the difference matrix was $d_F = 1.9 \times 10^{-5}$ and the maximum absolute difference was $d_{\max} = 1.5 \times 10^{-5}$. Similarly, for the Bid sequence, we observed $d_F = 1.8 \times 10^{-5}$ and $d_{\max} = 1.3 \times 10^{-5}$.

All these values are well below our conservative threshold of $10^{-4}$, classifying the matrices as numerically indistinguishable. Furthermore, the total transition counts differ only by exactly the expected number of cross-boundary transitions created by concatenation (3 pairs × 1 transition = 3), confirming data consistency. This extreme similarity indicates that our primary findings are not sensitive to the choice between the baseline aggregation window and longer, concatenated blocks.

### S4 (b4). Sensitivity to Equal versus Unequal Interval Segmentation

We evaluated whether our baseline choice of unequal-duration trading intervals introduces systematic bias by comparing it against a uniform 65-minute segmentation. We focused on three critical transitions from the neutral state (state 5): inertia ($p_{5,5}$), defensive positioning ($p_{5,4}$), and aggressive positioning ($p_{5,6}$).

Tables S9 and S10 present the interval-by-interval comparison for Amazon (AMZN) on November 7, 2018. The results indicate a high degree of robustness to the segmentation scheme. For the Ask side (Table S9), the transition probabilities estimated from unequal intervals closely track those from equal intervals. The maximum absolute difference observed for any single transition across all intervals was only 0.0087 (for $p_{5,5}$ during the midday period).

Similarly, the Bid side (Table S10) shows remarkable consistency. The mean absolute difference across all intervals for price inertia ($p_{5,5}$) was just 0.004. Crucially, after applying the Benjamini-Hochberg correction for multiple testing, *none* of the 36 compared pairs (18 for Ask, 18 for Bid) showed a statistically significant difference ($p_{\text{adj}} > 0.05$ in all cases).

This analysis confirms that the specific boundaries of our baseline intervals do not artificially generate the observed intraday patterns. The market microstructure dynamics captured by our model remain stable regardless of whether standard 65-minute or tailored 60–75 minute windows are employed.

**Table S9.** Sensitivity Comparison: Unequal vs. Equal Intervals (AMZN, Ask Side)

| Baseline Interval | Inertia ($p_{5,5}$) | | Defensive ($p_{5,4}$) | | Aggressive ($p_{5,6}$) | |
|---|---|---|---|---|---|---|
| | Uneq. | Equal | Uneq. | Equal | Uneq. | Equal |
| $T_1$ (09:30–10:30) | 0.683 | 0.683 | 0.097 | 0.097 | 0.202 | 0.202 |
| $T_2$ (10:30–11:30) | 0.486 | 0.485 | 0.149 | 0.149 | 0.344 | 0.343 |
| $T_3$ (11:30–12:45) | 0.506 | 0.507 | 0.140 | 0.141 | 0.342 | 0.341 |
| $T_4$ (12:45–14:00) | 0.472 | 0.480 | 0.137 | 0.132 | 0.377 | 0.373 |
| $T_5$ (14:00–15:00) | 0.463 | 0.464 | 0.139 | 0.139 | 0.378 | 0.375 |
| $T_6$ (15:00–16:00) | 0.561 | 0.554 | 0.111 | 0.112 | 0.287 | 0.294 |

Note: All differences are statistically non-significant after FDR correction.

**Table S10.** Sensitivity Comparison: Unequal vs. Equal Intervals (AMZN, Bid Side)

| Baseline Interval | Inertia ($p_{5,5}$) | | Defensive ($p_{5,4}$) | | Aggressive ($p_{5,6}$) | |
|---|---|---|---|---|---|---|
| | Uneq. | Equal | Uneq. | Equal | Uneq. | Equal |
| $T_1$ (09:30–10:30) | 0.665 | 0.665 | 0.210 | 0.210 | 0.104 | 0.104 |
| $T_2$ (10:30–11:30) | 0.454 | 0.460 | 0.366 | 0.362 | 0.142 | 0.140 |
| $T_3$ (11:30–12:45) | 0.490 | 0.492 | 0.342 | 0.340 | 0.140 | 0.139 |
| $T_4$ (12:45–14:00) | 0.472 | 0.467 | 0.376 | 0.380 | 0.136 | 0.136 |
| $T_5$ (14:00–15:00) | 0.454 | 0.454 | 0.388 | 0.387 | 0.144 | 0.145 |
| $T_6$ (15:00–16:00) | 0.490 | 0.480 | 0.349 | 0.358 | 0.136 | 0.139 |

Note: All differences are statistically non-significant after FDR correction.



\end{document}